\documentclass[12pt]{article}
\usepackage{cite}
\usepackage{amsmath,amsfonts,amsthm,amssymb,slashed,url,kbordermatrix}
\usepackage[mathscr]{euscript}
\usepackage{amscd}
\usepackage{color}

\usepackage{ifpdf}
\ifpdf
  \usepackage[pdftex]{graphicx}
  \usepackage{epstopdf}
\else
  \usepackage[dvips]{graphicx}
\fi
\parskip=\baselineskip
\begin{document}
\def\bar{\overline}
\def\hat{\widehat}
\def\tilde{\widetilde}
\def\sign{{\mathrm{sign}}}
\def\Hom{\mathrm{Hom}}
\def\coker{{\mathrm{coker}}}
\def\dim{{\mathrm{dim}}}
\def\R{{\mathbb{R}}}
\def\C{{\mathbb{C}}}
\def\Z{{\mathbb{Z}}}
\def\CP{{\mathbb{CP}}}
\def\TT{\mathfrak T}

% --- equations   ----
\def\be{\begin{equation}}
\def\ee{\end{equation}}
\def\beq{\be\begin{array}{c}}
\def\eeq{\end{array}\ee}
\def\nn{\nonumber}

%-----greek letters ------
\def\b{{\beta}}
\def\a{{\alpha}}
\def\g{{ \gamma}}
\def\d{{\delta}}
\def\e{{\epsilon}}

%----math -------
\def\p{\partial}
\def\<{\langle}
\def\>{\rangle}
\def\cO{\mathcal{O}}
\def\co{\cO}
\def\l{\ell}

%---- Sergei's definitions -------
\def\CC{\mathcal{C}}
\def\CO{\mathcal{O}}
\def\CN{\mathcal{N}}
\def\CT{\mathcal{T}}
\def\CH{\mathcal{H}}
\def\CM{\mathcal{M}}
\def\CI{\mathcal{I}}
\def\CL{\mathcal{L}}
\def\CR{\mathcal{R}}
\def\CS{\mathcal{S}}
\def\CQ{\mathcal{Q}}
\newtheorem{theorem}{Theorem}[section]
\renewcommand{\kbldelim}{(}% Left delimiter
\renewcommand{\kbrdelim}{)}% Right delimiter

\font\teneurm=eurm10 \font\seveneurm=eurm7 \font\fiveeurm=eurm5
\newfam\eurmfam
\textfont\eurmfam=\teneurm \scriptfont\eurmfam=\seveneurm
\scriptscriptfont\eurmfam=\fiveeurm
\def\eurm#1{{\fam\eurmfam\relax#1}}
\font\teneusm=eusm10 \font\seveneusm=eusm7 \font\fiveeusm=eusm5
\newfam\eusmfam
\textfont\eusmfam=\teneusm \scriptfont\eusmfam=\seveneusm
\scriptscriptfont\eusmfam=\fiveeusm
\def\eusm#1{{\fam\eusmfam\relax#1}}
\font\tencmmib=cmmib10 \skewchar\tencmmib='177
\font\sevencmmib=cmmib7 \skewchar\sevencmmib='177
\font\fivecmmib=cmmib5 \skewchar\fivecmmib='177
\newfam\cmmibfam
\textfont\cmmibfam=\tencmmib \scriptfont\cmmibfam=\sevencmmib
\scriptscriptfont\cmmibfam=\fivecmmib
\def\cmmib#1{{\fam\cmmibfam\relax#1}}
\numberwithin{equation}{section}
\def\neg{\negthinspace}

\begin{titlepage}
\begin{flushright}
CALT 2016-019
\end{flushright}
\vskip 1.5in
\begin{center}
{\bf\Large{RG Flows and Bifurcations}}
\vskip0.5cm
{\bf
Sergei Gukov$^{a,b}$}
\\
\vskip.5cm
$^a$ {\small{\it Walter Burke Institute for Theoretical Physics, California Institute of Technology, Pasadena, CA 91125 USA}}\\
$^b$ {\small{\it Max-Planck-Institut f\"ur Mathematik, Vivatsgasse 7, D-53111 Bonn, Germany}}
\end{center}
\vskip.5cm
\baselineskip 16pt
\begin{abstract}
Interpreting RG flows as dynamical systems in the space of couplings
we produce a variety of constraints, global (topological) as well as local.
These constraints, in turn, rule out some of the proposed RG flows
and also predict new phases and fixed points, surprisingly, even in familiar theories such as $O(N)$ model, QED$_3$, or QCD$_4$.
\end{abstract}
\date{August, 2016}
\end{titlepage}

%\newpage
\tableofcontents
%\newpage

\section{Motivation}
\label{sec:intro}

\subsection{Spectra and Flows}

Aiming for a non-perturbative description of RG flows \cite{Gross:1997bc}, it was proposed in \cite{Gukov:2015qea}
to view spectra of the UV and IR theories as measuring degrees of freedom,
in a way similar to the standard $C$-function~\cite{Zamolodchikov:1986gt,Cardy:1988cwa,Komargodski:2011vj}.
Hopefully, comparing the spectra of the UV and IR theories can teach us useful lessons about RG flows
and provide information not captured by the $C$-function.

Here, ``spectra'' could mean several different things, and all options are interesting.
Thus, in the context of supersymmetric theories, it is natural to consider spectra of supersymmetric operators or states,
such as chiral rings, BPS states, {\it etc}.
Even though all these candidates have been extensively studied in the past 20 years, surprisingly,
the question of comparing them in the UV and the IR has not been emphasized.
Moreover, apart from different types of spectra, one could explore different types of relation between UV and IR objects.
For example, applying this philosophy to chiral rings
\be
\CR_{\text{UV}} \quad \xrightarrow[~]{~~\text{RG flow}~~} \quad \CR_{\text{IR}}
\label{Rtheorem}
\ee
one could ask if $\dim \CR_{\text{UV}} \le \dim \CR_{\text{IR}}$ always holds or, if not,
what physical consequences of violating this bound are.
A stronger version of such ``$\CR$-theorem'' might look like $\CR_{\text{IR}} \subset \CR_{\text{UV}}$
or a similar relation that goes beyond numbers.

Similarly, and staying for a moment with supersymmetric theories, the ``spectrum'' could refer to the spectrum of states
annihilated by some supercharge $\CQ$ modulo $\CQ$-exact states (a.k.a. BPS states) on various branches of the superconformal theory.
Regarded as a characteristic of the superconformal theory itself, such BPS spectrum
is expected to ``loose'' states via a mechanism analogous to a spectral sequence \cite{Gukov:2015gmm}:
\be
\CH^{BPS}_{\text{UV}} \quad \xrightarrow[~~\text{spectral sequence}~~]{~~\text{?}~~} \quad \CH^{BPS}_{\text{IR}}
\label{Htheorem}
\ee
since the supersymmetry algebra and, as part of it, the supercharge $\CQ$ are deformed upon the RG flow.
Here, and also in the $\CR$-theorem \eqref{Rtheorem}, the most dramatic change is discrete and, in particular,
requires flowing to the deep IR where some states / operators decouple at the very last stage.
There are many concrete examples of SUSY theories where the BPS spectrum is known exactly,
{\it e.g.} many examples of two-dimensional SUSY theories with and without boundary considered in \cite{Gukov:2015gmm}
support this form of the ``$\CH$-theorem'' \eqref{Htheorem}.
Pursuing this direction quickly leads to other interesting questions,
such as the ``flow'' of walls of marginal stability that separate different chambers.
For example, the structure of walls and the spectra of BPS states in each chamber are known
for $A_2$ and $A_3$ Argyres-Douglas theories \cite{Alim:2011kw}.
Although it points in the general direction of ``loosing BPS states'' along the RG flow, it would be interesting
to explore more precise relations along the lines of \eqref{Htheorem}.

In this paper, we consider most general non-supersymmetric RG flows,
deferring the study of additional structures associated with supersymmetry to future work.
Typical examples of such flows --- which will also be our examples here --- include
the RG flow in the $O(N)$ model in $d$ dimensions as well as RG flows in
strongly coupled gauge theories, such as the four-dimensional QCD and three-dimensional QED
(often denoted QCD$_4$ and QED$_3$, respectively).

Without further assumptions about supersymmetry, our options are more limited and
the ``spectrum'' could simply stand for the spectrum of all operators (or states).
Since the latter is ordered by conformal dimension $\Delta$, it is natural to aim for a finite-dimensional
version that, on the one hand, could be sufficiently simple to deal with and, on the other hand,
would hopefully capture interesting information about the RG flow in question.
But where do we draw the line?
In other words, when we make a comparison of the UV and IR spectra below a certain cutoff $\Delta_0$,
what value of $\Delta_0$ should we choose?

\begin{figure}[t!]
\begin{center}
\includegraphics[width=0.4\textwidth]{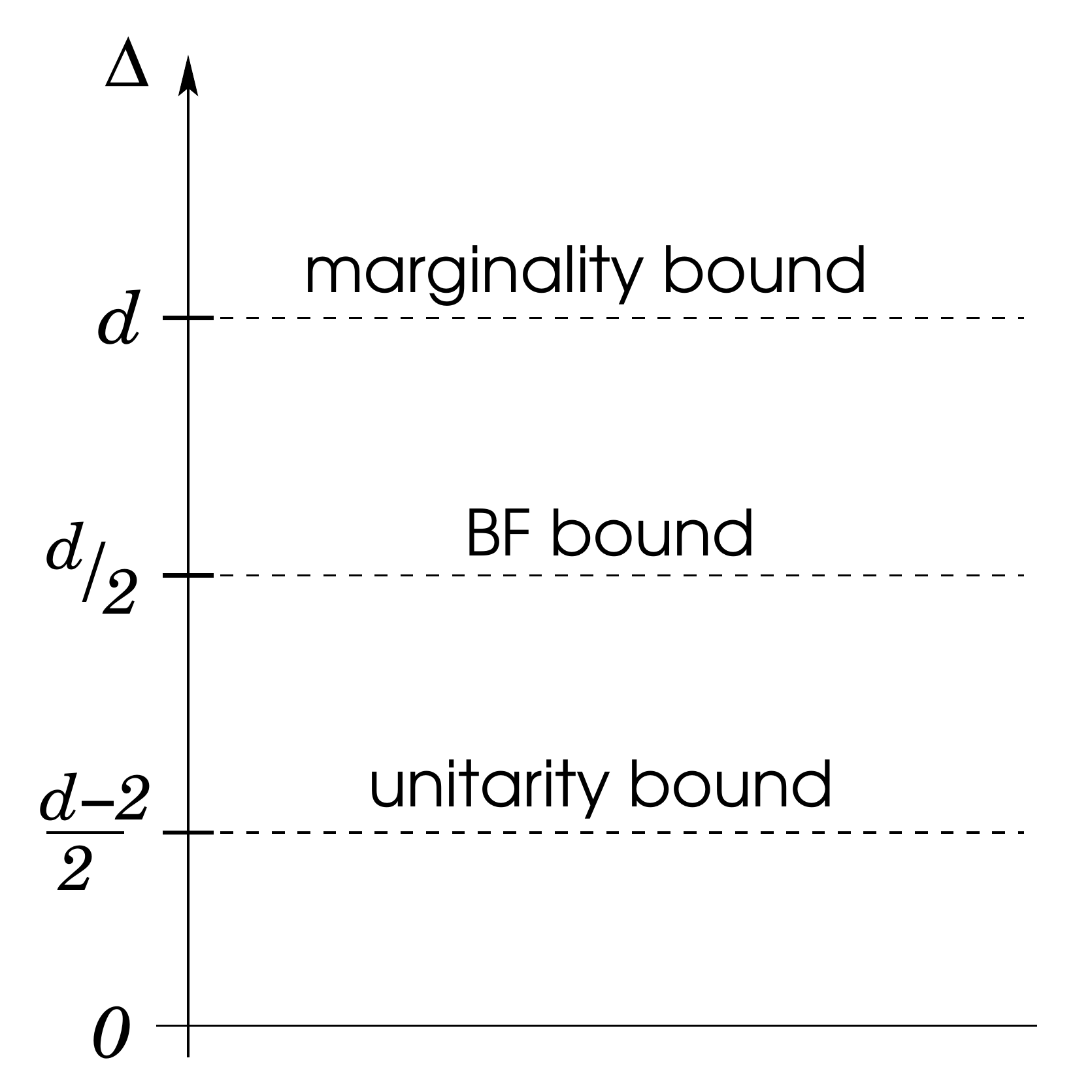}
\end{center}
\caption{\label{fig:scalardims} Various bounds on scaling dimensions for scalar operators in CFT$_d$.}
\end{figure}

On the scale of conformal dimensions, there are several natural benchmarks, illustrated in Figure~\ref{fig:scalardims}
for scalar operators of spin-0.
Starting with the lowest, $\Delta_{\text{min}} = \frac{d-2}{2}$ is the unitarity bound for scalar operators in $d$ space-time dimensions.
This cut-off is a bit too low for our purposes since in a unitary theory it would essentially lead to counting free fields.
The special value $\Delta = d$ is the ``marginality bound'' which will be our choice of the cutoff in this paper.
The scalar operators which are singlets (in theories with symmetries)
can be added to the Lagrangian without explicitly breaking any of the symmetries;
in particular, the operators with $\Delta (\CO) < d$ are relevant, while the operators with $\Delta (\CO) > d$ are irrelevant.
Hence, the part of the spectrum with $\Delta (\CO) < d$ can be conveniently characterized by the following quantity:
\be
\mu = \# \Big( \text{relevant spin-0 singlet}~\CO \Big)
\label{mudef}
\ee
which can be viewed as a measure of degrees of freedom in a CFT.
The remaining line in Figure~\ref{fig:scalardims}, namely $\Delta = \frac{d}{2}$, is what we call the BF bound because in a holographic dual
it would correspond to bulk scalar fields saturating the Breitenlohner-Freedman stability bound
$m^2 \ell^2 = \Delta (\Delta - d) \ge - \frac{d^2}{4}$.
{}From the CFT point of view, there is nothing special about operators with $\Delta (\CO) = \frac{d}{2}$,
except the fact that, in a weakly coupled theory, $\CO^2$ crosses the marginality precisely when $\CO$ reaches the BF bound.
This, however, effectively takes us back to the analysis of the spectrum below the marginality cutoff.

The operators below the marginality cutoff are also the most relevant ones from the Wilsonian point of view (no pun intended).
Indeed, since irrelevant operators do not destabilize a given conformal theory $T_*$ they can be integrated in or integrated out
without affecting the physics at the fixed point $T_*$ which, in turn, can be ``embedded'' in a larger field theory or even into string theory.
A similar sequence of embeddings is ubiquitous in holography, where a $(d+1)$-dimensional AdS dual arises
as a ``consistent truncation'' of ten- or eleven-dimensional supergravity which, in turn, is embedded in the full-fledged string theory.
This reasoning naturally leads to the idea of {\it universality} that turns out to be extremely useful in describing
real macroscopic systems whose physics is dominated by relevant operators and one has little control
over small effects due to irrelevant operators. From this perspective, a renormalization that violates the inequality
\be
\mu_{\text{UV}} \; > \; \mu_{\text{IR}}
\label{muuvir}
\ee
would almost undermine the ideas of universality and the Wilsonian approach because it would mean that some
irrelevant spin-0 singlet operators suddenly become relevant along the RG flow.
It has been argued that such peculiar RG flows can not be smooth \cite{Gukov:2015qea},
either themselves or in a larger family of flows.
Mathematically, the lack of smoothness is due to violation of the transversality (Morse-Smale) condition in the theory space $\CT$.
Physically, phenomena where certain quantities cease to be smooth are usually called {\it phase transitions}
and one of the main goals of this paper is to shed light on the nature of transitions that accompany marginality crossing.

In order to understand if there is anything special about RG flows that violate \eqref{muuvir} we need to examine carefully
simple concrete examples where this happens. We should have no difficulty finding such examples if marginality crossing
(a.k.a. dangerously irrelevant operators) is really abundant in quantum field theory.
Moreover, unless marginality crossing is inherent to free theories or theories with conformal manifolds (which is very hard to believe),
the simplest examples should be RG flows among isolated interacting CFTs ({\it cf.} minimal models in two dimensions)
where computing $\mu$ is especially clear and leads to a finite value.\footnote{In theories with conformal manifolds,
moduli spaces, or free fields the definition of $\mu$ requires extra care \cite{Gukov:2015qea}; a naive definition can give $\mu = \infty$.}
How easily does one find such examples? And how abandon they really are?

\subsection{Is marginality crossing difficult to find?}

On the one hand, irrelevant operators that upon renormalization become relevant
(also known as dangerously irrelevant operators) seem to be extremely rare,
which makes our task of finding a simple model that would shed light on their nature unexpectedly difficult.
Below we summarize their status in various dimensions and comment on the violation of \eqref{muuvir}.
Basically, the upshot is that a possibility of violating \eqref{muuvir} decreases in theories with
larger supersymmetry and larger values of $(d-4)^2$, where $d$ is the space-time dimension\footnote{The reader
may find it helpful to picture a distribution, such as a bell-shaped curve, centered around $d=4$ and with tails near $d=2$ and $d=6$.},
and the search for the {\it simplest} isolated interacting CFTs that are supposed to help us understand marginality crossing
takes us to strongly interacting theories on par with QCD$_4$:

\noindent
\underline{\bf d=2:}
This case is (by far) most well-understood.
In particular, both the weak version and the stronger version of the $C$-theorem are proved in two dimensions \cite{Zamolodchikov:1986gt},
and the strongest version is believed to hold \cite{Friedan:2009ik}.
There are no known RG flows among isolated interacting CFTs that violate \eqref{muuvir}.

\noindent
\underline{\bf d=3:}
This is one of the least understood cases, {\it e.g.} the proposed candidates for the $C$-function do not appear to be
stationary at the fixed points \cite{Klebanov:2012va} and a lot more work is needed before one can conclude whether
marginality crossing is easy to find in $2+1$ dimensions.

\noindent
\underline{\bf d=4:}
Four-dimensional theories and RG flows provide most interesting examples for our study.
In this case, the weak and the stronger versions of the $C$-theorem are known to hold \cite{Komargodski:2011vj}.
While supersymmetry helps to maintain analytical control over RG flows, it seems to suppress marginality crossing,
which is still possible in $\CN=1$ theories \cite{Gukov:2015qea},
but was conjectured not to exist in $\CN=2$ theories \cite{Argyres:2015ffa}.

\noindent
\underline{\bf d=5:}
This is another case where little is known.
In particular, we are not aware of any examples of marginality crossing in $4+1$ dimensions.

\noindent
\underline{\bf d=6:}
As we approach $d=6$, the structure of conformal theories becomes even more constrained and, in a way,
mirrors what happens at the lower end of $d$.
In fact, six-dimensional CFTs with $\CN=(0,1)$ SUSY or higher do not admit any relevant operators at all \cite{Cordova:2015fha}.
%As in the previous case, there are no relevant deformations in CFTs with $\CN=(0,1)$ SUSY or higher \cite{Cordova:2015fha}.
(There are, however, moduli-space flows in $d=6$.)

To summarize, using the field theory techniques, it seems that examples where irrelevant operators cross through marginality
are extremely rare and, roughly speaking, are centered around $d=4$ and low amount of supersymmetry.
In fact, no single weakly-coupled example of such phenomenon seems to be known, and all proposed candidates
rely on various assumptions, typically about the strongly-coupled dynamics, which, in turn, is more robust in supersymmetric theories.
Thus, a four-dimensional $\CN=1$ RG flow from a superconformal family of $A_3$ theories to $\CN=1$ SQCD
has a 4-quark operator that crosses through marginality \cite{Gukov:2015qea}.
However, for the purposes of understanding the physics of such phenomena
they are just as strongly interacting as ordinary, non-supersymmetric QED$_3$ or QCD$_4$
near the lower end of the conformal window, which we will use as our examples and where marginality crossing may indeed
be responsible for phase transitions and lead to dynamical symmetry breaking {\it a la} Nambu-Jona-Lasinio \cite{Nambu:1961tp,Nambu:1961fr}.

%%%%%%%%%%%%%%%%%%%%%%%%%%%%%%%%%%%%%%%%%%%%%%%%%%%%%%%%%%%%%%%%%%%%%%%%%%%%%%%%

\subsection{Is marginality crossing easy to find?}

\begin{figure}[t!]
\begin{center}
\includegraphics[width=0.6\textwidth]{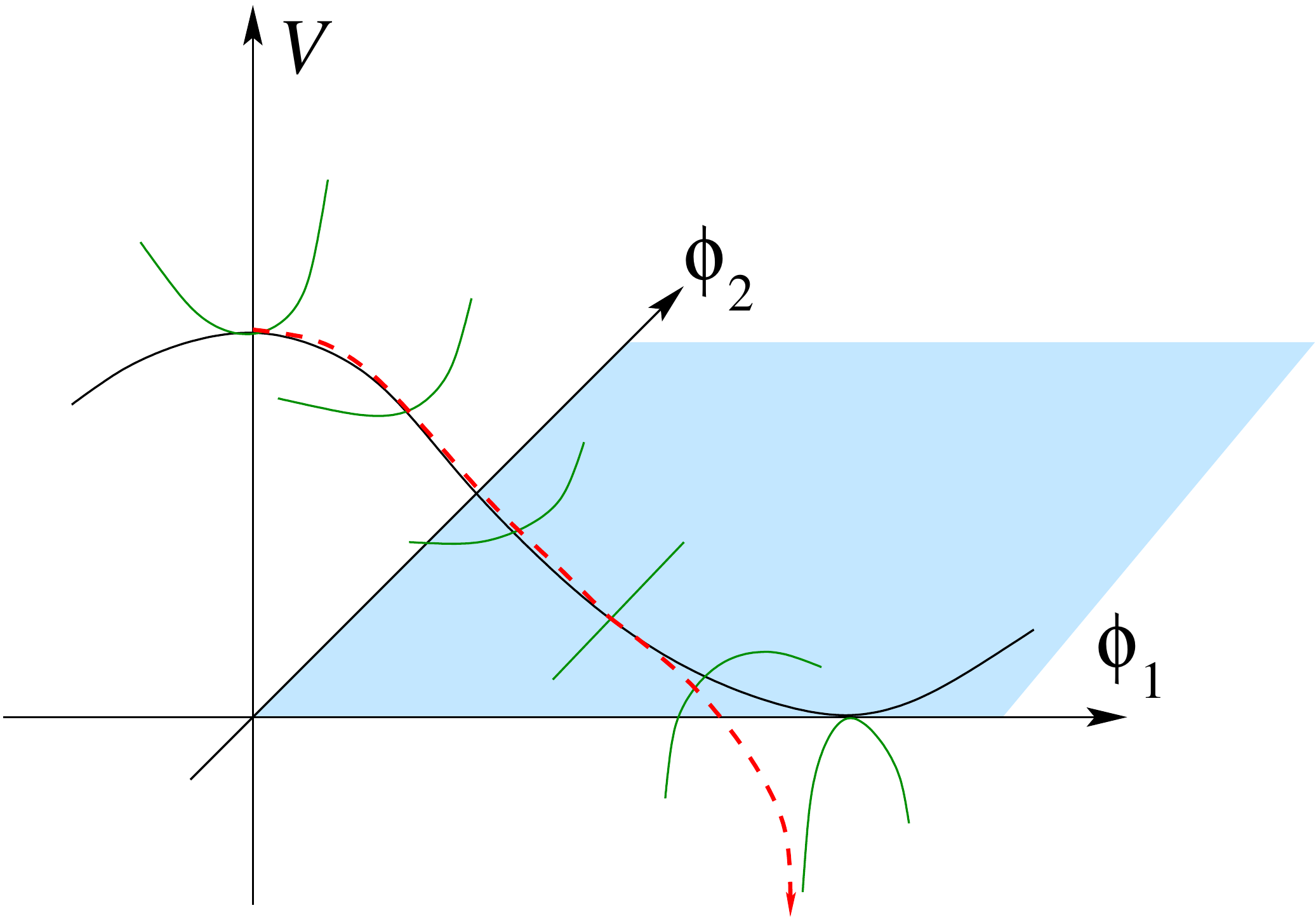}
\end{center}
\caption{\label{fig:gravityV} In holography, marginality crossing is realized by a model solution that
``rolls'' between two saddle points of a potential function $V(\phi_1, \phi_2)$.}
\end{figure}

On the other hand, from the holographic viewpoint, constructing RG flows with irrelevant operators
crossing through marginality appears to be incredibly easy (in any dimension and even in supersymmetric cases,
where field theory techniques tell us otherwise).
Indeed, in phenomenological models, including numerous applications to AdS/CMT,
one usually takes a $(d+1)$-dimensional gravity minimally coupled to scalar fields $\phi_i$
interacting via a potential $V(\phi)$:
\be
S \; = \; \int d^{d+1} x \sqrt{-g} \left( \frac{1}{4} R + \frac{1}{2} g^{\mu \nu} \partial_{\mu} \phi_i \partial_{\nu} \phi^i + V (\phi) \right)
\ee
The standard AdS/CFT dictionary \cite{Maldacena:1997re,Aharony:1999ti}
tells us that AdS vacua ({\it i.e.} critical points of the potential function with $V<0$)
correspond to conformal fixed points in the $d$-dimensional theory on the boundary,
mass eigenvalues of the scalar fields at the the critical point
determine the conformal dimensions of the corresponding primary operators, {\it etc.}
Therefore, in order to engineer a marginality crossing we only need to come up with a potential $V(\phi)$
such that the effective mass squared for one of the fields, say $\phi_2$, changes sign as the other field, say $\phi_1$,
``rolls'' between two vacua of $V(\phi)$, see Figure~\ref{fig:gravityV}:
\be
V (\phi) \; = \; V_0 + \frac{g}{4} (\phi_1^2 - a^2)^2 + (m^2 - C \phi_1) \phi_2^2 + \ldots
\ee
Here, $V_0$, $g$, $a$, $m$ and $C$ are some constants, such that $C, \, g>0$ and $0< \frac{m^2}{C} <a$.
The marginality crossing takes place at $\phi_1^{\text{crit}} = \frac{m^2}{C}$
when $\phi_1$ ``rolls'' from $\phi_1 = 0$ to another critical point $\phi_1 = a$.
(See also Figure~\ref{fig:MCTTT} for an illustration of this flow in the boundary theory.)

A reader may notice that we adopted the terminology as well as the form of this model potential from the hybrid inflation~\cite{Linde:2004kg},
where time evolution in the same theory of gravity coupled to scalar fields is used model the early universe cosmology
which ends abruptly with a {\it phase transition} and spontaneous symmetry breaking.
In our present context, the time evolution is replaced by a radial evolution --- that, in the context of AdS/CFT,
corresponds to RG flow of the boundary theory --- and the very rapid roll (``waterfall'') at $\phi_1^{\text{crit}} = \frac{m^2}{C}$
corresponds to marginality crossing in our holographic RG flow.
It is natural to expect, therefore, that a similar behavior in our context also means some kind of phase transition,
elucidating which will be one of our main motivations.

Scalar field potentials with the features described here appear to be ubiquitous in (super)gravity theories.
In theories without supersymmetry, there are virtually no constraints on $V(\phi)$, and one often takes it to be any
desired function, hoping that there exists an embedding into a consistent quantum theory.
Scalar field potentials in supergravity theories are usually more constrained, but there still seems to be a fairly large number
of potential candidates for marginality crossing.
For example, following \cite{Berg:2001ty},
in Table~\ref{tab:2dvacua} we list AdS vacua\footnote{To produce this list, one actually needs to correct a few small typos in \cite{Berg:2001ty}:
the potential in eq. (2.15) has to contain a term $16 \prod_i x_i y_i$ instead of $16 \prod_i x_i^2 y_i^2$,
and the critical point (b) in Table II should have $(x_1,x_2,y_1,y_2) =  (0,z_0,0,z_0)$ instead of $(z_0,0,z_0,0)$.
We thank M.~Berg and H.~Samtleben for correspondence and for the help in identifying these issues.}
in 3d $\CN=8$ gauged supergravity with gauge group $SO(4) \times SO(4)$.
Note, that a flow from the critical point denoted (b) in {\it loc. cit.} to the critical point (A.6), while consistent with the $c$-theorem,
has at least one irrelevant operator crossing through marginality.

\begin{table}
\begin{centering}
\begin{tabular}{|c|c|c|}
\hline
~fixed point~ & ~central charge~ & ~index~$\mu$ \tabularnewline
\hline
\hline
(a) & $c=1$ & $8$ \tabularnewline
\hline
(c) & $c=2/3$ & $7$ \tabularnewline
\hline
(d) & $c=1/2$ & $6$ \tabularnewline
\hline
(b) & $c=\sqrt{2}-1$ & $5$ \tabularnewline
\hline
(A.6) & $c=0.3790$ & $5$ \tabularnewline
\hline
(A.8) & $c=0.3765$ & $4$ \tabularnewline
\hline
(A.7) & $c=0.3762$ & $3$ \tabularnewline
\hline
\end{tabular}
\par\end{centering}
\caption{\label{tab:2dvacua} Vacua of 3d $\CN=8$ gauged supergravity.}
\end{table}

%%%%%%%%%%%%%%%%%%%%%%%%%%%%%%%%%%%%%%%%%%%%%%%%%%%%%%%%%%%%%%%%%%%%%%%%%%%%%%%%

\subsection{RG Flows and Dynamical Systems}

In the theory of dynamical systems, a compact space $\CT$ with a vector field $\beta$ is called, well, a {\it dynamical system}.

\begin{figure}[t!]
\begin{center}
\includegraphics[width=0.5\textwidth]{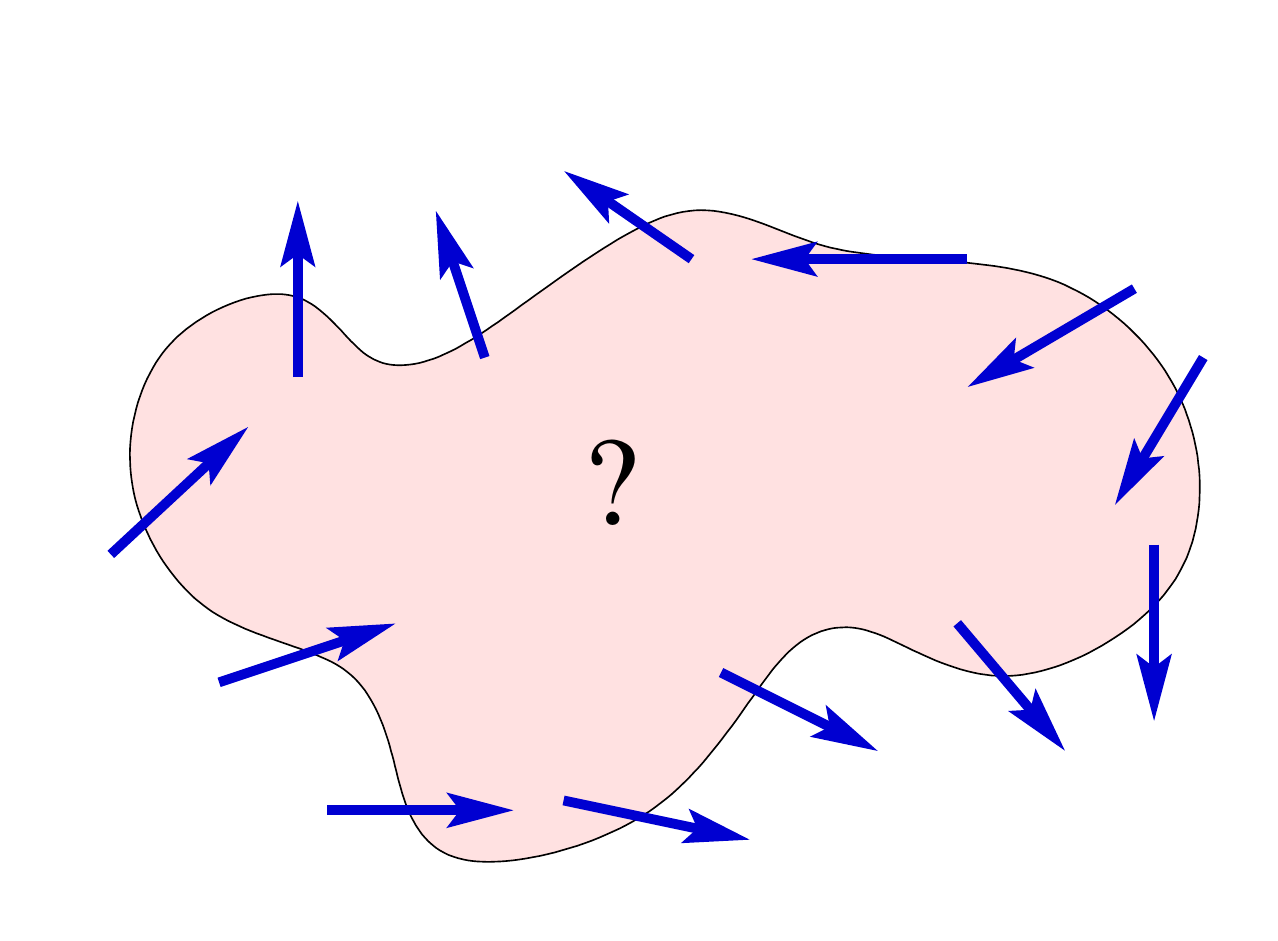}
\end{center}
\caption{\label{fig:winding} By knowing the flow at the boundary of a region $N$ one can deduce whether the RG flow should
have any fixed points in the interior of $N$. For example, is it possible that the RG flow shown in this figure has no
fixed points in $\text{Int} (N)$?}
\end{figure}

Therefore, whether we like it or not, our task of understanding RG flows and marginality crossing naturally belongs to
the domain of dynamical systems.
In particular, the space $\CT$ is what one often calls the ``theory space'', while the vector field $\beta$ is the beta-function.
The dictionary between RG flows and dynamical systems goes much deeper and, as a result, it is perhaps not too surprising
after all that powerful techniques developed in dynamical system can be successfully applied to RG flows.
As a prelude, consider a flow shown in Figure~\ref{fig:winding}; from the Poincar\'e-Hopf index theorem
it follows that it should have at least one fixed point in the interior of the region $N \subset \CT$.

As in dynamical systems, we define a {\it flow} on the space $\CT$ to be a continuous map $\beta : \R \times \CT \to \CT$ such that
\be
\beta (0,\lambda) = \lambda
\ee
\be
\beta (t,\beta (s,\lambda)) = \beta (t+s, \lambda)
\ee
where $t \in \R$ is the RG ``time'' and $\lambda \in \CT$ labels a point on the space of couplings $\CT$.
A {\it fixed point} or {\it equilibrium} is a point $\lambda \in \CT$ such that $\beta (\R,\lambda) = \lambda$.
In other words, these are conformal fixed points.
More generally, a set $S \subset \CT$ is called an {\it invariant set} for the flow $\beta$ if
\be
\beta (\R, S) \; := \; \bigcup_{t \in \R} \beta (t,S) = S
\label{firstlookatS}
\ee
This notion will play a key role in analyzing topology of the RG flows.
Note, $S$ does not need to consist entirely of fixed points, see {\it e.g.} Figure~\ref{fig:ONmodel}
for an illustration of fixed points and the invariant set $S$ in the $O(N)$ model.
One of the fundamental theorems in dynamical systems is the decomposition theorem of Conley
which states that any compact invariant set can be divided into its chain recurrent part and the rest.
Furthermore, on the latter part one can define a strictly decreasing Lyapunov function and has gradient-like dynamics.
In the context of RG flows, it means that the strongest form of the $C$-theorem holds on the latter part of $S$
and provides a candidate for the $C$-function.

This is a convenient place to remark that, in the study of both RG flows and dynamical systems,
one often makes a further assumption that $\CT$ is a locally compact metric space with metric $g$.
In the context of RG flows, the Zamolodchikov-type metric can be defined via two-point correlation functions
and without it the strongest form of the $C$-theorem would not even be a viable possibility.\footnote{Indeed,
$\lambda^i$ --- when interpreted as a coordinate on the coordinate patch of the space $\CT$ --- naturally carries
a contravariant index $i$. On the other hand, a gradient of the $C$-function is then a covariant object
(which carries a lower index) and requires a metric $g_{ij}$ or, rather, its inverse $g^{ij}$ to turn it
into a beta-function for $\lambda^i$.}
We will return to this point throughout the text, notably in section~\ref{sec:resurgence}.
Note, however, that interesting phenomena, such as violation of \eqref{muuvir}
or marginality crossing, do not necessarily require degeneration of the metric $g$.
In fact, many examples of such phenomena that we shall encounter in this paper occur at a perfectly regular point point on $\CT$
where the metric $g$ is positive and non-degenerate. In other words, the physics of such phenomena has little
to do with the regularity of the metric $g$ and, for this reason, in many of our model examples we simply
take $g$ to be a flat Euclidean metric~$g_{ij} = \delta_{ij}$.

In the course of applying the techniques from dynamical systems to RG flows we gradually develop the dictionary between
the two subjects and introduce standard notions from dynamical systems in the context of quantum field theory.
Although no prior familiarity with dynamical systems is required, a reader interested in further mathematical
details may find it helpful to consult the book by Charles Conley \cite{MR511133},
some of the relevant mathematics papers \cite{MR797044,MR1045282,MR1686770,MR1870126},
or applications to mechanics \cite{MR2989517}
(see also \cite{MR1675403} for a good introduction to the subject).
For introduction to dynamical systems and bifurcation theory
see {\it e.g.} \cite{MR947141,MR1733750,MR956468,MR709768,MR2071006}.

In this paper, when we talk about ``renormalizaiton'' we mostly mean renormalization
in the Wilsonian sense, which is most readily suited for the interpretation in the language of dynamical systems.
It should be interesting, though, to explore application of the techniques presented here to other
closely related problems, {\it e.g.} to the 1-PI effective action and various other questions that
are waiting to be translating from the language of QFT to dynamical systems or vice versa.

%%%%%%%%%%%%%%%%%%%%%%%%%%%%%%%%%%%%%%%%%%%%%%%%%%%%%%%%%%%%%%%%%%%%%%%%%%%%%%%%

\subsection{Organization and summary}

The rest of the paper is roughly divided into
a part devoted to general techniques and ideas (sections \ref{sec:Conley}, \ref{sec:bifurcations}, and \ref{sec:resurgence})
and a part illustrating how these tools and ideas can be applied in concrete examples to produce new results
(sections \ref{sec:ONmodel}, \ref{sec:QED}, and \ref{sec:QCD}).

In section \ref{sec:Conley} we start building a bridge between dynamical systems and RG flows.
Among other things, we introduce several tools that can help
in finding fixed points of an RG flow only from partial information about the flow, as in Figure~\ref{fig:winding}.
A typical situation where such tools can be useful is when complementary methods
({\it e.g.} perturbation theory, large $N$ techniques, {\it etc.}) can provide us
with the asymptotic behavior or various limits of the RG flow in space of couplings and/or parameters.
This is indeed a standard situation in non-supersymmetric theories, where exact analytical control over
RG flows away from fixed points is extremely limited, and we hope that it is in such situations
where the techniques from dynamical systems can be most helpful.

In section \ref{sec:bifurcations} we transition from the study of a fixed point set in a given theory
to questions that involve creation, annihilation, and collision of fixed points as the parameters of a system vary.
When a fixed point disappears or becomes unstable (while remaining in the bounded region of the coupling space),
does it necessarily require existence of another fixed point nearby?
Is the ``merger and annihilation'' of fixed points that already appeared in the CFT literature the only type
of generic behavior? Or, are there alternative ways in which fixed points can generically appear and disappear?
As we explain in section \ref{sec:bifurcations}, the answer to such questions depends very much on
the number of couplings in the RG flow and on the number of parameters.
The proper tool to answer these questions is called {\it bifurcation theory}, which roughly speaking
studies different ways in which fixed points can merge, appear, or disappear.
And, of all available possibilities, only the simplest ones (notably, the merger and annihilation scenario)
have been explored so far in the QFT literature, while a much longer list of interesting phenomena is waiting to be explored,
especially in RG flows with several couplings and parameters.

These general techniques and ideas can be applied in many concrete examples of RG flows.
Aiming to gain a better analytical control of non-supersymmetric RG flows,
in this paper we mainly consider three prominent examples:
\begin{itemize}

\item $O(N)$ model in three dimensions,

\item three-dimensional Quantum Electrodynamics (QED$_3$),

\item four-dimensional Quantum Chromodynamics (QCD$_4$).

\end{itemize}
\noindent
They all share certain similarities, including the existence of a ``conformal window'' in a certain range of parameters.
In each of these cases, the physics becomes strongly coupled near the lower end of the conformal window,
leaving us without any reliable tools or controlled approximations to analyze the system.
%and determine precisely the lower end of the conformal window.
We illustrate how the techniques from dynamical systems can fill this gap and work well in conjunction with other methods.

%And the loss of conformality at the lower end of this range has been rather challenging to understand in each of the three cases.
%Despite many attempts, we still do not know the precise value of the parameters where the transitions happen.
%Among other things, in this paper we try to shed light on the nature of the conformal phase transition in each of these
%systems by examining it with the tools from Conley index theory and from bifurcation theory.
%This analysis leads us to a concrete verifiable prediction about the conformal dimension of
%an operator that crosses through marginality at the transition point:

For example, bifurcation theory shifts the focus from the much-studied question about the position of the lower end
of the conformal window to the question: {\it What} happens near the lower end of the conformal window?
Moreover, it leads to concrete verifiable predictions for the scaling dimension of a nearly marginal operator,
which in our examples can be either a ``square root behavior''
%\begin{subequations}
\be
\Delta - d \; \sim \; \sqrt{N - N_{\text{crit}}}
\label{dsqroot}
\ee
or a ``quadratic behavior'' (with some $N$-independent constant $\Delta_0 > d$):
\be
\Delta - \Delta_0 \; \sim \; (N - N_{\text{crit}})^2
\label{dquadr}
\ee
%\end{subequations}
or a ``linear behavior''
\be
\Delta - d \; \sim \; N - N_{\text{crit}}
\label{dlinear}
\ee
Bifurcation analysis leads to precise criteria that, in conjunction with other methods, can uniquely
identify which type of the characteristic behavior takes place in a given system near the lower end of the conformal window.
And some of the results are rather interesting.
For example, the bifurcation analysis leads to interesting and somewhat surprising predictions in the case of QED$_3$,
which recently received a lot of attention due to numerous applications in condensed matter physics.
Contrary to some of the current scenarios, which are more likely to predict a linear behavior \eqref{dlinear},
if anything at all,\footnote{In most of the studies, the focus is usually on finding the best estimate for $N_{\text{crit}}$
rather than dependence of scaling dimensions on $N$.} in the case of QED$_3$ the bifurcation analysis
leads to the square root behavior \eqref{dsqroot} or even to the less familiar quadratic behavior \eqref{dquadr},
depending on the precise criteria that we spell out in section \ref{sec:QED}.
On the other hand, in QCD$_4$ where the square root behavior \eqref{dsqroot} is more in line with the existent scenarios,
ironically we find that a more complex behavior is possible
at special values of $N_c$ along the curve $N_f^{\text{crit}} (N_c)$.

Since \eqref{dsqroot}--\eqref{dlinear} are supposed to describe the behavior of conformal dimensions
near the lower end of the conformal window, where each of our examples is strongly coupled,
the only practical way to test such predictions at present is either by experimental studies
or in lattice simulations of these systems.\footnote{One of the simplest systems where
the square root behavior \eqref{dsqroot} can be observed and signals the merger of two fixed points
is the $q$-state Potts model in two dimensions.
Its thermal exponent and the latent heat both exhibit the characteristic behavior $\sim \sqrt{q - q_c}$
near the critical point $q_c = 4$ where the critical and tricritical Potts models ``annihilate'' \cite{Cardy:1980wm,Nienhuis:1984wm}.}
We could not find any experimental or lattice studies of the $O(N)$ model, QED$_3$, or QCD$_4$
that could verify the behavior of $\Delta$ as a function of $N$.

We also make some predictions for the $\epsilon$-expansion in the higher-dimensional version of the $O(N)$ model
and for the $N_f$-dependence of the $C$-function in 3d $\CN=2$ theories with many flavors.

%%%%%%%%%%%%%%%%%%%%%%%%%%%%%%%%%%%%%%%%%%%%%%%%%%%%%%%%%%%%%%%%%%%%%%%%%%%%%%%%%%%%%%%%%%%%%%%%%%%%%%%%%%%%%%%%%%%%%

\section{The Conley Index of RG Flows}
\label{sec:Conley}

The existence of conformal fixed points and RG flows connecting them is subject to certain topological constraints.
A simple illustration is the RG flow shown in Figure~\ref{fig:winding}, where the existence of a fixed point
can be inferred from very limited information in a completely different regime which in the space of couplings
may be very far from the fixed point in question.

\begin{figure}[t!]
\begin{center}
\includegraphics[width=0.6\textwidth]{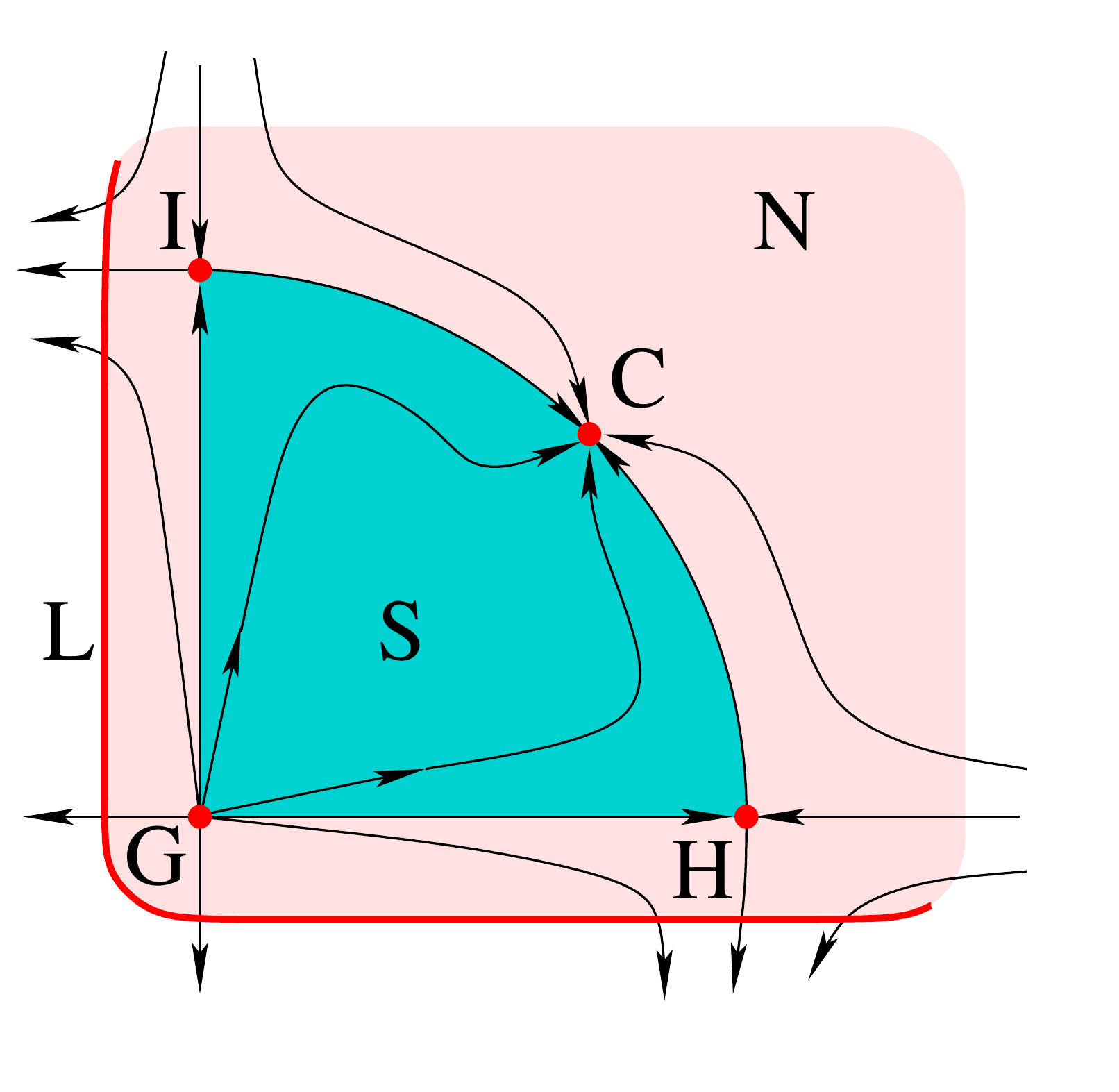}
\end{center}
\caption{\label{fig:ONmodel} An isolated invariant set $S$, an isolating neighborhood $N$, and the corresponding exit set $L$, all shown on the plot (from \cite{Cardy:1996xt}) of the RG flow in the $O(N)$ model. There are four fixed points: ($G$) Gaussian, ($H$) Wilson-Fisher, ($I$) Ising, and ($C$) Cubic with $\mu = 2$, $1$, $1$, and $0$, respectively.}
\end{figure}

Here, our goal is to develop this line of thought into a more elaborate and refined framework which then
can be applied to strongly coupled systems such as QCD$_4$ or QED$_3$.
In particular, we explain that the Conley index theory is an ideal tool for studying topology of RG flows.
In order to keep the discussion concrete and less formal, we introduce relevant mathematical techniques
through a familiar example of $O(N)$ model in $4 - \epsilon$ dimensions (see {\it e.g.} \cite{Cardy:1996xt}).
In the presence of a symmetry breaking quartic interaction, it exhibits a simple yet non-trivial flow diagram
shown in Figure~\ref{fig:ONmodel}, with several fixed points.
Moreover, the $O(N)$ model is also a good illustration of families of RG flows, to which we turn in section~\ref{sec:MCcrossing}.

\subsection{What's inside a black box?}

Suppose we are presented with a ``black box'', {\it i.e.} a compact set $N \subset \CT$ in our theory space $\CT$.
And suppose we only know what an RG flow $\beta$ is doing at the boundary\footnote{This problem can be generalized
to other ``black boxes'' which will be covered by the general framework outlined in this section.} of $N$,
just as in our toy example in Figure~\ref{fig:winding}.
Then, it may seem surprising at first that from such extremely limited information one can actually
infer what the RG flow is doing in the interior of $N$, in particular,
not only the existence but also some of the structure of the fixed points of $\beta$.
This can be done by computing the Conley index of the RG flow, and the main goal of the present section is to explain how to
carry out such calculations in practice, in concrete examples.

As we already mentioned, the input data is extremely limited: it consists of $N$ itself and the information about RG flow
at the boundary of $N$. Clearly, we couldn't even formulate the question (about fixed points inside $N$) without the former
and the latter does not seem like much information either: it basically tells us on which part of the boundary the RG flow
is entering (or, alternatively, exiting) the ``black box'' $N$.

Below we shall give a more formal definition of the {\it exit set} of $N$ where the RG flow exists $N$.
Denoting the exit set by $L$, one defines the pointed space\footnote{A pointed set $(X,x_0)$
is a topological space $X$ with a distinguished point $x_0 \in X$.} $(N/L,[L])$:
\be
N/L \; := \; (N \setminus L) \cup [L]
\label{NmodL}
\ee
where $[L]$ denotes the equivalence class of points in $L$ under the equivalence relation
$\lambda_1 \sim \lambda_2$ if and only if $\lambda_1 , \lambda_2 \in L$.
The Conley index essentially captures the topology of \eqref{NmodL}.

\begin{figure}[t!]
\begin{center}
\includegraphics[width=0.7\textwidth]{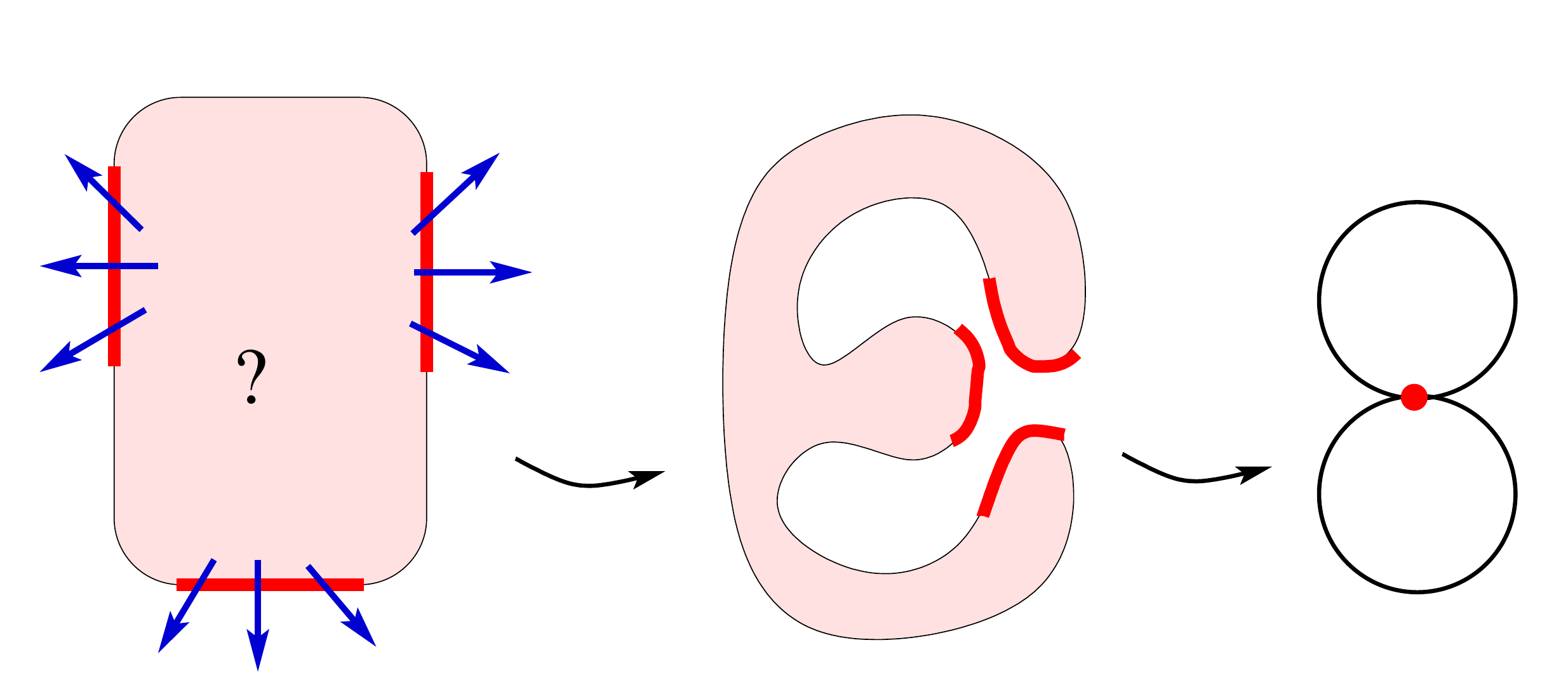}
\end{center}
\caption{\label{fig:hindexSS} Computing the Conley index of the RG flow shown on the left involves
identifying the points of the exit set $L$. The resulting pointed space $N/L$ has the homotopy type
of the wedge sum of two circles (shown on the right) \cite{MR518546}.}
\end{figure}

For example, for the RG flow in Figure~\ref{fig:winding}, the set $N$ is homeomorphic to a 2-dimensional disk
and the exit set $L$ consist of two disjoint arcs on its boundary. Identifying the points of these arcs,
we quickly learn that $N/L \cong S^1$ in this example (or, to be more precise, a circle with a marked point).
A slightly more interesting RG flow shown in Figure~\ref{fig:hindexSS} also has $N \cong D^2$, but this time
the exit set $L$ consists of three disjoint arcs.
Identifying the points of $L$ as shown in the center of Figure~\ref{fig:hindexSS} leads to a pointed space
homeomorphic to a bouquet of two circles,
\be
N / L \; \cong \; S^1 \vee S^1
\ee
These examples clearly illustrate that topology of $N/L$ can be non-trivial and probably tells us something about
the RG flow inside $N$, but how do we read off or ``decode'' this information from $N/L$ ?

Roughly, the topology of $N/L$ tells us about the topology of the invariant set \eqref{firstlookatS} in the interior of $N$.
In order to give a more precise answer, we need to introduce an important notion of {\it isolating neighborhoods}
which should come well motivated at this point.
Thus, an isolating neighborhood is a compact set $N \subset \CT$ such that
\be
\text{Inv} (N, \beta) \; := \; \{ \lambda \in N \vert \beta (\R, \lambda) \subset N \} \; \subset \; \text{int} (N)
\label{isolatingN}
\ee
where $\text{int}(N)$ denotes the interior of $N$.
Given an isolating neighborhood $N$, the invariant set $S = \text{Inv} (N)$ is called an {\it isolated invariant set}.

One of the most important properties of an isolated invariant set is that it is robust with respect to perturbations.
This stability\footnote{A more technical notion called the {\it structural stability} is going to enter the stage soon.}
plays an important role in our story.
We also note that the definitions of an isolating neighborhood and an isolated invariant set carry over to discrete dynamical systems,
which means we can study ``discrete RG flows'' where the RG time $t$ takes discrete values.

Every isolated invariant set $S$ has an {\it index pair}, that is a pair of compact sets $(N,L)$ such that $L \subset N$ and

\begin{enumerate}

\item
$S = \text{Inv} (\bar{N \setminus L})$ and $N \setminus L$ is a neighborhood of $S$.

\item
$L$ is {\it positively invariant} in $N$, that is $\lambda \in L$ and $\beta ([0,t], \lambda) \subset N$
imply $\beta ([0,t], \lambda) \subset L$.

\item
$L$ is an {\it exit set} for $N$, that is given $\lambda \in N$ and $t_1 > 0$ such that $\beta (t_1, \lambda) \notin N$,
then there exists $t_0 \in [0,t_1]$ for which $\beta ([0,t_0],\lambda) \subset N$ and $\beta (t_0,\lambda) \in L$.

\end{enumerate}

\noindent
Now we are finally ready to introduce the Conley index. There are two versions.
The {\it homotopy Conley index} of $S$ is basically what we saw before:
\be
h (S) \; = \; h (S,\beta) \; \sim \; (N/L, [L])
\label{Conleyhomotopy}
\ee
In particular, the Conley index is well-defined and does not depend on the choice of the index pair.
This version, however, is slightly more difficult to work with compared to another version called
the {\it homological Conley index}, defined by
\be
CH_* (S) \; := \; H_* (N/L, [L]) \; \cong \; H_* (N,L)
\label{Conleyhomology}
\ee
where $H_* (N,L) = (H_k (N,L))_{k \in \Z_{\ge 0}}$ denotes the relative homology groups\footnote{One usually
takes the coefficients in $\Z$ or in $\Z_2 = \Z / 2\Z$.}

In the Conley index theory, whether the primary role is played by an isolating neighborhood $N$
or by an isolated invariant set $S$ is somewhat analogous to the chicken and egg dilemma.
On the one hand, the definition of the Conley index \eqref{Conleyhomotopy} - \eqref{Conleyhomology} involves $N$.
On the other hand, it can be interpreted as an invariant of isolated invariant sets
in the sense that if $N$ and $N'$ are isolating neighborhoods for the flow $\beta$ and
\be
\text{Inv} (N, \beta) \; = \; \text{Inv} (N', \beta) \,,
\ee
then the Conley index of $N$ is the same as that of $N'$.

If the Conley index of $N$ is non-trivial,
\be
CH_* (\text{Inv} N) \; \not \cong \; 0 \,,
\ee
then $\text{Inv} (N) \ne \emptyset$.
A good illustration is an isolated conformal fixed point $S$ with $\mu$ relevant operators;
the Conley index of such theory is
\be
CH_k (S) \; \cong \;
\begin{cases}
\Z \,, & \text{if}~k=\mu \\
0 \,, & \text{otherwise}
\end{cases}
\ee
In dynamical system, it would be called a hyperbolic fixed point with an unstable manifold of dimension $\mu$.
In our example of the $O(N)$ model, there are four such points with indices
\begin{eqnarray}
CH_* (T_C) & = & (\Z,0,0,0,\ldots) \nonumber \\
CH_* (T_H) & = & (0,\Z,0,0,\ldots) \label{ONCHpoints} \\
CH_* (T_I) & = & (0,\Z,0,0,\ldots) \nonumber \\
CH_* (T_G) & = & (0,0,\Z,0,\ldots) \nonumber
\end{eqnarray}
where $T_G$ denotes the Gaussian CFT, $T_H$ denotes the Wilson-Fisher fixed point, {\it etc}.

The homology Conley index is additive under taking a disjoint union.
Namely, if $S_1$ and $S_2$ are disjoint and $S = S_1 \sqcup S_2$ is an isolated invariant set,
then
\be
CH_* (S) \; \cong \; CH_* (S_1) \oplus CH_* (S_2)
\label{CHsummation}
\ee
A typical application of this summation property is to establish the existence of flow lines between $S_1$ and $S_2$.
For example, in the $O(N)$ model we quickly deduce that $S$
is not just a set of four fixed points $\{ C, H, I, G \}$, so there must exist flows between these points,
{\it cf.} Figure~\ref{fig:ONmodel}.
Indeed, applying \eqref{CHsummation} to \eqref{ONCHpoints} we get
\be
CH_* (T_C \sqcup T_H \sqcup T_I \sqcup T_G) \; = \; (\Z,\Z \oplus \Z,\Z,0,0, \ldots)
\label{ONcomplex}
\ee
On the other hand, since $N$ is topologically a disk and the exit set $L$ has a single component ($\cong$ interval on the boundary of $N$),
it follows from \eqref{Conleyhomology} that $CH_* (S) = 0$.
This example illustrates a general qualitative pattern that we shall explore in detail later: when the topology of
the exit set $L$ is trivial, so is the Conley index.\footnote{In most of our applications, $N$ is topologically trivial
and, therefore, the topology of the pointed space \eqref{NmodL} is determined by the exit set $L$.}
But, in such situation, if conformal fixed points are found in the interior of $N$, then they necessarily must be connected by RG flows.

The power of the Conley index, though, has its limitations.
In particular, without additional information it is not very sensitive to the nature of the isolated invariant set $S$.
For example, $S \cong S^1$ can be a circle on which two hyperbolic fixed points are connected by two heteroclinic orbits,
or it can be a hyperbolic periodic orbit, or it can consist entirely of fixed points (in which case $S$ is a {\it conformal manifold}).
In all of these cases,
\be
CH_k (S) \; \cong \;
\begin{cases}
\Z & \text{if}~k=\mu \,, \mu +1 \\
0 & \text{otherwise}
\end{cases}
\ee
where, as usual, $\mu$ is the number of unstable (relevant) directions from $S$.
However, supplying additional information about the RG flow can break the tie.
For example, if $S$ has the Conley index of a periodic orbit and the isolating neighborhood possesses
a Poincar\'e section, then $S$ must indeed contain a periodic orbit.
Another theorem \cite{MR1001276}
relevant to the strongest form of the $C$-conjecture says that if $S$ be an isolated invariant set for a Morse-Smale gradient flow $\beta$,
then the Morse homology computed from the set of all critical points and flow lines in $S$
is isomorphic to the reduced homology of the Conley index $h (S,\beta)$.

Relegating a more detailed analysis of families of RG flows to section~\ref{sec:MCcrossing},
here we briefly mention one property that can be especially useful in relating a flow of interest to a simpler RG flow.
Suppose we have a family of RG flows $\beta (x)$ parametrized by a continuous parameter $x \in [a,b]$.
For example, $x = \frac{N_f}{N_c}$ in the Veneziano limit of QCD$_4$.
If $N$ is an isolating neighborhood for the entire family, that is
\be
\text{Inv} (N, \beta (x)) \; \subset \; \text{int} (N) \,, \qquad x \in [a,b]
\label{Invxfamilies}
\ee
then the Conley index of $N$ under $\beta (a)$ is the same as the Conley index of $N$ under $\beta (b)$.

%%%%%%%%%%%%%%%%%%%%%%%%%%%%%%%%%%%%%%%%%%%%%%%%%%%%%%%%%%%%%%%%%%%%%%%%%%%%%%%%%%%%%%%%%%%%%%%%%%%%%%%%%%%%%%%%%%%%%

\subsection*{Two-dimensional flows}

In many physical systems, just like in real life, there are two main characters, namely, two ``relevant'' coupling constants
that we denote $\lambda_1$ and $\lambda_2$. We put ``relevant'' in quotes because here it is used not in the technical sense,
but rather to indicate that $\lambda_1$ and $\lambda_2$ are essential for a given physical problem,
whereas other couplings have negligible effect and can be ignored.
Various potential candidates of marginality crossing,
such as the $O(N)$ model, QED$_3$ and QCD$_4$ in the conformal window are essentially of this type.

\begin{table}
\begin{centering}
\begin{tabular}{|c|c|}
\hline
~ $\phantom{\int^{\int^\int}}$ fixed point $\phantom{\int_{\int}}$ ~& ~matrix of anomalous dimensions~$J$~ \tabularnewline
\hline
\hline
saddle & $\phantom{\int^{\int^\int}} \det (J) < 0 \phantom{\int_{\int}}$ \tabularnewline
\hline
stable node & $\phantom{\int^{\int^\int}} \text{Tr} (J) < 0$ ~~and~~ $0 < \det (J) < \frac{1}{4} (\text{Tr} J)^2 \phantom{\int_{\int}}$
\tabularnewline
\hline
unstable node & $\phantom{\int^{\int^\int}} \text{Tr} (J) > 0$ ~~and~~ $0 < \det (J) < \frac{1}{4} (\text{Tr} J)^2 \phantom{\int_{\int}}$ \tabularnewline
\hline
stable spiral & $\phantom{\int^{\int^\int}} \text{Tr} (J) < 0$ ~~and~~ $\frac{1}{4} (\text{Tr} J)^2 < \det (J) \phantom{\int_{\int}}$
\tabularnewline
\hline
unstable spiral & $\phantom{\int^{\int^\int}} \text{Tr} (J) > 0$ ~~and~~ $\frac{1}{4} (\text{Tr} J)^2 < \det (J) \phantom{\int_{\int}}$ \tabularnewline
\hline
center & $\phantom{\int^{\int^\int}} \text{Tr} (J) = 0$ ~~and~~ $\det (J) > 0 \phantom{\int_{\int}}$
\tabularnewline
\hline
~~star / degenerate node~~ & $\phantom{\int^{\int^\int}} \det (J) = \frac{1}{4} (\text{Tr} J)^2 \phantom{\int_{\int}}$
\tabularnewline
\hline
fixed line & $\phantom{\int^{\int^\int}} \det (J) = 0 \phantom{\int_{\int}}$ \tabularnewline
\hline
\end{tabular}
\par\end{centering}
\caption{\label{tab:zoo} Classification of fixed points in a two-coupling system.}
\end{table}

Up to quadratic order, an RG flow with two coupling constants $\lambda_1$ and $\lambda_2$ looks like:
\beq
\label{toy_13}
\dot \lambda_1 \; = \;  \beta_1 \; = \; (d -\Delta_1) \lambda_1  - C_{111} \lambda_1^2 -C_{122} \lambda_2^2 -2 C_{112} \lambda_1\lambda_2 + \CO (\lambda^3) \\
\dot \lambda_2 \; = \;  \beta_2 \; = \; (d -\Delta_2) \lambda_2   -2C_{212}\lambda_1  \lambda_2 -C_{222}\lambda_2^2 -C_{211} \lambda_1^2 + \CO (\lambda^3)
\eeq
This fairly simple class of flows may have different types of behavior, that depends in a rather complicated way on
the values of conformal dimensions $\Delta_i$ and the OPE coefficients $C_{ijk}$.
Even finding critical points directly, by solving this system of second order equations
is a rather non-trivial problem.\footnote{This problem is equivalent to finding intersection
points of two arbitrary quadrics in $\R{\bf P}^2$.}

Let us see how the Conley Index theory can help. In order to find the Conley index of a flow \eqref{toy_13}
we need to know the isolating invariant neighborhood $N$ and the exit set $L$.
The former is just a disk, like in our other examples, including the $O(N)$ model in Figure~\ref{fig:ONmodel}.
So, we only need to find the exit set $L$, which is also easy.
If we denote by $\vec n$ the unit normal vector to the boundary of $N$ (pointing outward),
then the exit set $L$ is a set of points where $\vec \beta \cdot \vec n$ is positive,
\be
L \; := \; \{ \, \lambda \in \partial N \; | \; \vec \beta (\lambda) \cdot \vec n (\lambda) > 0 \, \}
\label{Lexitvianb}
\ee
Specifically, in our class of flows \eqref{toy_13}, we can choose $N \cong D^2$ to be a disk of radius $r$
in the two-dimensional $\lambda$-plane, and parametrize its boundary circle $\partial N \cong S^1$ by the angle $\phi \in (0, 2\pi]$.
Then, $\vec \beta \cdot \vec n$ is a cubic polynomial in $\cos \phi$ and $\sin \phi$ with real coefficients.\footnote{Its
explicit form is easy to write (but we won't actually need it here):
\begin{eqnarray}
\vec \beta \cdot \vec n & = & (d -\Delta_1) r \cos^2 \phi + (d -\Delta_2) r \sin^2 \phi
- r^2 C_{111} \cos^3 \phi
- r^2 C_{222} \sin^3 \phi \\
&&
-r^2 (C_{211} + 2C_{112}) \cos^2 \phi \sin \phi
-r^2 (C_{122} + 2C_{212}) \cos \phi  \sin^2 \phi \nonumber
\end{eqnarray}}
In particular, it can have an even number of real solutions (that is, values of $\phi$ for which $\vec \beta \cdot \vec n = 0$)
which by degree counting is no greater than 6.
Hence, we conclude that for a general class of flows \eqref{toy_13} the exit set $L$ can be one of the following:

\begin{itemize}

\item $L = \emptyset$ can be realized {\it e.g.} by a flow with $\Delta_1 ,\, \Delta_2 > d$ and $C_{ijk}=0$;

\item $L = S^1 \cong \partial N$ can be realized {\it e.g.} by a flow with $\Delta_1 ,\, \Delta_2 < d$ and $C_{ijk}=0$;

\item $L = I \subset S^1$ is realized in the $O(N)$ model (see Figure~\ref{fig:ONmodel});

\item $L = I \sqcup I$ is realized in a flow illustrated in Figure~\ref{fig:winding};

\item $L = I \sqcup I \sqcup I$ is realized in the example shown in Figure~\ref{fig:hindexSS}.

\end{itemize}

\begin{figure}[h]
\centering
\begin{minipage}{0.45\textwidth}
\centering
\includegraphics[width=0.8\textwidth, height=0.24\textheight]{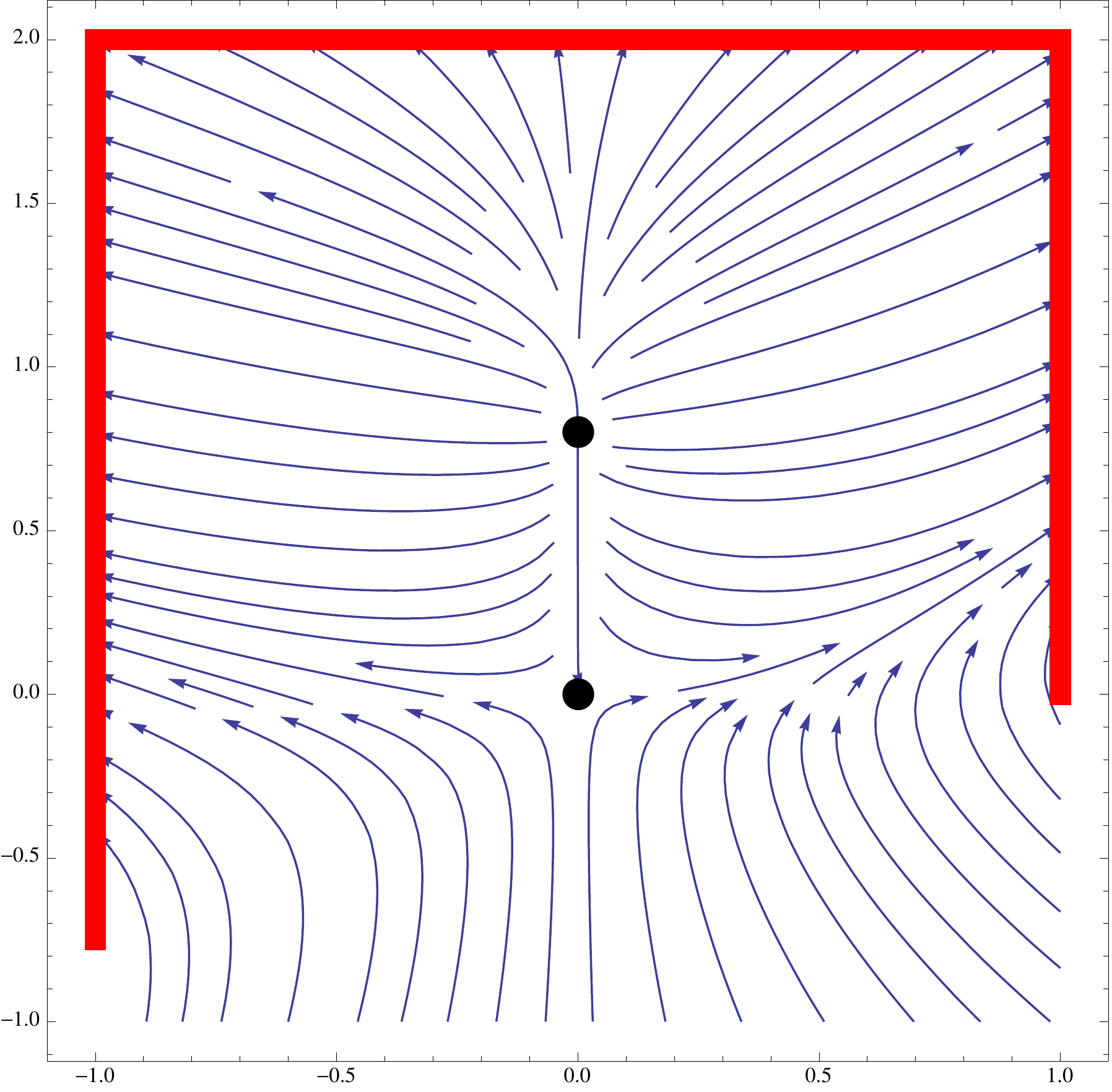}
\caption{An RG flow with $L=I$ and $CH_* (S) = 0$ can have one source and one saddle.}
\label{fig:Type I flow}
\end{minipage}
\qquad
\begin{minipage}{0.45\textwidth}
\centering
\includegraphics[width=0.8\textwidth, height=0.24\textheight]{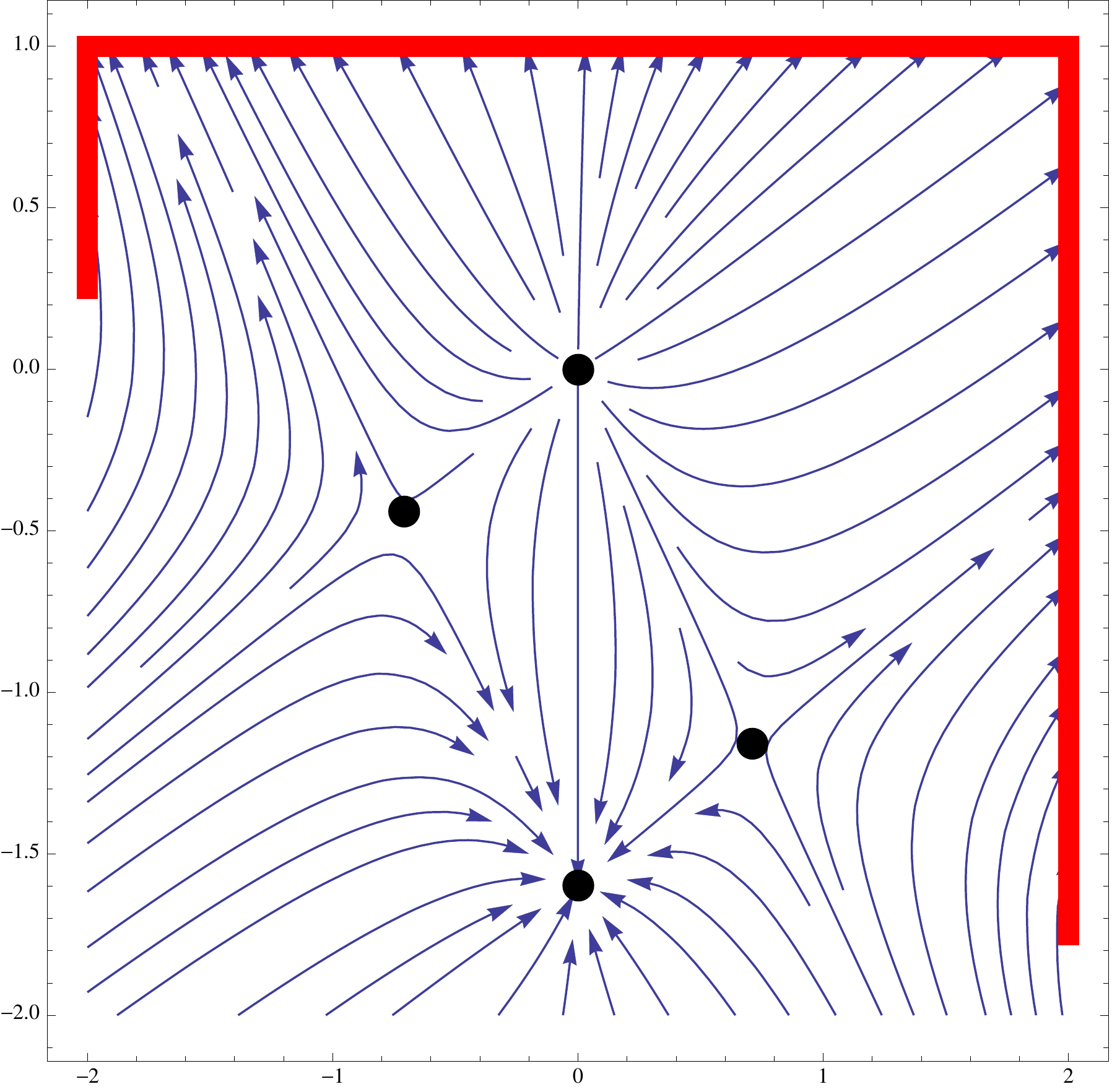}
\caption{Another example of RG flow with $L=I$ and $CH_* (S) = 0$
that has one sink, one source, and two saddles.}
\label{fig:Type II flow}
\end{minipage}
\end{figure}

\begin{figure}[h]
\centering
\begin{minipage}{0.45\textwidth}
\centering
\includegraphics[width=0.8\textwidth, height=0.24\textheight]{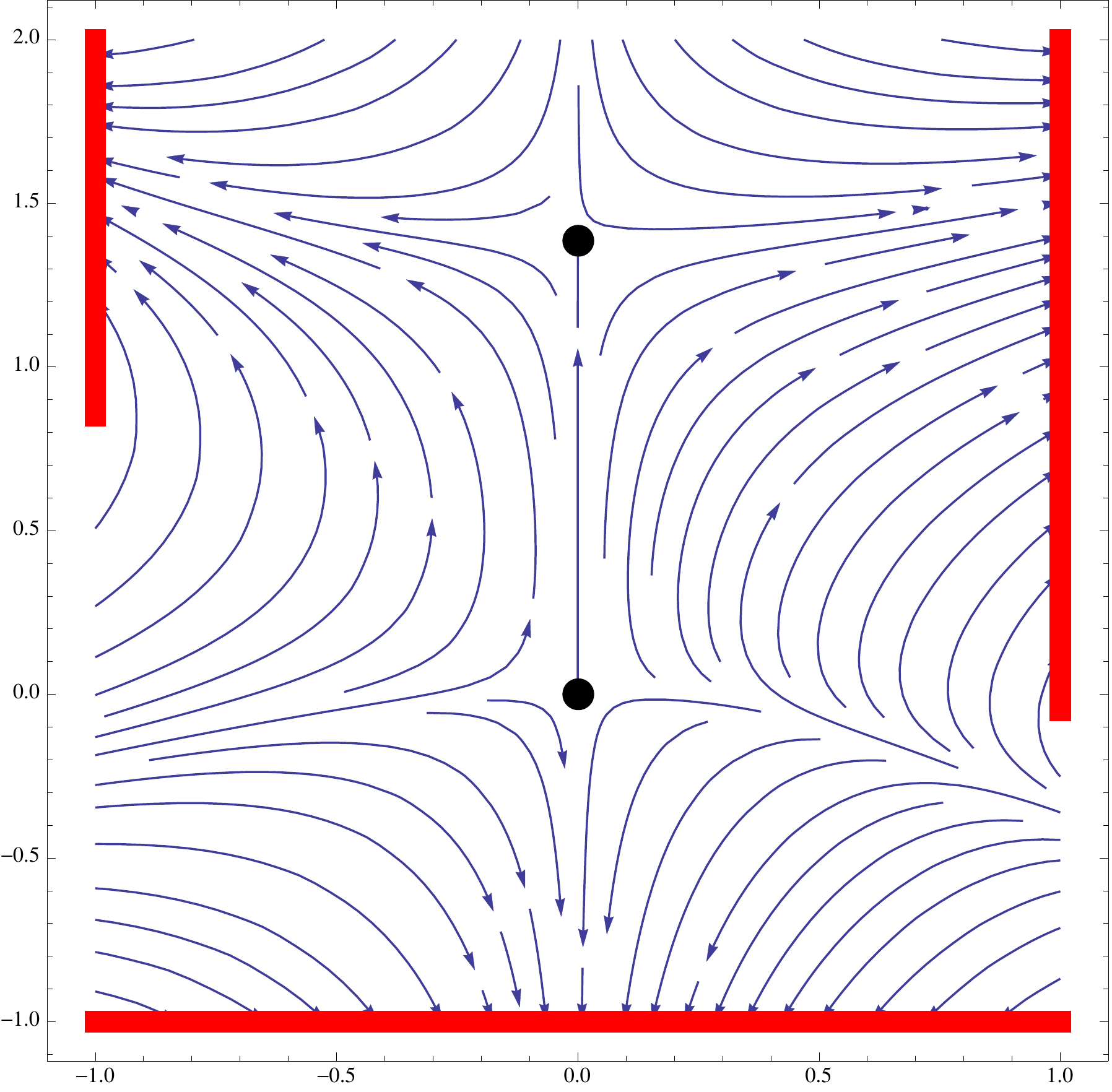}
\caption{An RG flow with $L=I \sqcup I \sqcup I$ and $CH_*(S) = \Z [1] \oplus \Z [1]$
can have two saddles.}
\label{fig:Type III flow}
\end{minipage}
\qquad
\begin{minipage}{0.45\textwidth}
\centering
\includegraphics[width=0.8\textwidth, height=0.24\textheight]{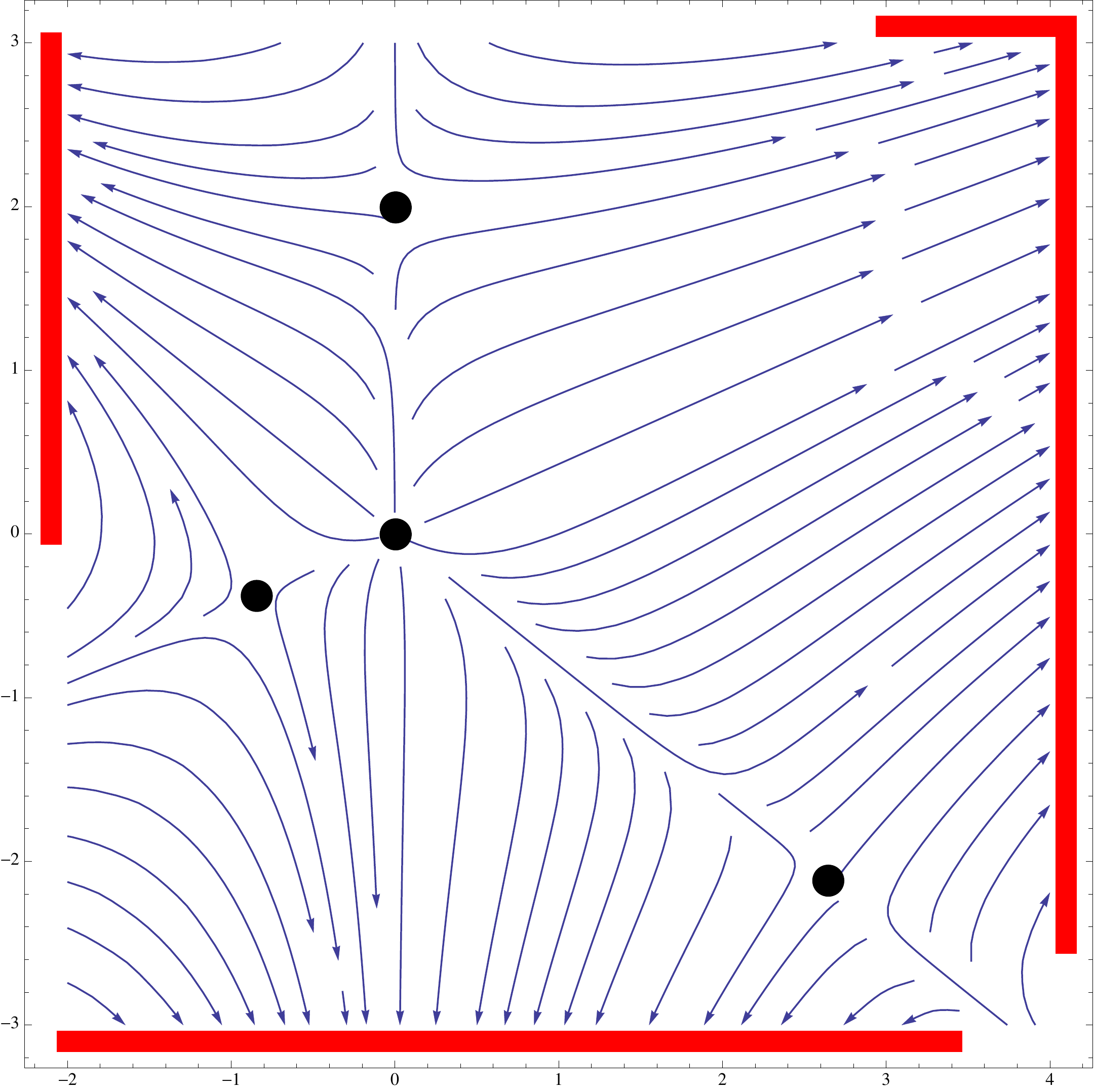}
\caption{Another realization of RG flow with $L=I \sqcup I \sqcup I$ and $CH_* (S) = \Z [1] \oplus \Z [1]$
that has a source and three saddles.}
\label{fig:Type IV flow}
\end{minipage}
\end{figure}

\noindent
In Table~\ref{tab:2dflowsCH} we summarize the Conley index in each of these cases.
As we pointed out earlier, however, the Conley index does not uniquely determine
the structure of the invariant set $S$.
For example, the third case in Table~\ref{tab:2dflowsCH} can be realized by two different RG flows,
illustrated in Figures \ref{fig:Type I flow} and \ref{fig:Type II flow},
one of which has a source and a saddle connected by a flow line,
while the other has four fixed points connected by flows.
In fact, the latter is another representation of a flow in the $O(N)$ model, {\it cf.} Figure~\ref{fig:ONmodel}.
Similarly, Figures \ref{fig:Type III flow} and \ref{fig:Type IV flow} illustrate two different RG flows that have $CH_* (S) = \Z [1] \oplus \Z [1]$
and realize the last case listed in Table~\ref{tab:2dflowsCH}.
The flow in Figure \ref{fig:Type III flow} is an example of the marginality crossing:
it violates \eqref{muuvir} since both fixed points have $\mu =1$.
While this flow looks perfectly smooth, there is indeed something special about it,
as will become evident shortly, in section~\ref{sec:MCcrossing}.

\begin{table}
\begin{centering}
\begin{tabular}{|c|c|c|}
\hline
~~exit set $L$~~ & ~~~~$N / L$~~~~ & ~~~~$\phantom{\int^{\int^\int}} CH_* (S) \phantom{\int_{\int}}$~~~ \tabularnewline
\hline
\hline
$\phantom{\int^{\int^\int}} \emptyset \phantom{\int_{\int}}$ & $D^2 \sqcup \{ \text{pt} \}$ & $\Z [0]$ \tabularnewline
\hline
$\phantom{\int^{\int^\int}} S^1 \phantom{\int_{\int}}$ & $S^2$ & $\Z [2]$ \tabularnewline
\hline
$\phantom{\int^{\int^\int}} I \phantom{\int_{\int}}$ & $D^2$ & $0$ \tabularnewline
\hline
$\phantom{\int^{\int^\int}} I \sqcup I \phantom{\int_{\int}}$ & $S^1$ & $\Z [1]$ \tabularnewline
\hline
$\phantom{\int^{\int^\int}} I \sqcup I \sqcup I \phantom{\int_{\int}}$ & $S^1 \vee S^1$ & $\Z [1] \oplus \Z [1]$ \tabularnewline
\hline
\end{tabular}
\par\end{centering}
\caption{\label{tab:2dflowsCH}
%Different types of RG flows that can be realized within a family \eqref{toy_13}.
%As explained in the main text, the information about the flow near the boundary of a given region (namely, the structure of the exit set $L$)
%can tell us about fixed points in the interior of that region.
Listed here are the homotopy Conley index and the homological Conley index for different exit sets $L$
that can be realized in the family of flows \eqref{toy_13}.
We use the standard notation from homological algebra, where $\Z [\mu]$ denotes a copy of $\Z$ placed in degree $\mu$.
Non-vanishing of the Conley index in all but one case implies that RG flow must have at least one fixed point.}
\end{table}

Note, the RG flows in Figures \ref{fig:Type I flow} and  \ref{fig:Type III flow} have the same number of critical points
connected by a flow line, but the types of critical points are different. This difference is detected by the Conley index.
Similarly, a pair of RG flows shown in Figures~\ref{fig:Type II flow} and \ref{fig:Type IV flow} has four critical points each,
but the structure of flow lines and the types of critical points are not the same.
Again, this difference is detected by the Conley index.
In fact, the Conley index can recognize even a more subtle phenomenon: two RG flows with the same number of critical points
{\it and} the same types of critical points may have different Conley index if they are connected by RG flows differently.
This leads us to the notion of a {\it connection matrix} that, roughly speaking, serves as a bridge connecting
such delicate information and $CH_* (S)$.

%%%%%%%%%%%%%%%%%%%%%%%%%%%%%%%%%%%%%%%%%%%%%%%%%%%%%%%%%%%%%%%%%%%%%%%%%%%%%%%%%%%%%%%%%%%%%%%%%%%%%%%%%%%%%%%%%%%%%

\subsection{Homological algebra of RG flows: connection matrices}

In our favorite example of the $O(N)$ model in Figure~\ref{fig:ONmodel}, we already noted that vanishing of the Conley index, $CH_* (S) = 0$,
implies the existence of RG flows connecting conformal fixed points. Indeed, the Conley index can be computed
just from the exit set $L$, without any information about RG flows in the interior of $N$.
And, if we happen to know about the four fixed points $\{ C, H, I, G \}$,
then \eqref{CHsummation} immediately tells us that $S$ can not be merely a set of these points
and must contain RG trajectories connecting them.

In this section, we explain how more detailed information about the connecting RG flows can be
deduced from the algebraic conditions obtained by interpreting \eqref{ONcomplex} as a chain complex
with a boundary map $\Delta$ which, on the one hand, packages information about RG flows
and, on the other hand, has homology equal to the Conley index $CH_* (S) = H_* (N,L)$:
\be
\frac{\text{ker} \, \Delta}{\text{im} \, \Delta} \; \cong \; CH_* (S)
\label{Deltahomology}
\ee
In particular, as a boundary map, $\Delta$ must square to zero,
\be
\Delta \circ \Delta \; = \; 0
\label{squarestozero}
\ee
which, together with \eqref{Deltahomology}, provides a set of constraints on the entries of $\Delta$.
The latter, in turn, ``count'' RG flows with $\Delta \mu = -1$.

To summarize, the data of connecting RG flows is packaged into an upper triangular {\it connection matrix} $\Delta$
whose precise definition will follow shortly and which satisfies \eqref{Deltahomology} and \eqref{squarestozero}.
For example, as the reader might have guessed by now, in our example of the $O(N)$ model the connection matrix looks like
\be
\Delta \; = \; \kbordermatrix{
& C & H & I & G \\
C & 0 & 1 & 1 & 0 \\
H & 0 & 0 & 0 & 1 \\
I & 0 & 0 & 0 & 1 \\
G & 0 & 0 & 0 & 0
}
\label{ONconnmatrix}
\ee
and, regarded as a differential acting on the complex \eqref{ONcomplex}, its cohomology is indeed trivial.
The $(p,q)$-entry in this matrix counts the number of RG flows from a fixed point $T_p$ to the fixed point $T_q$,
such that $\mu (T_p) = \mu (T_q) + 1$.

The technology of connection matrices can be viewed as a generalization of the Morse theory that does not
rely on the existence of a Morse function and works in much greater generality.
In particular, the flow $\beta$ does not need to be a gradient flow and the generators of the chain complex
do not need to be isolated fixed points, as in our example of the $O(N)$ model.
In fact, they don't even need to be conformal manifolds;
they only need to be isolated invariant subsets which, as we explained above, is a much weaker notion.
Thus, extrapolating Morse theory terminology to our dynamical system $(\CT,\beta)$,
we introduce a Morse decomposition of an isolated invariant set $S \subset \CT$ into
a finite collection of disjoint isolated invariant subsets $S_p$
labeled by elements of $(P,>)$, a partially ordered set (a.k.a. poset),
\be
\CM (S) \; = \; \{ \, S_p \, \vert \, p \in P \, \}
\ee
such that for every theory $T \in S \setminus \bigcup_{p \in P} S_p$ there exist $p,q \in P$ which satisfy
$p > q$ and $T \in \text{Con} (S_p, S_q)$ (= set of heteroclinic connections from $S_p$ to $S_q$).
For example, in the flow of Figure~\ref{fig:ONmodel}, there are four equilibrium points, which we can label by
elements of the set $P := \{ C, H, I, G \}$ and take $S_p$ to be the equilibrium $p$.

Before we proceed with the definition of $\Delta$, let is pause to remark that there can be several
natural orders on the index set $P$. The most natural is the flow induced order $>_{\beta}$:
\be
p >_{\beta} q  \qquad \Leftrightarrow \qquad \text{Con} (S_p , S_q) \ne \emptyset
\ee
For example, in the flow of Figure~\ref{fig:ONmodel}, we have
\be
G \; >_{\beta} \; H \,, I \; >_{\beta} \; C
\ee
Sometimes there exists a useful function $C : \CT \to \R$ and one can define an order $>_C$ induced by it:
\be
p >_C q \,, ~\text{iff}~ C(T_1) > C(T_2) ~\text{for all}~ T_1 \in S_p ~\text{and}~ T_2 \in S_q
\ee
Most of the time, in this paper we use the flow induced order.

Now we come to the main point of this subsection: the definition of the connection matrix $\Delta$.
Introduce a collection of abelian groups
\be
\CC_* \; = \; \bigoplus_{p \in P} CH_* (S_p)
\ee
A theorem of \cite{MR511133,MR978368,MR1162405} states that, given a Morse decomposition of $S$,
there exists a (not necessarily unique) linear map $\Delta: \CC_* \to \CC_*$ represented by
a matrix with $(p,q)$-entries:
\be
\Delta (p,q) : CH (S_p) \to CH (S_q) \qquad p,q \in P
\ee
such that
\begin{itemize}

\item $\Delta$ is strictly upper triangular, {\it i.e.} $\Delta (p,q) \ne 0$ implies $p>q$;

\item $\Delta$ is a boundary map, {\it i.e.} it is a homomorphism of degree $-1$
that maps $\CC_{\mu}$ to $\CC_{\mu-1}$, and $\Delta^2 = 0$;

\item
The cohomology of the chain complex $(\CC_* , \Delta)$ is the Conley index of $S$:
\be
H_* (\CC_* , \Delta) \; \cong \; CH_* (S)
\label{connvsCHS}
\ee

\end{itemize}

\noindent
As we already mentioned earlier, the main application of the connection matrix is to determine the existence of connecting orbits.
Thus, $\Delta (p,q) \ne 0$ implies the existence of an RG flow from $S_p$ to $S_q$.

Now, let's come back to our examples and revisit RG flows shown in Figures \ref{fig:Type I flow} -- \ref{fig:Type IV flow}.
For the RG flow in Figure~ \ref{fig:Type II flow} (same as in Figure~\ref{fig:ONmodel}),
we already wrote the connection matrix in \eqref{ONconnmatrix} and verified \eqref{connvsCHS}.
The RG flow in Figure~\ref{fig:Type III flow} has two saddle points connected by a flow line,
but since both fixed points have the same value of $\mu$, the connection matrix $\Delta$ is completely trivial. (All of its entries are zero.)
Therefore, in this case, cohomology of $\Delta$ is the same as the complex $\CC_{*}$, which
is generated by two saddle points with $\mu=1$. This agrees with the Conley index, $CH_* (S) \cong \Z [1] \oplus \Z [1]$,
computed earlier in a different way and listed in Table~\ref{tab:2dflowsCH}.

The RG flow in Figure~\ref{fig:Type I flow} is similar to the RG flow in Figure~\ref{fig:Type III flow}
in a sense that both have complex $\CC_*$ generated by two fixed points and in both cases there is one flow line connecting the two fixed points.
However, unlike our previous example, the fixed points in Figure~\ref{fig:Type I flow} have $\mu =2$ and $\mu =1$,
so that the connection matrix in this case is non-trivial:
\be
\Delta \;=\;
\begin{pmatrix}
0 & 1 \\
0 &  0
\end{pmatrix}
\ee
Acting on the complex $\CC_* = \Z [1] \oplus \Z [2]$, it has trivial cohomology, in agreement with $CH_* (S) = 0$
tabulated in the third line of Table~\ref{tab:2dflowsCH}.

Finally, the RG flow in Figure~\ref{fig:Type IV flow} has four fixed points, much as the RG flow in the $O(N)$ model,
but the types of fixed points and the connecting orbits are different.
In particular, the connection matrix for the RG flow in Figure~\ref{fig:Type IV flow}
looks like, {\it cf.} \eqref{ONconnmatrix}:
\be
\Delta  \; = \;
\begin{pmatrix}
0 & 0 & 0 & 1 \\
0 & 0 & 0 & 1 \\
0 & 0 & 0 & 1 \\
0 & 0 & 0 & 0
\end{pmatrix}
\ee
Acting on the chain complex $\CC_* =  \Z [1] \oplus \Z [1] \oplus \Z [1] \oplus \Z [2]$
it yields cohomology $H_* (\CC_* , \Delta ) \cong \Z [1] \oplus \Z [1]$,
in agreement with the Conley index computed earlier via a different method and listed in Table~\ref{tab:2dflowsCH}.

%%%%%%%%%%%%%%%%%%%%%%%%%%%%%%%%%%%%%%%%%%%%%%%%%%%%%%%%%%%%%%%%%%%%%%%%%%%%%%%%%%%%%%%%%%%%%%%%%%%%%%%%%%%%%%%%%%%%%

\subsection{Marginality crossing and transitions}
\label{sec:MCcrossing}

\begin{figure}[t!]
\begin{center}
\includegraphics[width=0.4\textwidth]{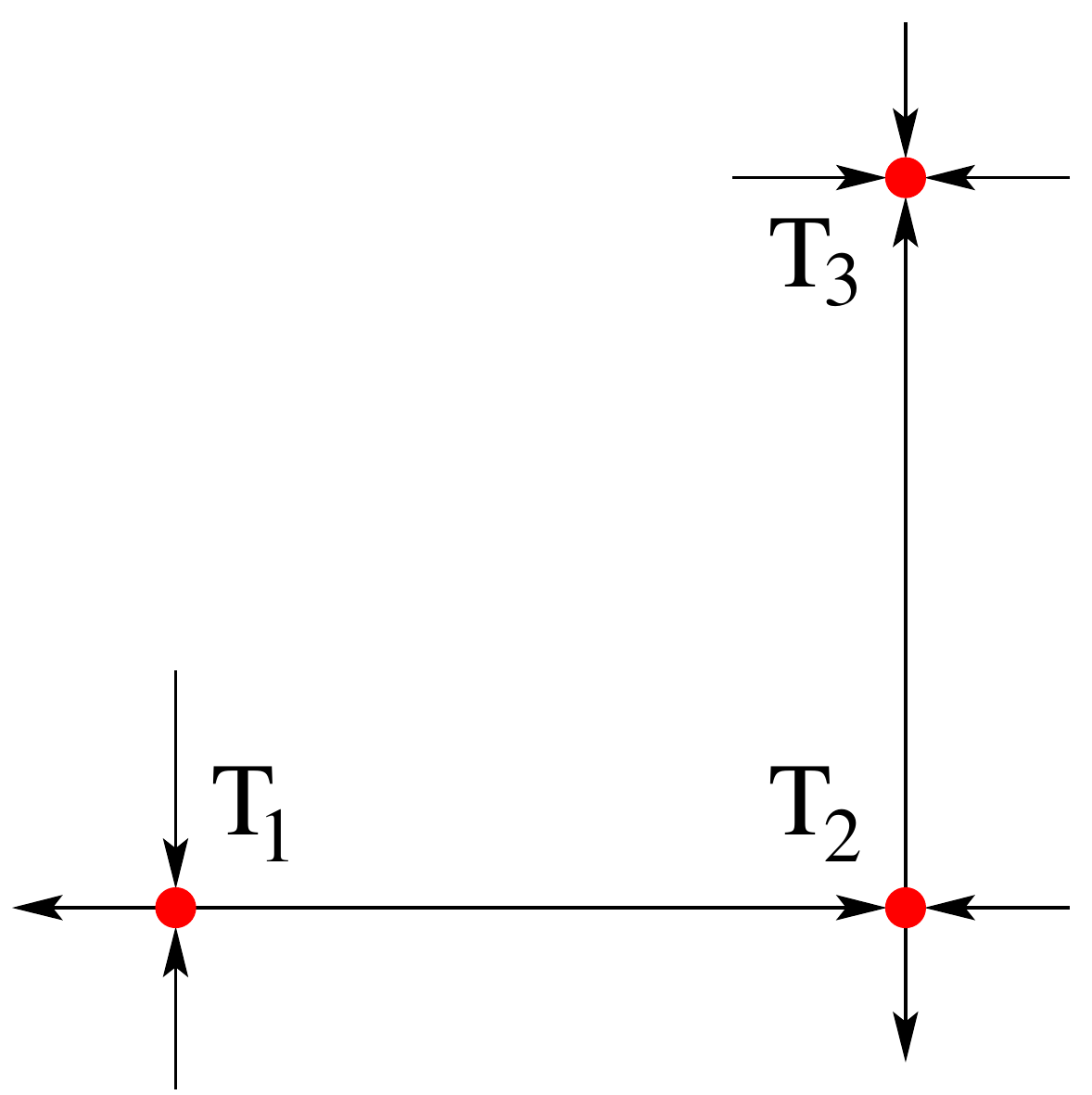}
\end{center}
\caption{\label{fig:MCTTT} A typical example of RG flow exhibiting ``marginality crossing'' along the segment from a fixed point $T_1$ to another fixed point $T_2$. Both $T_1$ and $T_2$ have $\mu=1$ in this example.}
\end{figure}

Now we are ready to take our first look at the RG flows with irrelevant operators crossing through marginality.
We already came across an example of such flow in Figure~\ref{fig:Type III flow}, which for convenience we reproduce
again in Figure~\ref{fig:MCTTT} showing only the essential flow lines, and with an extra stable IR fixed point added:
\be
T_1 \to T_2 \to T_3 \,, \qquad \mu(T_1) = \mu (T_2) = 1 \,, \qquad \mu (T_3) = 0
\label{TTTMCflow}
\ee
In particular, a separatrix from $T_1$ to $T_2$ shown in Figure~\ref{fig:MCTTT} gives a classic illustration of
a non-transverse flow $\beta$ that violates $\mu_{\text{UV}} > \mu_{\text{IR}}$.
Here, the flow-defined order is
\be
T_1 > T_2 > T_3
\ee

\begin{figure}[t!]
\begin{center}
\includegraphics[width=0.9\textwidth]{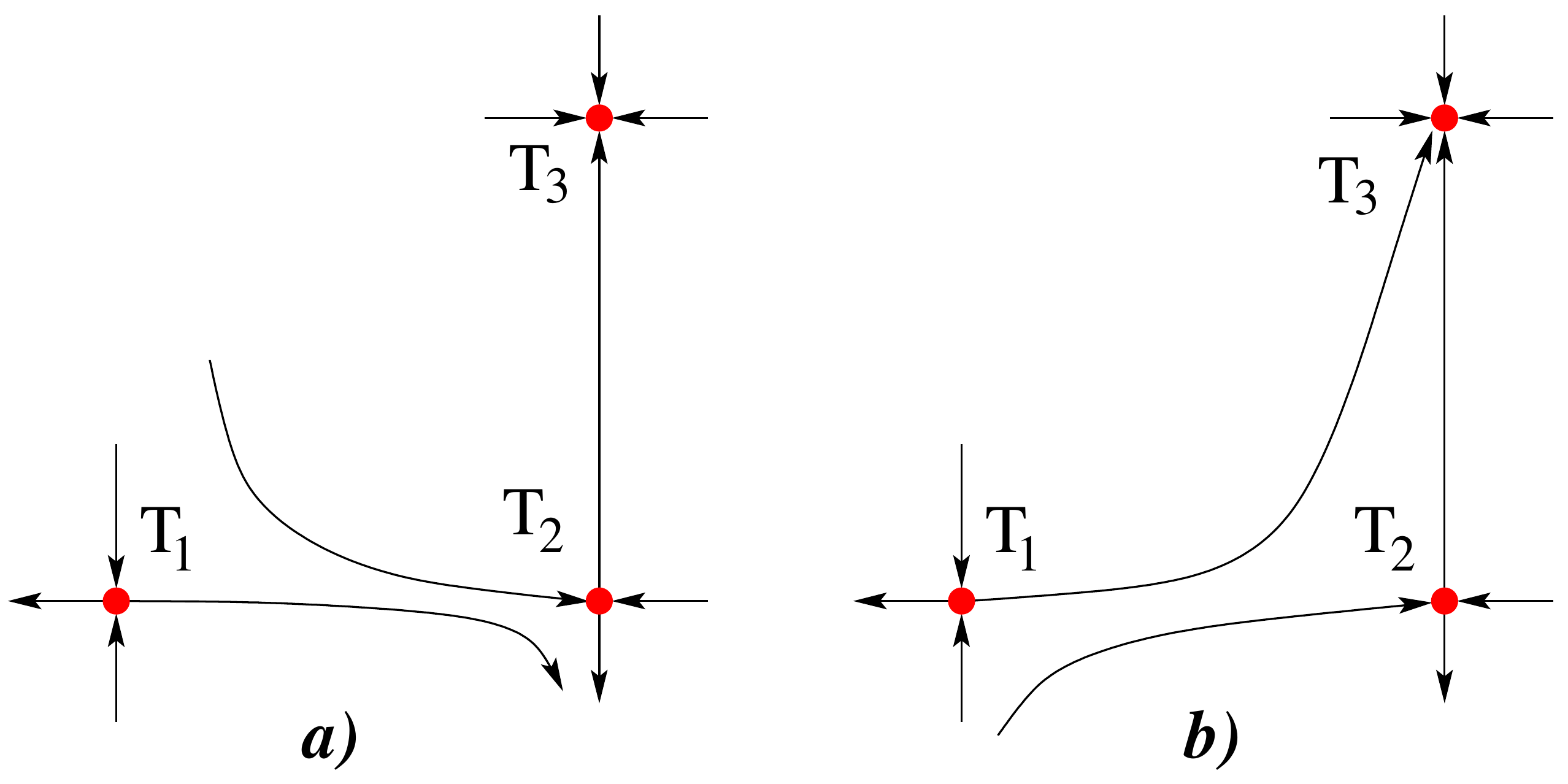}
\end{center}
\caption{\label{fig:MCTTTpert} Small perturbations of the marginality crossing flow in Figure~\ref{fig:MCTTT}.}
\end{figure}

The flow described here has a property which is a general feature of {\it any} RG flow that violates $\mu_{\text{UV}} > \mu_{\text{IR}}$:
it is not {\it structurally stable}. In other words, it requires a certain degree of fine-tuning (that we quantify below)
that, furthermore, needs to be ``stabilized'', much like in the hierarchy problem of particle physics.
By definition, a flow (or, as we later say, a phase portrait) is structurally stable
if its topology can not be changed by an arbitrarily small perturbation of the vector field.
This is clearly not the case for the flow \eqref{TTTMCflow} in Figure~\ref{fig:MCTTT}
since arbitrarily small perturbations destroy a non-generic trajectory from $T_1$ to $T_2$
and lead to one of the two scenarios, shown in Figure~\ref{fig:MCTTTpert}.
One has the flow-defined order $T_2 > T_3$ and no relation to $T_1$ since the perturbed flows to / from $T_1$ ``decouple''.
The corresponding connection matrix looks like:
\be
\Delta_a \; = \; \kbordermatrix{
& T_3 & T_2 & T_1 \\
T_3 & 0 & 1 & 0 \\
T_2 & 0 & 0 & 0 \\
T_1 & 0 & 0 & 0
}
\ee
The second perturbation has the flow-defined order $T_1 > T_3 < T_2$ and the connection matrix
\be
\Delta_b \; = \; \kbordermatrix{
& T_3 & T_2 & T_1 \\
T_3 & 0 & 1 & 1 \\
T_2 & 0 & 0 & 0 \\
T_1 & 0 & 0 & 0
}
\ee
which is easy to read off the Figure~\ref{fig:MCTTTpert}$b$ by applying the steps of the previous section.
At this point, it is natural to ask: Is there a simple relation between topology of the original flow in Figure~\ref{fig:MCTTT}
and its perturbations in Figure~\ref{fig:MCTTTpert}?
In other words, if we know two out of three, can we determine the remaining one?

These questions can be answered in the affirmative with the help of connection matrices and their analogues,
called {\it transition matrices}, that encode information about extra flow lines
which generically should not be present\footnote{since they are structurally unstable}
and only appear ``momentarily'' in transitions between topologically different RG flows.
Specifically, if $\Delta_b$ and $\Delta_a$ are the connection matrices ``before'' and ``after'' the transition,
then in general the relation has the form
\be
\Delta_a T \; = \; T \Delta_b
\label{TDDrel}
\ee
where $T$ is the transition matrix. Its diagonal entries are all equal to 1, and off-diagonal entries
``count'' the unstable flow trajectories with $\Delta \mu = 0$, much like connection matrices count flows with $\Delta \mu = -1$.
Note, since $T$ is invertible, we can also write this relation as $T^{-1} \Delta_a = \Delta_b T^{-1}$
which can be interpreted as a reverse transition.
In particular, in our example of the transition between flows in Figure~\ref{fig:MCTTTpert}
it is easy to verify that the above $\Delta_a$ and $\Delta_b$ satisfy \eqref{TDDrel} with the transition matrix:
\be
T \; = \;
\begin{pmatrix}
1 & 0 & 0 \\
0 & 1 & 1 \\
0 & 0 & 1
\end{pmatrix}
\ee
whose non-zero off-diagonal entry indicates that there must be an RG flow from $T_1$ to $T_2$
at the ``phase transition'' between flows described by $\Delta_a$ and $\Delta_b$.
This is another version of the ``black box'' problem where we can reconstruct what happens in the middle from the boundary data.

In general, the topological transition matrices are degree 0 maps.
In other words, $T_{pq}$ can only be non-zero if there is some $\mu$ for which $CH_{\mu} (S_p)$ and $CH_{\mu} (S_q)$ are both non-trivial.
Then, if we also recall that $\Delta_a$ and $\Delta_b$ both square to zero, the condition \eqref{TDDrel} which we used to
determine the transition matrix can be equivalently expressed as $\Delta^2 = 0$ for a ``connection matrix'' of a larger system:
\be
\Delta \; = \;
\begin{pmatrix}
\Delta_a & T \\
0 & - \Delta_b
\end{pmatrix}
\label{DDD}
\ee
In fact, this interpretation of \eqref{TDDrel} is more than just a mathematical trick.

As in \eqref{Invxfamilies}, let $\beta (\lambda ; x)$ be a parametrized family of flows on $\CT$ with parameter space $X = \R$.
The parametrized system
\be
\dot \lambda \; = \; \beta (\lambda ; x)
\ee
can be viewed as a flow $\hat \beta$ on $\CT \times X$ governed by the flow equations
\begin{eqnarray}
\dot \lambda & = & \beta (\lambda ; x) \\
\dot x & = & 0 \nonumber
\end{eqnarray}
such that $\hat \beta (\lambda, x) \; = \; (\beta (\lambda) , x)$.
Because of this interpretation of the parametrized flows, which we shall adopt in what follows,
there is often no harm in omitting the ``hat'' when we talk about the flow trivially extended to $\CT \times X$.
The latter, in turn, can be regarded as a limit $\epsilon \to 0$ of the {\it transition system}:
\begin{eqnarray}
\dot \lambda & = & \beta (\lambda; x) \\
\dot x & = & \epsilon (x-a) (x-b) \nonumber
\end{eqnarray}
The connection matrix for this larger system is precisely \eqref{DDD}, where the entries $\Delta_a$ and $\Delta_b$
are the familiar connection matrices for $\CM (S_{x=a})$ and $\CM (S_{x=b})$, respectively, and $T$ is a degree-0 isomorphism
\be
T : \quad \bigoplus_{p \in P} CH_* (S_p (x=b)) \; \to \; \bigoplus_{p \in P} CH_* (S_p (x=a))
\ee
which has the properties described above and gives a more formal definition of the transition matrix.
In particular, it clarifies the elegant interpretation of \eqref{TDDrel} as the condition $\Delta^2 = 0$.

In fact, in this one-parameter family of flows, the transition illustrated in Figure~\ref{fig:MCTTTpert}
is what in dynamical systems is known as the {\it heteroclinic saddle bifurcation}.

Note, even though many of our RG flows here (and in the following sections) exhibit non-trivial
topology --- captured {\it e.g.} by the Conley index or connection matrices --- the theory space $\CT$
and the isolating neighborhood $N \subset \CT$ are topologically trivial in these examples.
This does not need to be the case and was only assumed for simplicity of the exposition;
many of the techniques discussed here and below extend to $N$ and $\CT$ which {\it e.g.}
may not be connected or simply-connected. In fact, one of the main ideas in \cite{Gukov:2015qea}
was that topology of $\CT$ can be studied with the invariants such as the index $\mu$ or the Conley index.

%%%%%%%%%%%%%%%%%%%%%%%%%%%%%%%%%%%%%%%%%%%%%%%%%%%%%%%%%%%%%%%%%%%%%%%%%%%%%%%%%%%%%%%%%%%%%%%%

\section{Bifurcations of RG flows}
\label{sec:bifurcations}

In section \ref{sec:MCcrossing} we described situations where the structure of the RG flows changes,
but the fixed point set remains unchanged under the variation of the parameters.
(In particular, if fixed points are non-degenerate critical points, their $\mu$-index \eqref{mudef} remains unchanged.)
Here we consider a more dramatic change where fixed points (or periodic orbits, if they exist) of the flow $\beta$
change themselves or change their stability properties, as parameters of the system are varied.
In dynamical systems, these changes are called {\it bifurcations} and parameters are often called {\it control parameters}.
As before, we denote the control parameters by $x \in X$.

What are the different ways in which fixed points can appear or disappear? And, can one classify them?
Bifurcation theory is precisely the right tool to address this type of questions.
Moreover, just like in section \ref{sec:Conley}, it can make the best use of topology to predict
what type(s) of phase transitions the system should undergo as the parameters vary,
based only on symmetries and partial information about the RG flow.

\subsection{Different types of critical behavior}

In bifurcation theory, one often divides bifurcations into two general classes: {\it local} and {\it global}.
The former can be detected entirely by the stability analysis of the equilibria (fixed points),
whereas the latter take place when larger invariant sets of the system `collide' with fixed points or with each other.
In particular, local bifurcations can be always confined within a bounded isolating neighborhood $N$ and,
therefore, do not change the exit set $L$ and the Conley index of the system.

A more detailed classification of bifurcations depends on the dimension of space $\CT$ in which the flow is defined
({\it i.e.} on the number of coupling constants $\lambda_i$) and also on the type of flow.
For example, existence of Lyapunov functions highly restricts the types of bifurcations.
Note, in particular, that no oscillations are possible for such flows or if the flow is one-dimensional,
that is when there is only one participating coupling constant $\lambda$.
The latter exhibit very simple types of bifurcation: saddle-node bifurcation, transcritical bifurcation,
pitchfork bifurcation, and imperfect bifurcation, all of which will be described below and can be found in higher-dimensional flows as well.
Many interesting RG flows, even in simple systems such as $O(N)$ model involve at least two relevant\footnote{figuratively
speaking and also in a technical sense of this term} coupling constants, and the structure of bifurcations can be much richer,
possibly producing {\it chaotic dynamics}.

\begin{figure}[h]
\centering
\begin{minipage}{0.45\textwidth}
\centering
\includegraphics[width=0.9\textwidth, height=0.3\textheight]{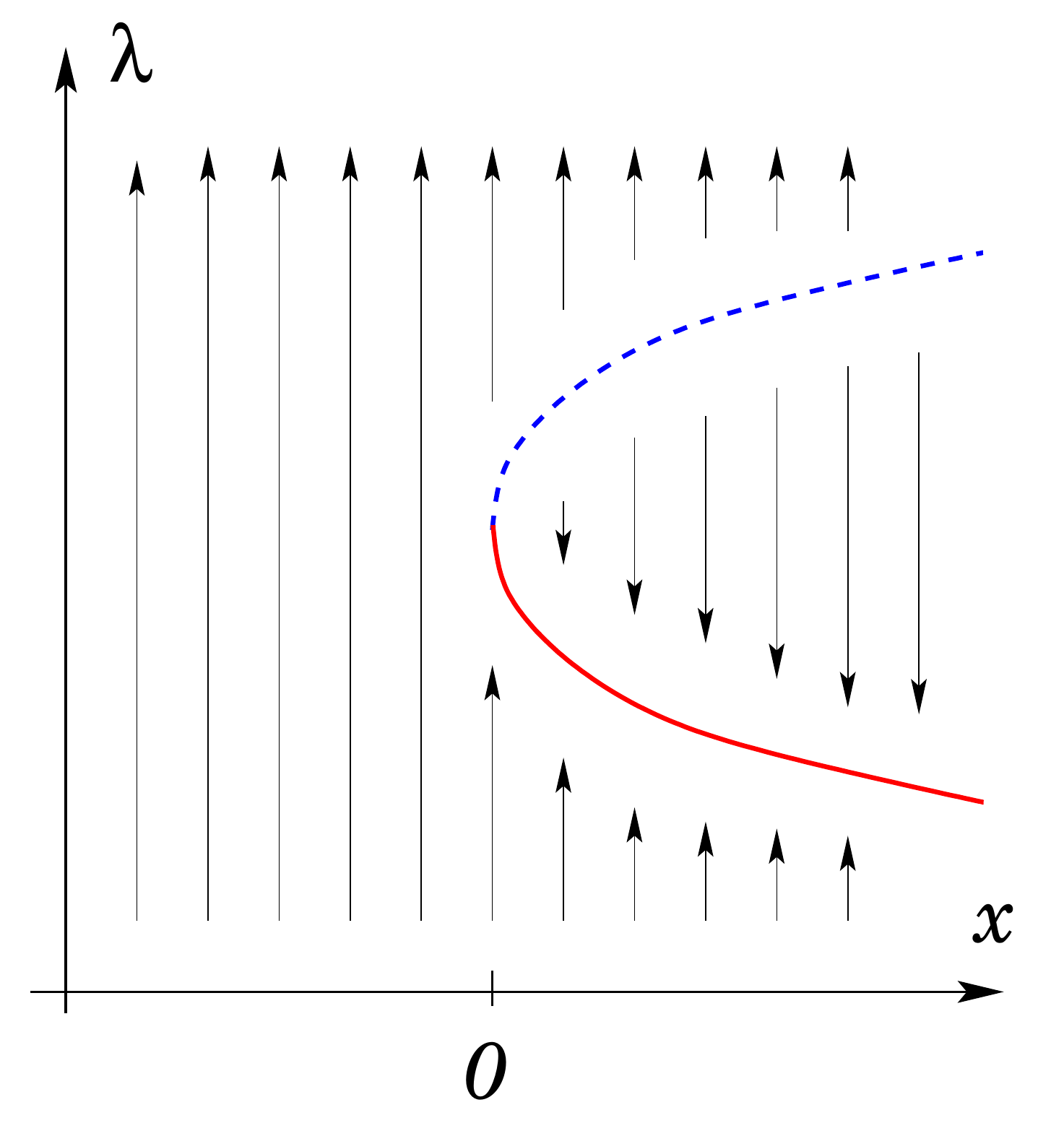}
\caption{(Extended) phase portrait for saddle-node bifurcation.}
\label{fig:SNphase}
\end{minipage}
\qquad
\begin{minipage}{0.45\textwidth}
\centering
\includegraphics[width=0.9\textwidth, height=0.3\textheight]{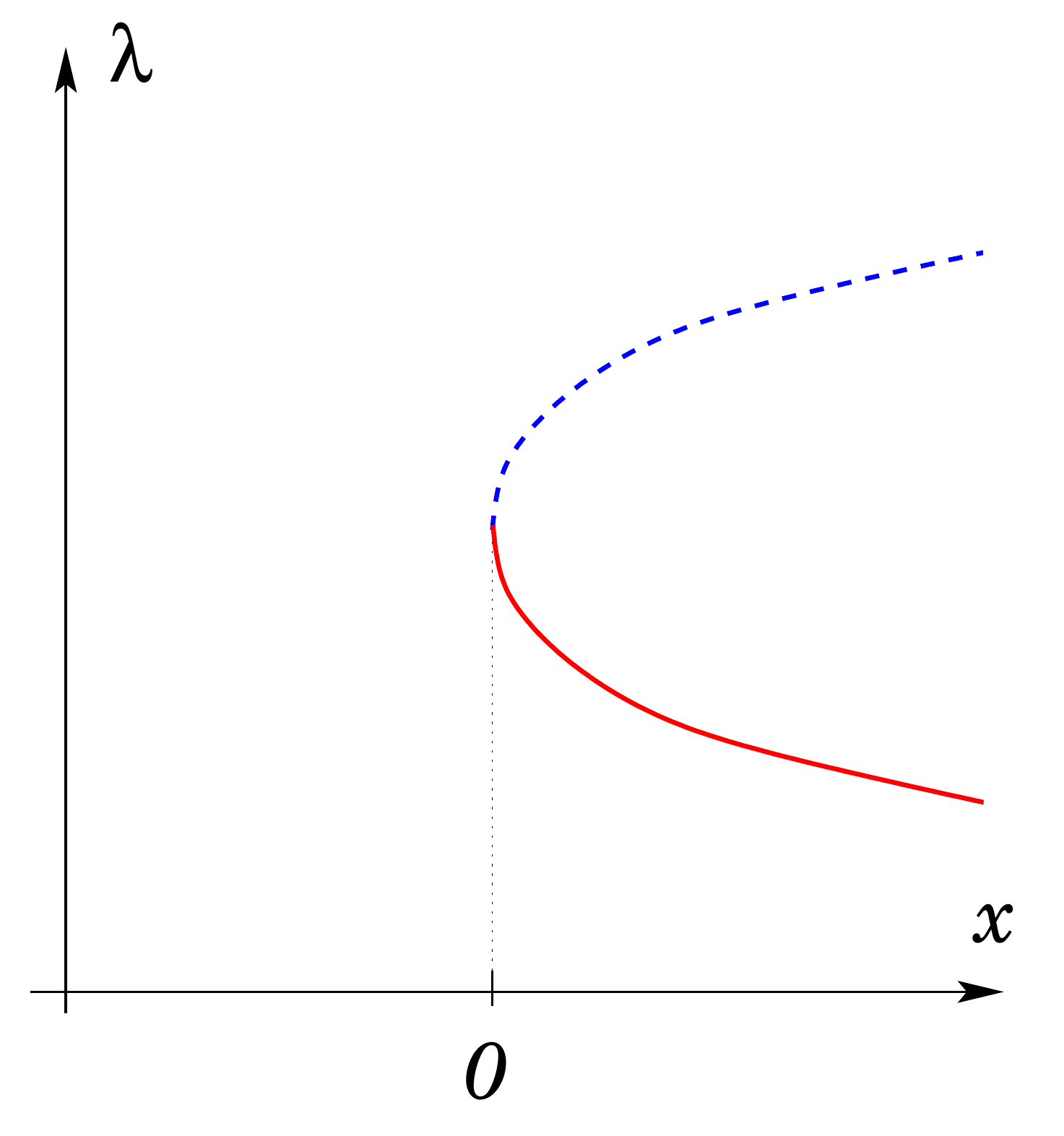}
\caption{Bifurcation diagram for saddle-node bifurcation.}
\label{fig:SNbifurcation}
\end{minipage}
\end{figure}

Bifurcations are often described with the help of either a {\it phase portrait} or a {\it bifurcation diagram}.
The former comprises all trajectories of a dyncamical system --- though, of course,
in practice one shows only representative trajectories and the equilibrium points --- whereas the latter
shows only fixed points, periodic orbits, or {\it chaotic attractors} of the flow as a function of the bifurcation parameter.
It is customary to represent stable points (attractors) with a solid line and unstable points (repellers) with a dashed line.
For example, Figures \ref{fig:SNphase} and \ref{fig:SNbifurcation} show, respectively,
the phase portrait and the bifurcation diagram of the simplest bifurcation type that will be discussed
in great detail below and will play an important role in applications to RG flows.
A bifurcation is called {\it supercritical} (resp. {\it subcritical}) if the new branch(es) is stable (resp. unstable).
Switching from one to the other is usually achieved by changing the sign of the control parameter.

\begin{figure}[t!]
\begin{center}
\includegraphics[width=0.45\textwidth]{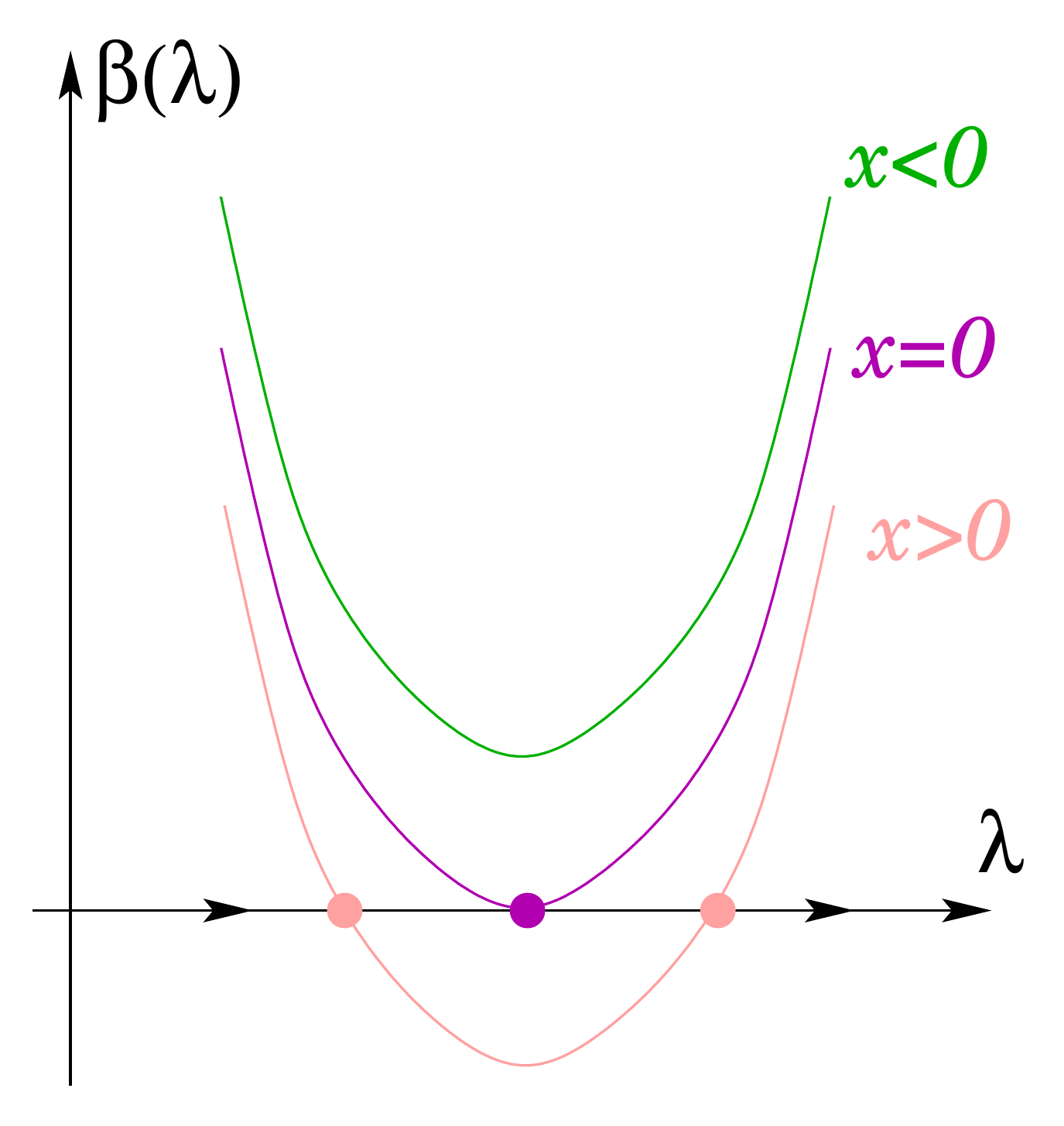}
\end{center}
\caption{\label{fig:SNbeta}As the parameter $x$ changes from $x>0$ to $x<0$, the flow $\dot \lambda = \beta (\lambda)$ undergoes
a saddle-node bifurcation: two fixed points collide and annihilate each other.}
\end{figure}

In the previous section, we already encountered a notion of the structural stability in the context of flows
that violate $\mu_{\text{UV}} > \mu_{\text{IR}}$ and saw that such flows are structurally unstable. Closely related to it
is the notion of {\it codimension}, which, in a way, quantifies the structural stability (or, rather, instability) of the flow.
Namely, the codimension of a bifurcation is the number of parameters that must be adjusted for the bifurcation to occur.
For example, in a one-dimensional flow
\be
\frac{d \lambda}{dt} \; = \; \beta (\lambda)
\label{1dflow}
\ee
the derivative $\beta' (\lambda)$ is in general non-zero when $\beta (\lambda)$ itself vanishes.
Indeed, two independent equations $\beta' (\lambda)=0$ and $\beta (\lambda)=0$ form an overdetermined
system for a single variable $\lambda$ and in general have no solutions.
However, in the presence of parameters they generically do have solutions, {\it e.g.} if $\beta (\lambda; x)$
depends on a parameter $x$ the system of two equations $\beta=\beta'=0$ in general has solutions for
isolated values of $\lambda$ and $x$, which are precisely the points where bifurcations take place.

A simple example illustrating this can be obtained by taking $\beta (\lambda; x) = \lambda^2 - x$
in our one-dimensional flow \eqref{1dflow}:
\be
\frac{d \lambda}{dt} \; = \; \lambda^2 - x
\label{saddlenode1d}
\ee
This flow has the so-called {\it saddle-node bifurcation} at $x=0$ (and $\lambda=0$).
Indeed, $\beta=0$ has two solutions $\lambda = \pm \sqrt{x}$ when $x>0$,
and no solutions ({\it i.e.} no fixed points) when $x<0$.
As the parameter $x$ varies from $x>0$ to $x<0$ the two fixed points coalesce and annihilate each other at $x=0$.
Note, using the language of dynamical systems, we can rephrase the proposal of \cite{Gies:2005as,Braun:2005uj,Braun:2006jd,Kaplan:2009kr}
by saying that the saddle-node bifurcation takes place at the lower end of the conformal window in QCD$_4$;
in what follows we revisit this proposal more carefully once we master other tools from bifurcation theory.

What we just presented is a standard argument showing that saddle-node bifurcation is of codimension 1
and that, in a system with one parameter $x$, it occurs at points in the parameter space $X$.
However, if the parameter space $X$ is $n$-dimensional, then the same argument implies that
a saddle-node bifurcation occurs on an $(n-1)$-dimensional hypersurface in $X$.
In other words, it is a codimension-1 bifurcation in any dimension.
Bifurcations of codimension $\ge 2$ are usually called {\it degenerate}; in a general system such bifurcations
may be encountered only ``by chance'' since additional conditions need to be satisfied.

\begin{figure}[h]
\centering
\begin{minipage}{0.45\textwidth}
\centering
\includegraphics[width=0.9\textwidth, height=0.35\textheight]{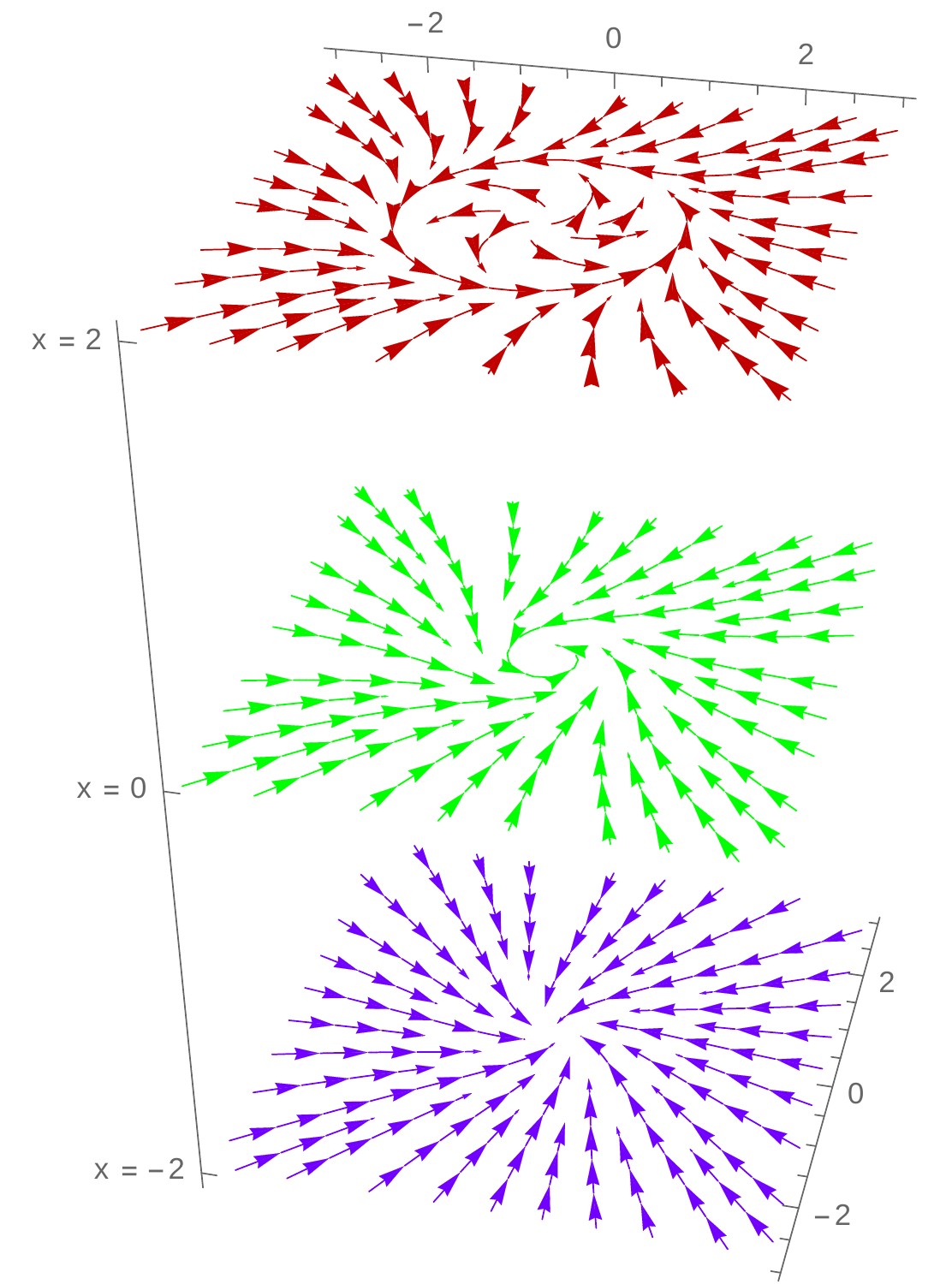}
\caption{Supercritical Hopf bifurcation.}
\label{fig:Hopf}
\end{minipage}
\qquad
\begin{minipage}{0.45\textwidth}
\centering
\includegraphics[width=0.9\textwidth, height=0.25\textheight]{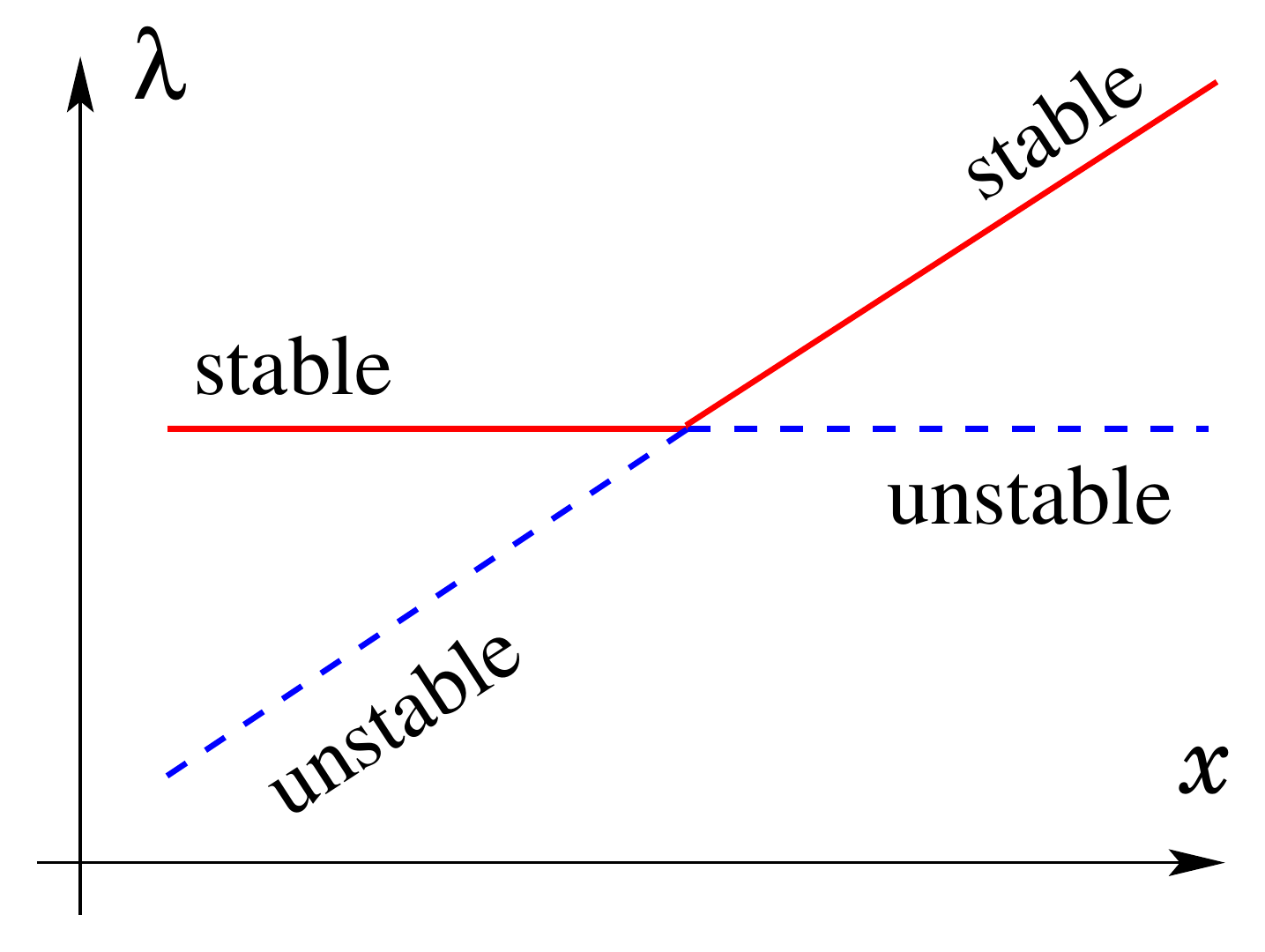}
\caption{Transcritical bifurcation.}
\label{fig:transcritical}
\end{minipage}
\end{figure}

To summarize, bifurcations can be local or global, subcritical or supercritical, and of various codimension.
Now we go more systematically trough the standard textbook list of bifurcations, and describe each one in turn
paying special attention to codimension, which will be important in applications to RG flows.
We start with the simplest ones and then gradually build our way up.
Among local bifurcations, there are only two which are truly of codimension-one,
namely the saddle-node bifurcation and the Hopf bifurcation:

\begin{itemize}

\item
{\bf Saddle-node (fold) bifurcation} is one of the simplest and most common types of bifurcation in which two fixed points
collide and annihilate each other.
Since the Conley index must remain invariant in local bifurcations, we immediately conclude that the fixed points involved
in the saddle-node bifurcation must have index \eqref{mudef} equal to $\mu$ and $\mu+1$. In particular, in two-dimensional flows one of the fixed
points must be a saddle and the other a node (either an attractor or a repellor).
The {\it normal form} of this bifurcation can be obtained from its one-dimensional variant \eqref{saddlenode1d} that we discussed earlier
by adding a decoupled equation $\dot \lambda_2 = - \lambda_2$:
\begin{eqnarray}
\dot \lambda_1 & = &  \lambda_1^2 - x \label{saddlenode2d} \\
\dot \lambda_2 & = &  - \lambda_2 \nonumber
\end{eqnarray}
The saddle-node bifurcation can be found in many models of population dynamics, {\it e.g.} in dynamics of the constantly harvested population.

\item
{\bf Hopf bifurcation} (a.k.a. Andronov-Hopf or Poincar\'e-Andronov-Hopf bifurcation) is a birth of a stable limit
cycle from a fixed point which looses its stability, see Figure~\ref{fig:Hopf}.
The normal form
\begin{eqnarray}
\dot \lambda_1 & = &  \lambda_1 (x - \lambda_1^2 - \lambda_2^2) - \lambda_2 \label{Hopf2d} \\
\dot \lambda_2 & = &  \lambda_2 (x - \lambda_1^2 - \lambda_2^2) + \lambda_1 \nonumber
\end{eqnarray}
is easy to understand in polar coordinates $\lambda_1 +i \lambda_2 = r e^{i \theta}$
where it becomes $\dot r = r (x - r^2)$ and $\dot \theta =1$.
For $x<0$ the fixed point at the origin is a stable focus (spiral point) and for $x>0$ it is an unstable focus;
in addition, for $x>0$ there is a stable limit cycle at $r = \sqrt{x}$.
Although this bifurcation has many applications, we do not expect to see it in unitary RG flows since the Jacobian
matrix of the linearization at the fixed point has complex eigenvalues $x \pm i$; it may play an important role, however, in non-unitary theories.

\end{itemize}

\begin{figure}[h]
\centering
\begin{minipage}{0.45\textwidth}
\centering
\includegraphics[width=0.9\textwidth, height=0.25\textheight]{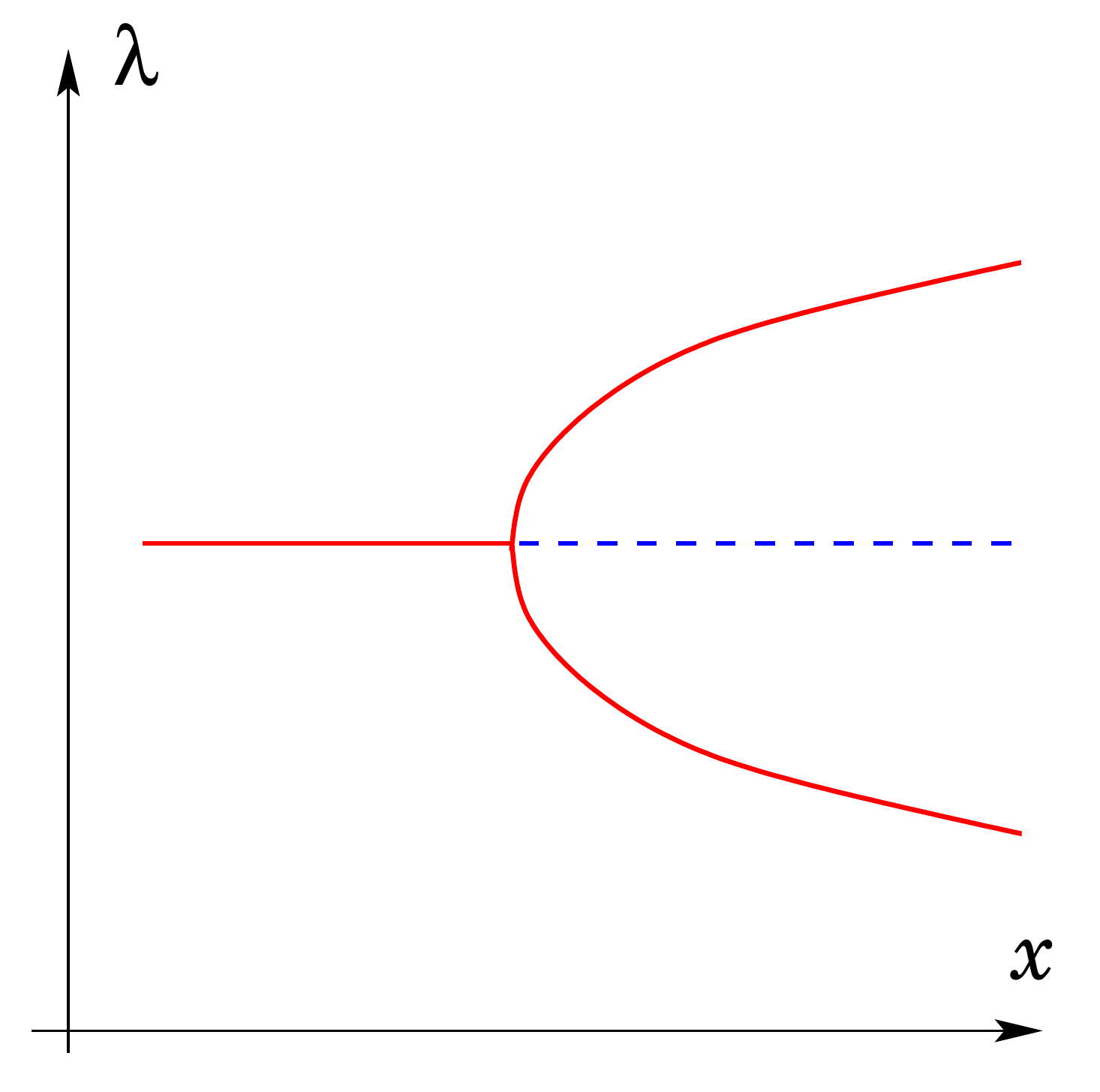}
\caption{Supercritical pitchfork bifurcation.}
\label{fig:pitchfork}
\end{minipage}
\qquad
\begin{minipage}{0.45\textwidth}
\centering
\includegraphics[width=0.9\textwidth, height=0.25\textheight]{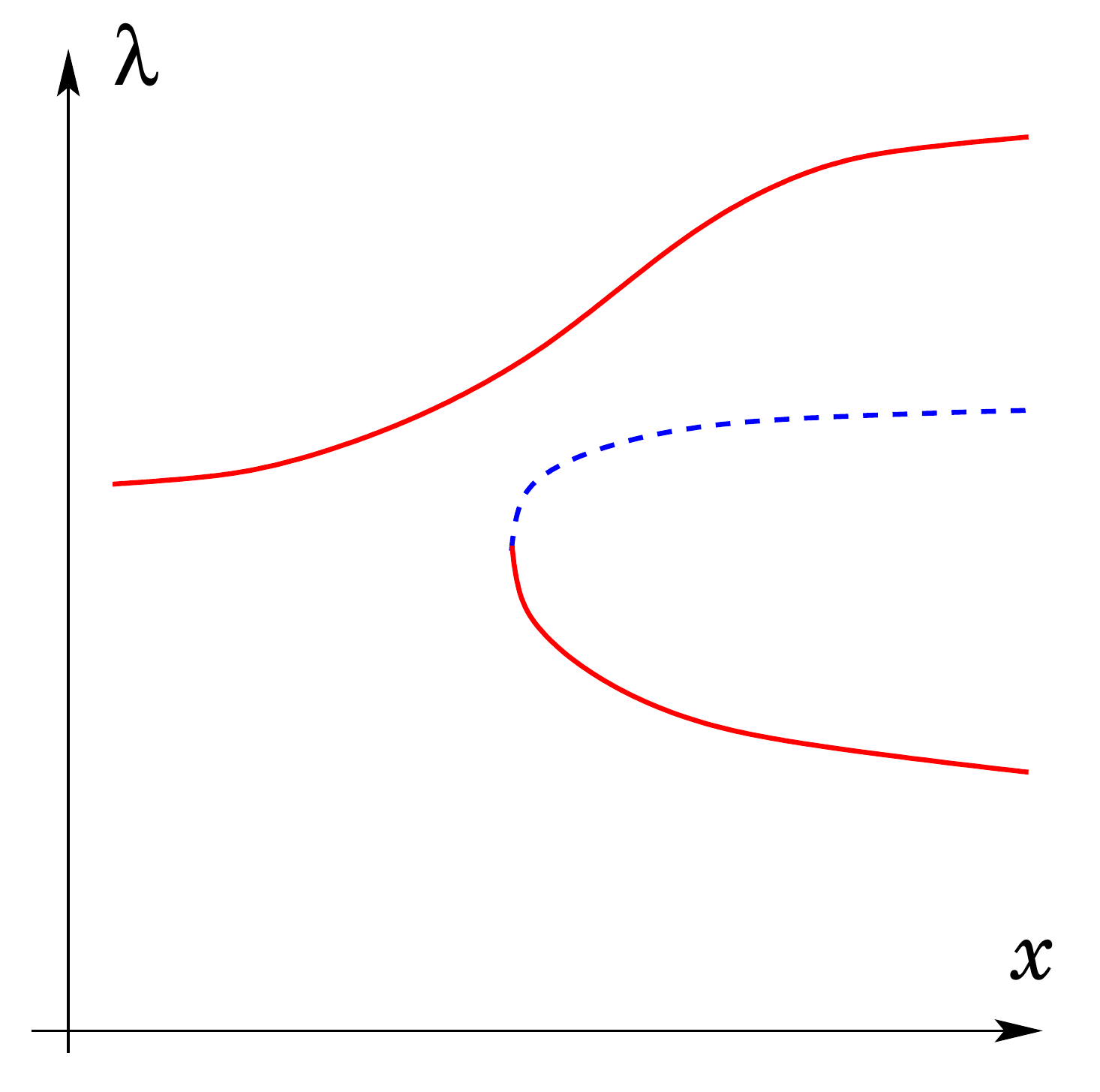}
\caption{Imperfect pitchfork bifurcation.}
\label{fig:imperfect}
\end{minipage}
\end{figure}

Even though we do not expect to see them in RG flows,
for completeness we briefly summarize more complex\footnote{They can occur only in three- or higher-dimensional continuous dynamical systems.}
local codimension-one bifurcations that involve periodic orbits:

\begin{itemize}

\item
{\bf Period-doubling (flip) bifurcation} often appears in discrete-time dynamical systems
and refers to an appearance of a new periodic orbit with double the period of the original orbit.
For example, the iterated logistic map on the interval $\lambda \in [0,1]$,
\be
\beta :~ \lambda \; \mapsto \; x \lambda (1 - \lambda) \,, \qquad (0 < x \le 4)
\ee
exhibits an entire cascade of period-doubling bifurcations when $x > 1 + \sqrt{6} \approx 3.449499$
followed by a transition to chaos at $x \ge 3.569946$.
The bifurcation diagram of a period-doubling is similar to that of a pitchfork bifurcation, {\it cf.} Figure~\ref{fig:pitchfork}.

\item
{\bf Neimark-Sacker bifurcation} (a.k.a. secondary Hopf or torus bifurcation) is a Hopf bifurcation of a periodic solution
when two complex conjugate Floquet multipliers cross the unit circle.\footnote{If one Floquet multiplier crosses
the unit circle along the negative real axis, then a period-doubling bifurcation occurs.
On the other hand, a real multiplier crossing at $+1$ can give rise to three different bifurcations,
depending on the non-linear nature of the system: saddle-node, transcritical, or pitchfork bifurcation.}
Depending on the ratio of the two new frequencies, the bifurcating solution can be periodic or quasi-periodic.
The latter almost covers a torus in a theory space.
A supercritical Neimark-Sacker bifurcation in which a new stable quasi-periodic solution appears
can be found {\it e.g.} in large-amplitude vibrations of circular cylindrical shells.

\end{itemize}

Continuing with local bifurcations, we now turn to bifurcations of higher codimension (a.k.a. degenerate bifurcations):

\begin{itemize}

\item
{\bf Transcritical bifurcation} requires three conditions to be satisfied, $\beta = \partial_{\lambda} \beta = \partial_{x} \beta = 0$.
Its normal form is $\beta (\lambda; x) = x \lambda - \lambda^2$ or, in a two-dimensional flow:
\begin{eqnarray}
\dot \lambda_1 & = &  x \lambda_1 - \lambda_1^2 \label{transcritical2d} \\
\dot \lambda_2 & = &  - \lambda_2 \nonumber
\end{eqnarray}
It describes two fixed points which exist for all values of the control parameter $x \ne 0$
and exchange their stability properties at $x=0$, as illustrated in the bifurcation diagram in Figure~\ref{fig:transcritical}.
A good example for a transcritical bifurcation is a laser at the threshold, where $\lambda$ is the photon density.

\item
{\bf Pitchfork bifurcation} requires four conditions to be satisfied,
$\beta = \partial_{\lambda} \beta = \partial_{x} \beta = \partial^2_{\lambda} \beta = 0$,
and is usually found in systems with a symmetry $\lambda \to - \lambda$.
This implies that more terms need to vanish in the Taylor series expansion of $\beta (\lambda;x)$,
compared to the transcritical bifurcation \eqref{transcritical2d}.
Thus, in a two-dimensional system, the normal form of a supercritical pitchfork bifurcation is
\begin{eqnarray}
\dot \lambda_1 & = &  x \lambda_1 - \lambda_1^3 \label{pitchfork2d} \\
\dot \lambda_2 & = &  - \lambda_2 \nonumber
\end{eqnarray}
The bifurcation diagram is shown in Figure~\ref{fig:pitchfork}.
Changing the sign of a cubic term we obtain a subcritical pitchfork bifurcation.
The pitchfork bifurcation occurs {\it e.g.} in dissipative magnetization dynamics.

\item
{\bf Imperfect bifurcation} is a version of a pitchfork bifurcation with a symmetry-breaking term
(external magnetic field in applications to magnetization dynamics):
\begin{eqnarray}
\dot \lambda_1 & = &  x_0 + x_1 \lambda_1 - \lambda_1^3 \label{imperfect2d} \\
\dot \lambda_2 & = &  - \lambda_2 \nonumber
\end{eqnarray}

\end{itemize}

Bifurcations where stable fixed points continue to exist before and after the transition are called {\it safe} (or {\it soft}) bifurcations.
On the other hand, when stable fixed points disappear and can be found only before or after the bifurcation,
such bifurcations are called {\it dangerous} (or {\it hard}).
Simple examples of soft bifurcations are transcritical bifurcation and supercritical pitchfork bifurcation,
whereas examples of hard ones are saddle-node and subcritical pitchfork bifurcation.

We now briefly review some of the global bifurcations:

\begin{itemize}
\item
{\bf Homoclinic bifurcation} is a codimension-one bifurcation that occurs when a limit cycle is destroyed by colliding with a saddle point.
This may happen {\it e.g.} if one of the flow trajectories leaving the saddle circles the spiral point
and returns back to the saddle. This special trajectory is called a {\it homoclinic cycle} and takes an infinite time to complete.
The normal form is
\begin{eqnarray}
\dot \lambda_1 & = &  \lambda_2 \label{homoclinic2d} \\
\dot \lambda_2 & = &  - x - \lambda_1 + \lambda_1^2 - \lambda_1 \lambda_2 \nonumber
\end{eqnarray}
The period of traversing the limit cycle is of the order of $\log (x)$.

\item
{\bf Heteroclinic (saddle) bifurcation} is precisely the transition illustrated in Figure~\ref{fig:MCTTTpert}
that we discussed in section~\ref{sec:MCcrossing} in the context of marginality crossing.
Now we can give it a proper name and identify it as a codimension-one global bifurcation.
In a way, this entire paper grew out of the attempt to understand RG flows that exhibit a heteroclinic bifurcation~\cite{Gukov:2015qea}.

%\item
%{\bf Heteroclinic bifurcation} is a collision of a limit cycle with two or more saddle points.

\item
{\bf SNIPER (Saddle-Node Infinite PERiod)}, also known as {\it Andronov bifurcation}
or {\it saddle-node homoclinic bifurcation}, occurs when a stable node and a saddle collide on a closed trajectory.
In polar coordinates $\lambda_1 +i \lambda_2 = r e^{i \theta}$ which we also used in discussing \eqref{Hopf2d},
the normal form looks like:
\begin{eqnarray}
\dot r & = &  r (1-r^2) \label{SNIPER2d} \\
\dot \theta & = & x + 1 + \cos \theta \nonumber
\end{eqnarray}
The limit cycle created in this bifurcation has a slow phase in the vicinity of the former fixed points
(sometimes called {\it ghosts of the fixed points}).
As a result, the period of traversing the limit cycle is of the order of $1 / \sqrt{x}$.

\item
{\bf Blue sky catastrophe} is a typical phenomenon in slow-fast dynamical systems
where a periodic orbit ``vanishes into the blue sky'' without loss of stability.
This is a codimension-one bifurcation in at least three-dimensional phase space,
such that both the period and the length of the periodic orbit exhibit unbounded growth as the control parameter
approaches its critical value, while the entire orbit remains in the bounded region of the phase space.
%in which a limit cycle collides with a non-hyperbolic cycle.
Examples of the blue sky bifurcation can be found in fluid dynamics
and in computational/mathematical neuroscience, {\it e.g.} in Hodgkin-Huxley models.

\end{itemize}

Prominent examples which combine both local and global bifurcations include:

\begin{itemize}

\item
{\bf Bogdanov-Takens bifurcation} is a codimension-2 bifurcation where saddle-node bifurcation,
Andronov-Hopf bifurcation, and a homoclinic bifurcation all meet at the same time. It has the normal form:
\begin{eqnarray}
\dot \lambda_1 & = &  \lambda_2 \\
\dot \lambda_2 & = &  x_1 + x_2 \lambda_1 + \lambda_1^2  \pm \lambda_1 \lambda_2  \nonumber
\end{eqnarray}

\item
{\bf Dumortier-Roussarie-Sotomayor bifurcations} are degenerate codimension-3 versions of Bogdanov-Takens bifurcations.
They have normal form:
\begin{eqnarray}
\dot \lambda_1 & = &  \lambda_2 \\
\dot \lambda_2 & = &  x_1 \lambda_2 \lambda_1 + x_2 \lambda_1^2 + x_3 \lambda_1^3 + x_4 \lambda_2 \lambda_1^2 \nonumber
\end{eqnarray}
A constant and coefficients of linear terms in the second equation are called unfolding parameters
whose general definition will come shortly. Turning on these parameters one finds that the above system
represents a codimension-3 point where three lines of codimension-2 bifurcations meet:
subcritical Bogdanov-Takens bifurcation, supercritical Bogdanov-Takens bifurcation, and a generalized Hopf (a.k.a. Bautin) bifurcation.

\end{itemize}

Now, once we familiarized ourselves with different types of bifurcations, a natural question is: Which RG flows realize these bifurcations?
Clearly, the simpler types, of lower codimension will be easier to find and, not surprisingly, the saddle-node bifurcation
and the transcritical bifurcation will show up in many simple examples with one control parameter, as we shall see below.
More interesting systems, however, with several parameters may exhibit more sophisticated bifurcations.
It would be interesting to produce a list of RG flows that realize different types of bifurcations;
in this work we only make a few initial steps in this direction.

%%%%%%%%%%%%%%%%%%%%%%%%%%%%%%%%%%%%%%%%%%%%%%%%%%%%%%%%%%%%%%%%%%%%%%%%%%%%%%%%%%%%%%%%%%%%%%%%

\subsection{Stability and unfolding}
\label{sec:unfolding}

In our previous discussion we already came across the question of stability of the fixed points and RG flows,
which is indeed a very important question that determines the fate of the system.
In particular, we already saw the notion of structural stability which refers to the property of the RG flow (resp. bifurcation)
to be immune to small perturbations. And, in case of bifurcations, it is related to the codimension.

For example, both transcritical bifurcation and the pitchfork bifurcation need multiple conditions to be satisfied.
In other words, these are {\it not} codimension-1 bifurcations.
Therefore, in one-parameter system such bifurcations can be found either if there is a certain symmetry
of the system (that leads to structural stability) or these higher-codimension bifurcations are degenerate,
in which case even arbitrarily small perturbations will change bifurcation diagram qualitatively.
This is called {\it unfolding} of degenerate bifurcations.

Thus, a pitchfork bifurcation is not structurally stable and under a small perturbation breaks into
a saddle-node bifurcation and an extra fixed point, as we saw in Figure~\ref{fig:imperfect}.
Completing \eqref{pitchfork2d} by lower-degree terms gives the deformed equation
$\beta = x_0 + x_1 \lambda + x_2 \lambda^2 - \lambda^3$, where the new parameters $x_0$ and $x_2$
are usually called the {\it unfolding parameters}.
Values of these parameters determine the structure of the deformed bifurcation, which can be
conveniently presented on a {\it unfolding diagram}.

\begin{figure}[h]
\centering
\begin{minipage}{0.45\textwidth}
\centering
\includegraphics[width=0.9\textwidth, height=0.25\textheight]{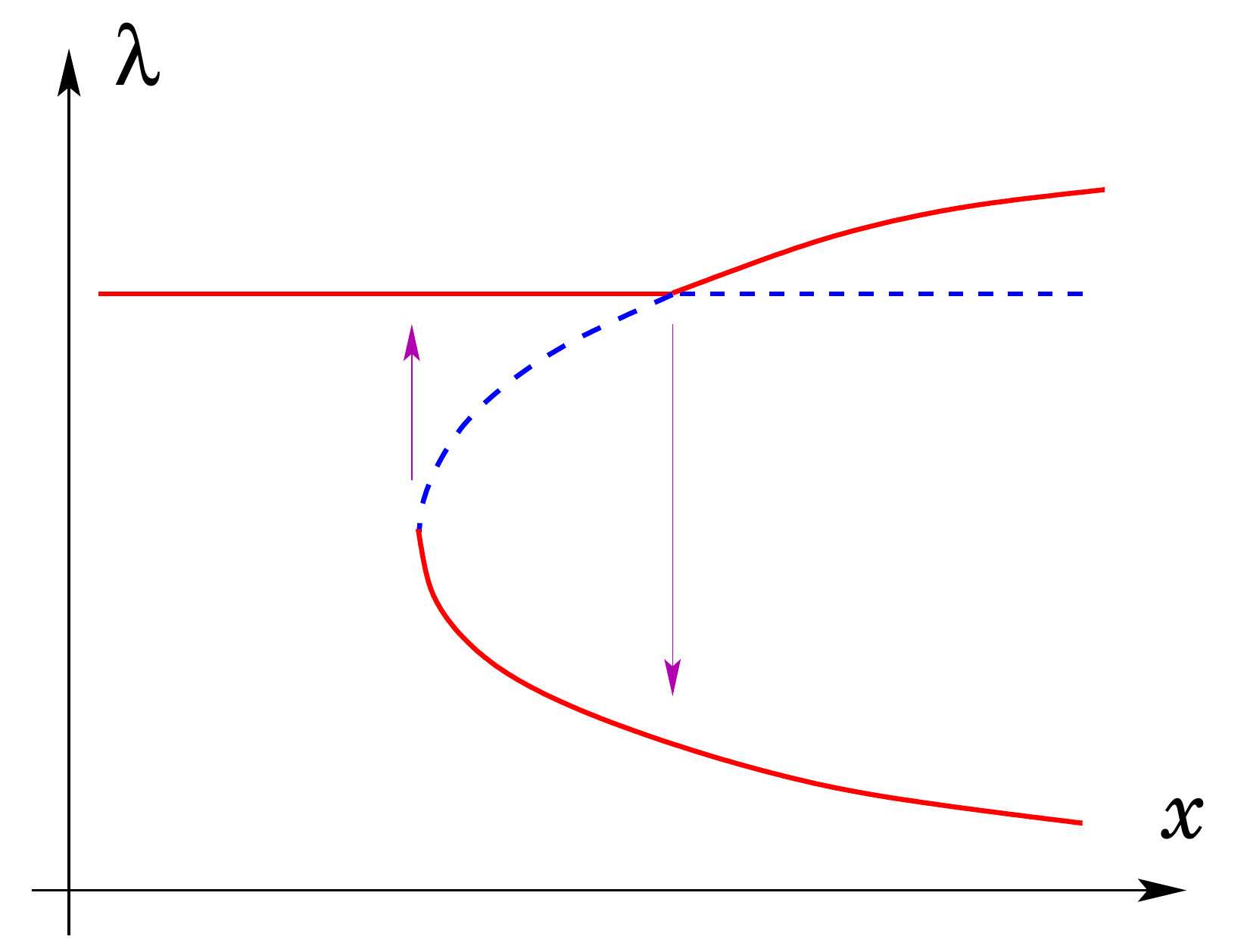}
\caption{Partial unfolding of the pitchfork. Equivalently,
a transcritical bifurcation with higher-order terms, $\beta = x \lambda - \lambda^2 - \lambda^3$.}
\label{fig:hysttrans}
\end{minipage}
\qquad
\begin{minipage}{0.45\textwidth}
\centering
\includegraphics[width=0.9\textwidth, height=0.25\textheight]{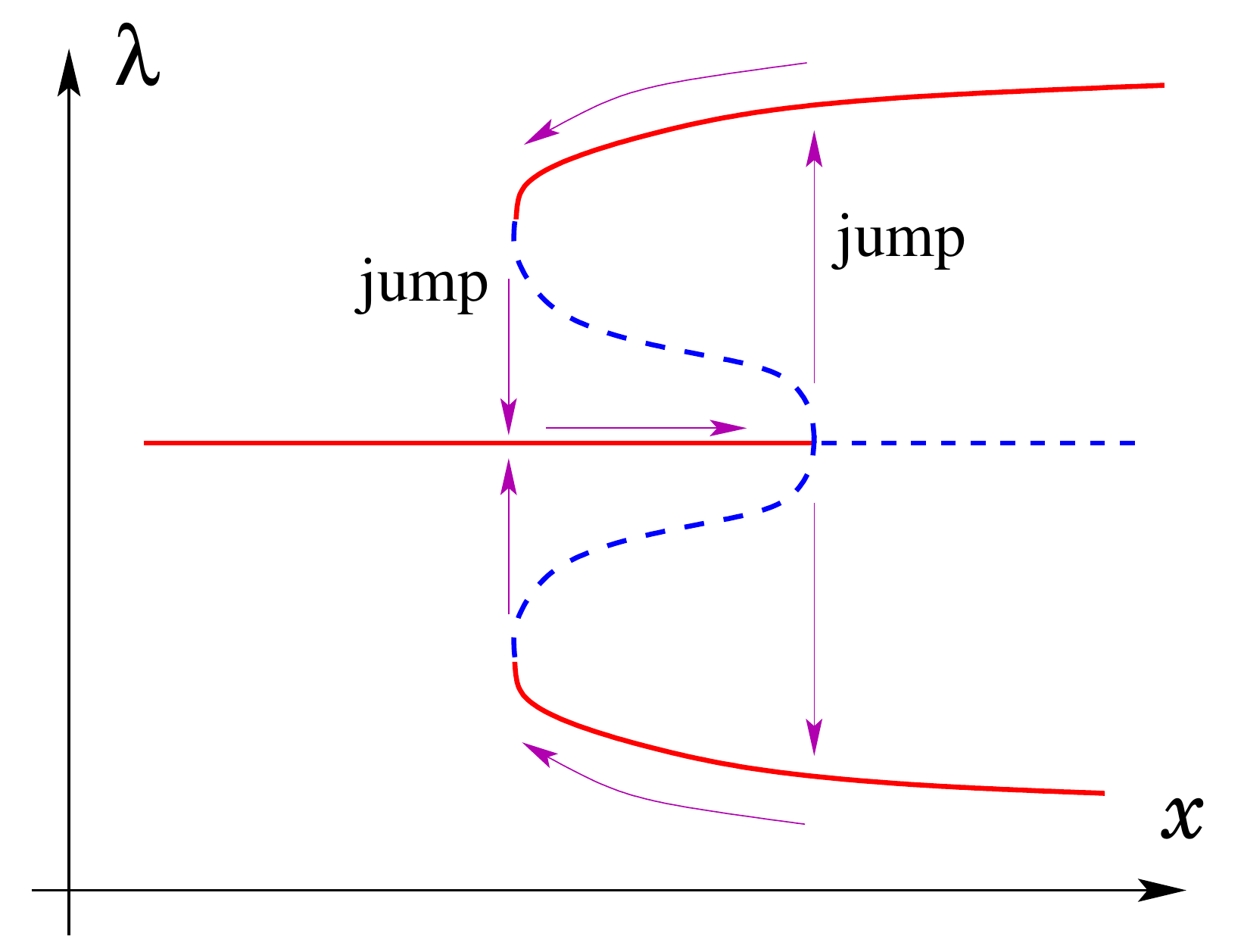}
\caption{Higher-order terms stabilize the subcritical pitchfork and lead to hysteresis.}
\label{fig:hystpitchfork}
\end{minipage}
\end{figure}

For the pitchfork bifurcation, the unfolding diagram consists of the $(x_0,x_2)$ divided by
the curves $x_0 = 0$ and $x_0 = x_2^3 / 27$.
In the regions $0< x_0 < x_2^3 / 27$ and $x_2^3/27 < x_0 < 0$ one finds three saddle-node bifurcations,
while for other values of the unfolding parameters there is only one.
For $x_0=0$ the leading behavior of $\beta (\lambda;x)$ coincides with \eqref{transcritical2d}
and so one finds transcritical bifurcation along this line,
{\it cf.} Figure~\ref{fig:hysttrans}.
Specializing further to $x_0=x_2=0$ gives the original pitchfork bifurcation.

Note, the other special case $x_2=0$ leads to the imperfect bifurcation \eqref{imperfect2d},
which was indeed introduced as a deformation (or, unfolding) of the pitchfork bifurcation with two parameters $(x_0,x_1)$.
Depending on the values of these parameters, the system has
\be
\text{3 fixed points when}~x_1 >0~\text{and}~x_0 \in \Big[ - \frac{2}{3\sqrt{3}} (x_1)^{3/2} , + \frac{2}{3\sqrt{3}} (x_1)^{3/2}\Big]
\ee
or one fixed point otherwise.
As illustrated in Figure~\ref{fig:imperfect},
a saddle-node bifurcation takes place at $x_0 = \pm \frac{2}{3\sqrt{3}} (x_1)^{3/2}$.
Keeping $x_1 > 0$ fixed and changing the value of $x_0$, the system exhibits the phenomenon
of {\it hysteresis}, {\it i.e.} an irreversible behavior as $x_0$ is ramped up and down.
On the other hand, for $x_1 < 0$ the behavior is completely reversible and the system simply retraces its path.

The starting point of any stability analysis is the {\it linear stability analysis} near each fixed point.
It is determined by the Jacobian of $\beta$, {\it i.e.} the matrix of partial
derivatives $\partial_i \beta_j$ with respect to $\lambda_i$:
\be
J \; = \;
\begin{pmatrix}
\partial_1 \beta_1 & \partial_2 \beta_1 & \cdots \\
\partial_1 \beta_2 & \partial_2 \beta_2 &  \\
\vdots &  & \ddots
\end{pmatrix}
\ee
Sometimes the Jacobian matrix is called the stability matrix.
If real parts of eigenvalues of this matrix are all non-zero, then the fixed point in question is called hyperbolic.
Since these eigenvalues are precisely the values of $d - \Delta_i$, {\it cf.} \eqref{toy_13},
we conclude that hyperbolic fixed points correspond to CFTs without marginal deformations.
According to Hartman-Grobman theorem, the local phase portrait near such a fixed point
is topologically equivalent to phase portrait of its linearized system,
\be
\frac{d \vec \lambda}{dt} = J \cdot \vec \lambda
\ee
When some couplings are marginal at the fixed point, the Jacobian matrix has zero eigenvalues and,
in the language of dynamical systems, we deal either with non-isolated fixed points (when the couplings are exactly marginal)
or with higher order fixed points. In either case, the analysis of such fixed points requires extra care, {\it cf.} \cite{Gukov:2015qea}.

In unitary theories all conformal dimensions are real.
And, since in this paper we are mainly interested in RG flows between unitary CFTs,
we can safely assume throughout that the eigenvalues of the Jacobian matrix are all real.
Then, continuing with the dictionary, we also learn that
stability in the sense of dynamical systems means that a fixed point has no relevant deformations.
Indeed, the fixed point is generally considered unstable if there are relevant operators which are singlets under global symmetries.
Likewise, in dynamical systems, a fixed point is called stable if all eigenvalues of the Jacobian matrix are negative.

Note, that many bifurcations require the determinant of the Jacobian matrix to vanish at the bifurcation point $x = x_{\text{crit}}$.
Moreover, as $x$ approaches $x_{\text{crit}}$, the rate of vanishing is different for different types of bifurcations
and, therefore, can be used as a ``fingerprint'' helping to identify the bifurcation in question.
In Table~\ref{tab:Jacobian} we summarize the order of the vanishing of $\det (J)$ for different types of local bifurcations.
For example, the Andronov-Hopf bifurcation occurs when a pair of complex conjugate eigenvalues crosses the imaginary axis,
and so the determinant of the Jacobian matrix remans non-zero at the bifurcation point.

\begin{table}
\begin{centering}
\begin{tabular}{|c|c|}
\hline
~ $\phantom{\int^{\int^\int}}$ Bifurcation $\phantom{\int_{\int}}$ ~& ~Behavior of~$\Delta - d$ \tabularnewline
\hline
\hline
Saddle-node & $\phantom{\int^{\int^\int}} \sqrt{|x - x_{\text{crit}}|} \phantom{\int_{\int}}$ \tabularnewline
\hline
Andronov-Hopf & $\phantom{\int^{\int^\int}} \text{const} \ne 0 \phantom{\int_{\int}}$ \tabularnewline
\hline
Transcritical & $\phantom{\int^{\int^\int}} x - x_{\text{crit}} \phantom{\int_{\int}}$ \tabularnewline
\hline
Pitchfork & $\phantom{\int^{\int^\int}} x - x_{\text{crit}} \phantom{\int_{\int}}$ \tabularnewline
\hline
\end{tabular}
\par\end{centering}
\caption{\label{tab:Jacobian} Behavior of the determinant of the Jacobian matrix or, equivalently, conformal dimension
of the operator nearest to marginality for different types of local bifurcations.}
\end{table}

As a first simple application of this formalism, we can clarify and formalize an expectation from the early days of the subject
that an irrelevant four-fermion operator should acquire large anomalous dimensions and cross through marginality
exactly at the lower end of the conformal window (see {\it e.g.} \cite{Fomin:1984tv,Bardeen:1985sm,Appelquist:1988sr,Appelquist:2004ib,Appelquist:1997fp}).
For simplicity, let us assume that the lower end of the conformal window is described either by a saddle-node
or transcritical bifurcation, an assumption that, on the one hand will be justified in many of the examples below and,
on the other hand, easy to relax and generalize. Then, from the above discussion ({\it cf.} Table~\ref{tab:Jacobian})
it follows that:

\begin{theorem}
\label{MCtheorem}
If the loss of conformality at the lower end of the conformal window is either due to annihilation of the IR stable fixed point with
another fixed point (``merger and annihilation'' scenario) or due to exchange of stability with another fixed point (so that the two
fixed points ``go through each other''), then at least one irrelevant operator should cross through marginality precisely at the transition point.
\end{theorem}

Now let us briefly discuss the role of the higher-order terms, which also affect stability.
For example, in order to stabilize the subcritical pitchfork bifurcation one often uses fifth-order terms.
This adds two saddle-node bifurcations to the pitchfork bifurcation:
\be
\dot \lambda \; = \; x \lambda + \lambda^3 - \lambda^5
\ee
Then, as $x$ varies, one finds three regions, with one, five, and three fixed points, respectively,
{\it cf.} Figure~\ref{fig:hystpitchfork}.
In particular, in the region $x \in [- \tfrac{1}{4} , 0]$ the system exhibits the famous {\it hysteresis effect}:
starting at a stable fixed point and, say, increasing $x$, the fixed point becomes unstable
causing the system to ``jump'' to the other branch at the same value of $x$ upon an arbitrarily small perturbation.
Then, decreasing $x$, the system remains on the second branch, thus showing an irreversible behavior.

Another example illustrating the influence of higher-order terms is the following flow:
\begin{eqnarray}
\dot \lambda_1 & = &  - \lambda_2 + x \lambda_1 (\lambda_1^2 + \lambda_2^2) \\
\dot \lambda_2 & = &  \lambda_1 + x \lambda_2 (\lambda_1^2 + \lambda_2^2) \nonumber
\end{eqnarray}
where the linear stability analysis leads to a wrong conclusion when $x \ne 0$:
a center at $(\lambda_1,\lambda_2)=(0,0)$ instead of a stable ($x<0$) or an unstable spiral ($x>0$).
This is easy to see in polar coordinates $\lambda_1 + i \lambda_2 = r e^{i \theta}$,
where the system is simply $\dot \theta = 1$ and $\dot r = x r^3$.

%%%%%%%%%%%%%%%%%%%%%%%%%%%%%%%%%%%%%%%%%%%%%%%%%%%%%%%%%%%%%%%%%%%%%%%%%%%%%%%%%%%%%%%%%%%%%%%%

\subsection{Application to the $O(N)$ model}
\label{sec:ONmodel}

The standard lore\footnote{It goes back to \cite{Aharony:1973zz}; see \cite{Pelissetto:2000ek} for a nice review and comparative analysis.}
says that the $O(N)$ model in three dimensions undergoes a transition at some vale of $N$, usually called $N_{\text{crit}}$,
in which the Wilson-Fisher fixed point and the cubic fixed point exchange their stability properties.
In the language of dynamical systems, it can be neatly summarized by saying that the RG flow has a transcritical bifurcation
at $N_{\text{crit}}$, modulo one small caveat ... this type of behavior is not to be found in a system with only one parameter!

\begin{figure}[h]
\centering
\begin{minipage}{0.45\textwidth}
\centering
\includegraphics[width=0.9\textwidth, height=0.25\textheight]{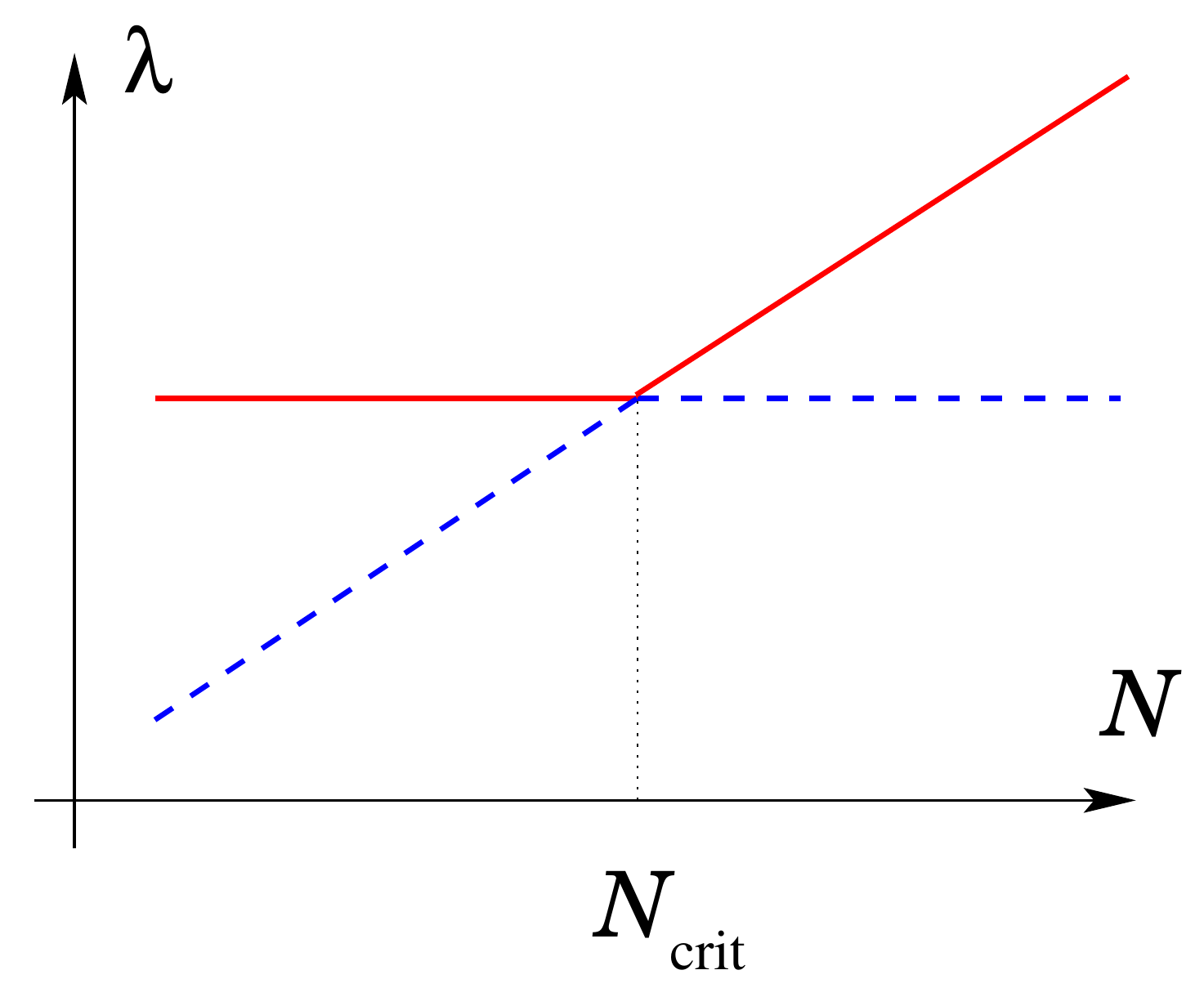}
%\caption{Comment.}
%\label{fig:original}
\end{minipage}\\
\begin{minipage}{0.45\textwidth}
\centering
\includegraphics[width=0.85\textwidth, height=0.24\textheight]{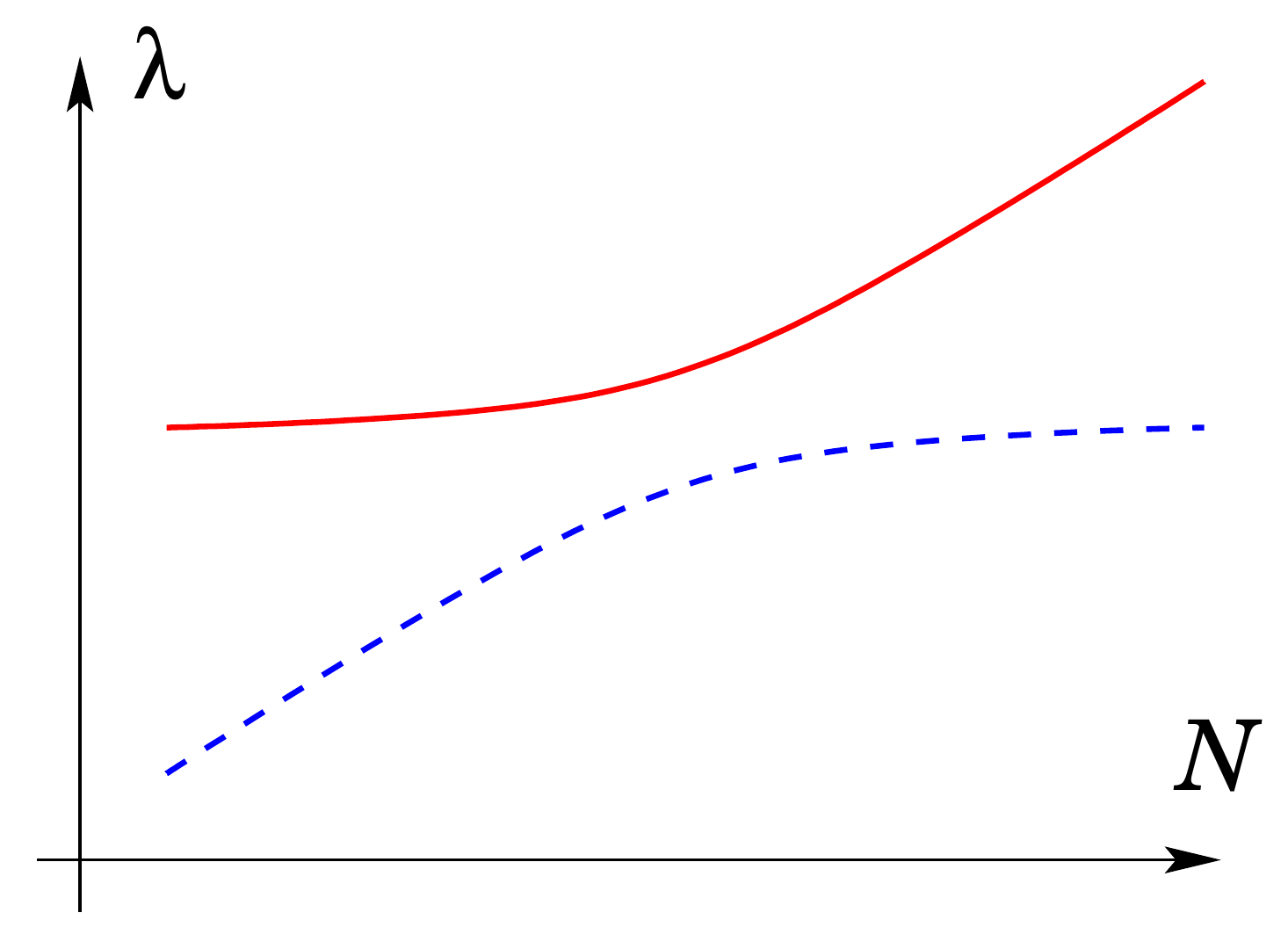}
%\caption{Comment.}
%\label{fig:resolved}
\end{minipage}
\qquad
\begin{minipage}{0.45\textwidth}
\centering
\includegraphics[width=0.9\textwidth, height=0.26\textheight]{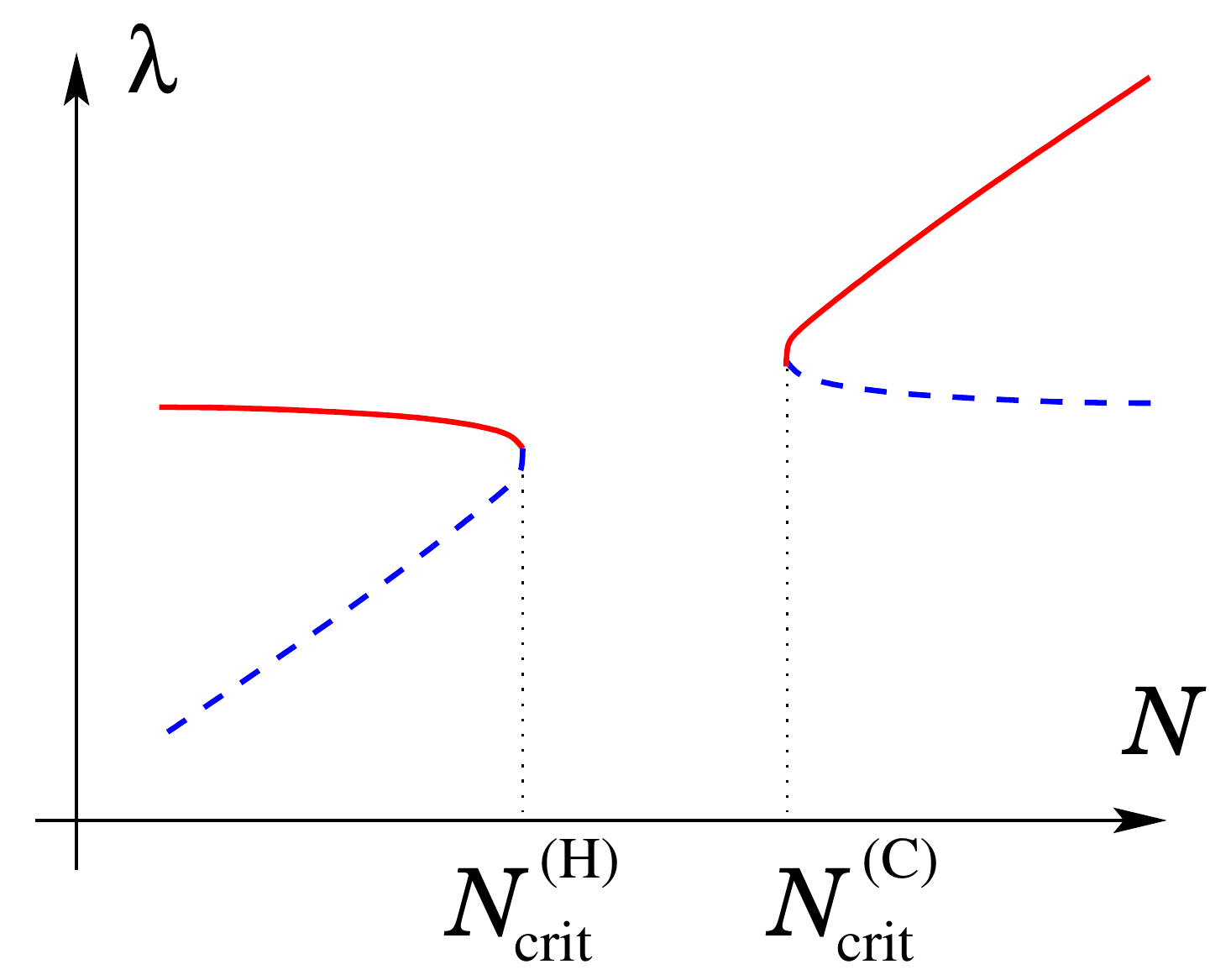}
%\caption{Comment.}
%\label{fig:deformed}
\end{minipage}
\caption{Unfolding transcritical bifurcation.}
\label{fig:deformed}
\end{figure}

Indeed, as we now know, the transcritical bifurcation is not of codimension-1 and,
as such, can not occur in a one-parameter system unless there is a fine-tuning and,
in addition, a symmetry (or a similar mechanism) protecting the fine-tuning from perturbations.
Otherwise, an arbitrarily small perturbation will destabilize the transcritical bifurcation
transforming it either into a pair of two independent saddle-node bifurcations or
into two smoothly changing branches of fixed points without any bifurcation, as illustrated in Figure~\ref{fig:deformed}.
A string theorist might call these two ways of unfolding the transcritical bifurcation
a {\it resolution} and {\it deformation}, respectively.

\begin{figure}[t!]
\begin{center}
\includegraphics[width=0.45\textwidth]{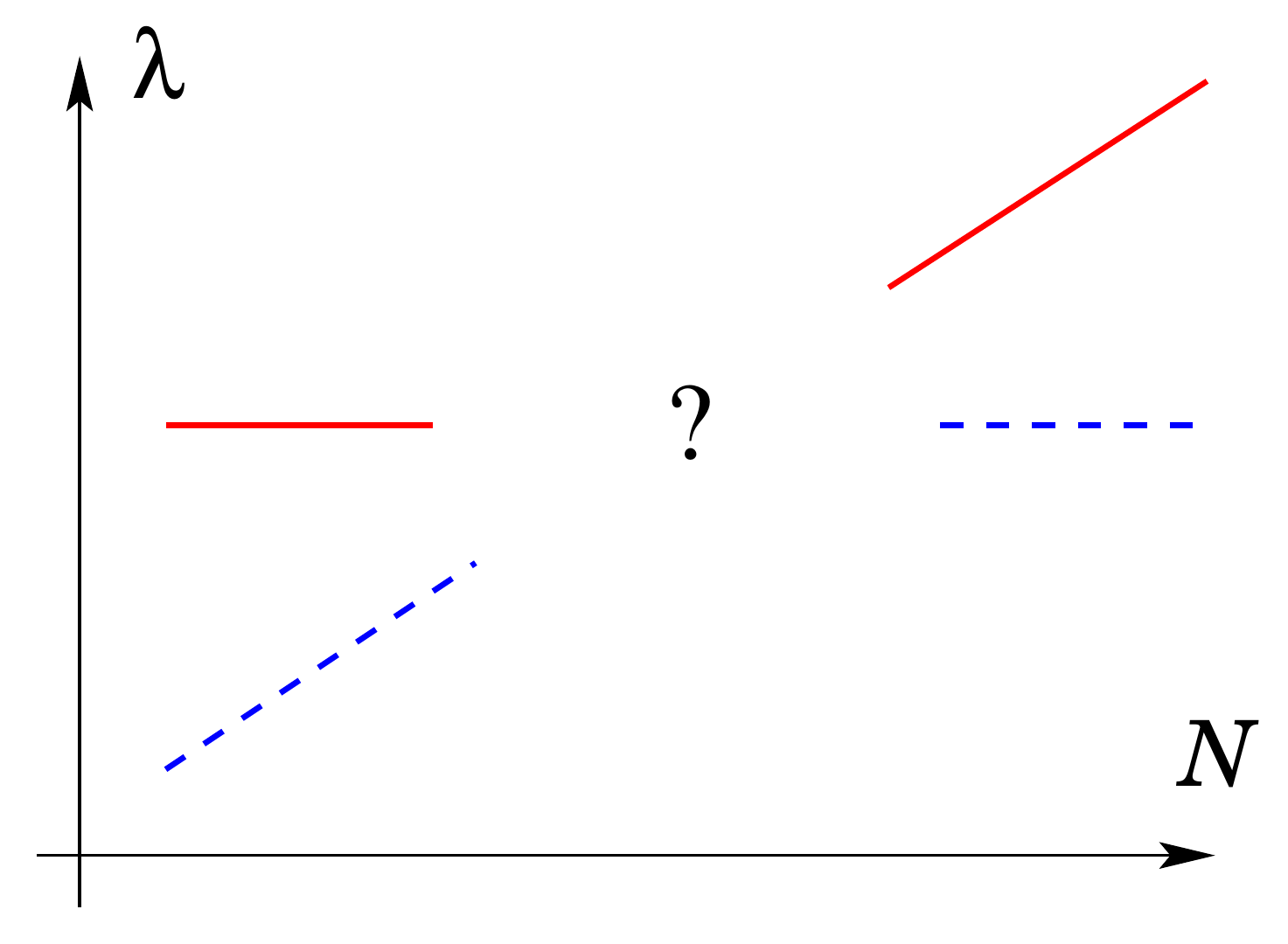}
\end{center}
\caption{\label{fig:transcritpuzzle} If we are presented with a ``black box'' and asked to fill in the intermediate
region of a {\it structurally stable} bifurcation diagram compatible with the boundary conditions (at small $N$ and large $N$),
we would produce the lower panels of Figure~\ref{fig:deformed}, but not the top one.}
\end{figure}

Another way to explain this phenomenon is to imagine that --- as was often the case in sections \ref{sec:Conley} and \ref{sec:bifurcations} ---
we know the limiting behavior of the system for small values of $N$ and for large $N$,
but need to determine what happens in the intermediate regime.
This situation, illustrated in Figure~\ref{fig:transcritpuzzle},
is in fact a fairly accurate summary of numerical simulations and experimental measurements in the 3d $O(N)$ model.
There are three ways to complete the partial bifurcation diagram in Figure~\ref{fig:transcritpuzzle},
which are precisely the possibilities shown in Figure~\ref{fig:deformed}.
Two of these possibilities (namely, the two lower panels) represent generic behavior and do not require any fine tuning,
whereas the third possibility (shown in the top panel) can be viewed as a special case of the lower
panels where one has to arrange the two curves meet at a point.
This is the reason why transcritical bifurcation has codimension 2 and is structurally unstable in a theory with one control parameter.

Nevertheless, there is a simple and instructive reason why
it is the latter possibility (represented by the top panel in Figure~\ref{fig:deformed})
which is realized in the 3d $O(N)$ model.
First, since numerical evidence rather clearly shows that the Wilson-Fisher fixed point is stable for small $N$
and the cubic fixed point is stable for large $N$,
\begin{eqnarray}
N=2: \quad & \det (J) \vert_{\text{Wilson-Fisher}} \; > \; 0 \,, & \quad \det (J) \vert_{\text{Cubic}} \; < \; 0   \\
N=4: \quad & \det (J) \vert_{\text{Wilson-Fisher}} \; < \; 0 \,, & \quad \det (J) \vert_{\text{Cubic}} \; > \; 0  \nonumber
\end{eqnarray}
it immediately rules out the possibility (shown in the lower left panel of Figure~\ref{fig:deformed})
that the transcritical bifurcation is ``deformed'' into two smoothly changing branches of fixed points without any bifurcation.
Normally, {\it i.e.} in the absence of fine tuning and symmetries, this would be the end of the story,
leaving us with only one option, illustrated in the lower right panel of Figure~\ref{fig:deformed}.

However, in the 3d $O(N)$ model the story is a little more interesting because the Wilson-Fisher fixed point has $O(N)$ symmetry,
whereas the cubic fixed point enjoys only a part of this symmetry given by
the semi-direct product of the symmetric group $S_N$ with $(\Z_2)^N$.
This is precisely the symmetry that, in the 3d $O(N)$ model, prevents the unfolding of the transcritical bifurcation,
at least to all orders in perturbation theory.\footnote{It is a pleasure to thank I.~Klebanov
and V.~Rychkov for useful discussions on this point.}
In general, if the operator crossing through marginality in a transcritical bifurcation preserves
the full symmetry of the system, then nothing prevents the ``unfolding'' shown in the lower panels of Figure~\ref{fig:deformed}.
This will be indeed the situation in some other examples,
such as the higher-dimensional version of the $O(N)$ model or QED$_3$,
where the transcritical bifurcation will show up again.
However, if the marginality crossing involves an operator that breaks the symmetry of the stable fixed point,
then it prevents the unfolding and protects the transcritical bifurcation.
This is precisely what happens in the 3d $O(N)$ model, where the Wilson-Fisher fixed point
and the cubic fixed point have different symmetries.

This behavior can be also verified directly, by examining the perturbative RG flow in the 3d $O(N)$ model.
Namely, one can check that including the higher-loop terms does not affect the structure of the transcritical bifurcation:
\begin{eqnarray}
\dot \lambda_1 & = & (d-4) \lambda_1 + \frac{9 \lambda_1^2 + (N-1)\lambda_2^2}{8 \pi^2}
- \frac{51 \lambda_1^3 + 5(N-1) \lambda_1 \lambda_2^2 + 4 (N-1) \lambda_2^3}{64 \pi^4} + \ldots  \nonumber \\
\dot \lambda_2 & = & (d-4) \lambda_2 + \frac{6 \lambda_1 \lambda_2 + (N+2)\lambda_2^2}{8 \pi^2}
- \frac{15 \lambda_1^2 \lambda_2 + 36 \lambda_1 \lambda_2^2 + 9 (N-1) \lambda_2^3}{64 \pi^4} + \ldots
\end{eqnarray}
Written here is a 2-loop RG flow (see {\it e.g.} \cite{Kleinert:1994td,Fei:2015oha}) and one can verify that
truncating it to 1-loop terms or, in the opposite direction, including 3-loop corrections, does not unfold the transcritical
bifurcation where the cubic fixed point and the Wilson-Fisher fixed point exchange their stability properties.

Note, that all three bifurcation diagrams shown in Figure~\ref{fig:deformed} belong to the family
\be
\dot \lambda \; = \; u + x \lambda - \lambda^2
\label{transunfold}
\ee
where we added the unfolding parameter $u$ to the normal form of the transcritical bifurcation \eqref{transcritical2d}.
Here, $u>0$ and $u<0$ correspond to the two topologically distinct ways of unfolding
the original transcritical bifurcation \eqref{transcritical2d} which, in turn, corresponds to $u=0$.
In all of these cases, one can read off the scaling dimensions of the nearly marginal operators
at the two fixed points of the RG flow equation \eqref{transunfold}:
\be
\Delta - d \; \sim \; \pm \sqrt{4u + x^2}
\label{dtransunfold}
\ee
In our applications, the control parameter $x = N - N_{\text{crit}}$.
And, since \eqref{dtransunfold} is supposed to describe scaling dimensions only in the vicinity of $N_{\text{crit}}$,
we can focus only on the leading behavior, which turns out to be either square root \eqref{dsqroot},
or quadratic \eqref{dquadr}, or linear \eqref{dlinear}, depending on whether $u<0$, $u>0$, or $u=0$, respectively:
\begin{subequations}\label{dtransthree}
\begin{eqnarray}
u<0: & ~~ & \Delta - d \; \sim \; \sqrt{N - N_{\text{crit}}} \\
u>0: & ~~ & \Delta - \Delta_0 \; \sim \; (N - N_{\text{crit}})^2 \\
u=0: & ~~ & \Delta - d \; \sim \; N - N_{\text{crit}}
\end{eqnarray}
\end{subequations}
We can summarize this by saying that the scaling dimension of a slightly irrelevant operator
can be used as a diagnostic tool for each of the three types of behavior in Figure~\ref{fig:deformed}.
In other words, measuring $\Delta$ as a function of $N$ can unambiguously determine topology of
the bifurcation diagram and, conversely, merely from topology of the bifurcation one can predict
the shape of $\Delta (N)$ near the critical value $N_{\text{crit}}$.

\begin{figure}[h]
\centering
\begin{minipage}{0.45\textwidth}
\centering
\includegraphics[width=0.9\textwidth, height=0.25\textheight]{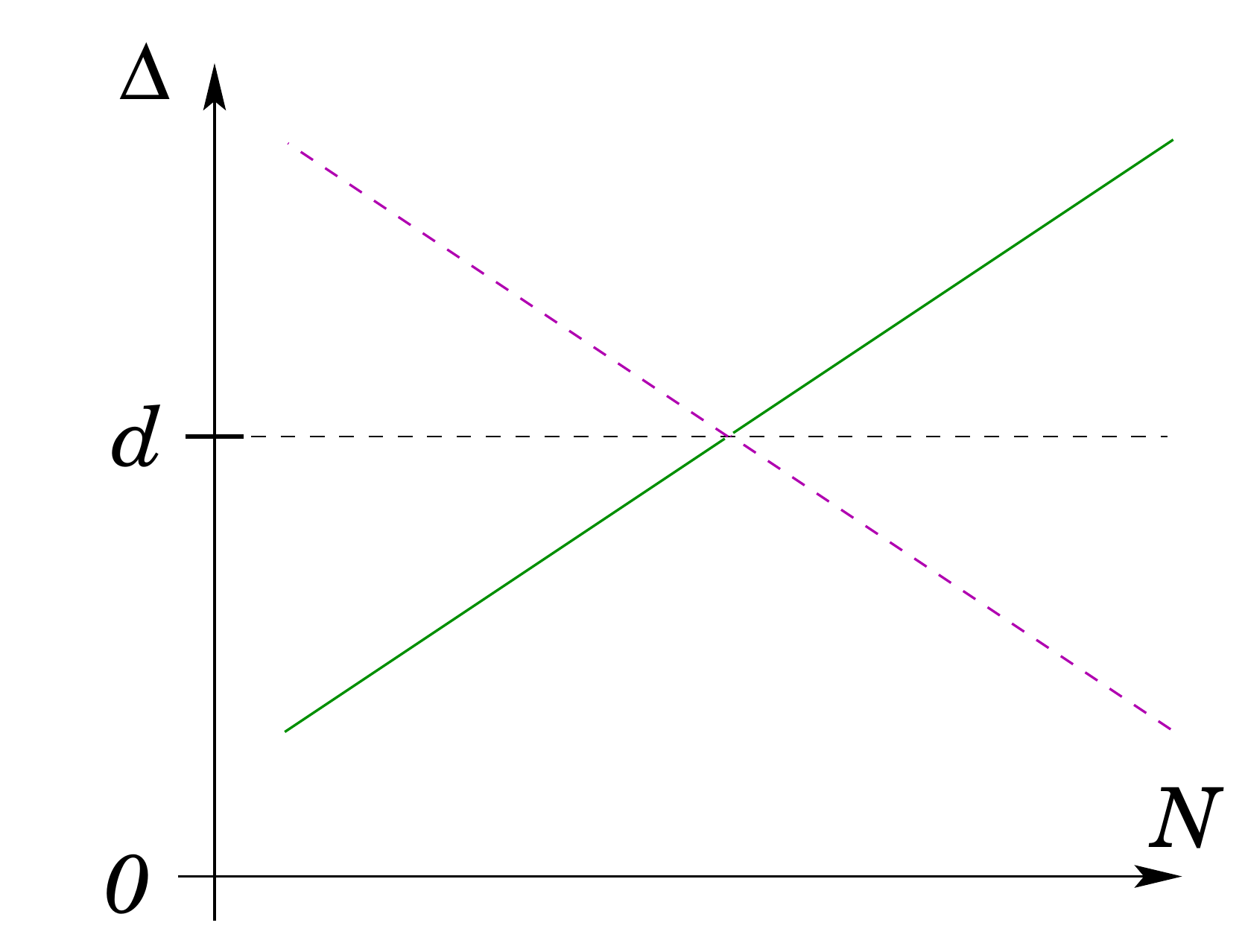}
\end{minipage}
\qquad
\begin{minipage}{0.45\textwidth}
\centering
\includegraphics[width=0.9\textwidth, height=0.25\textheight]{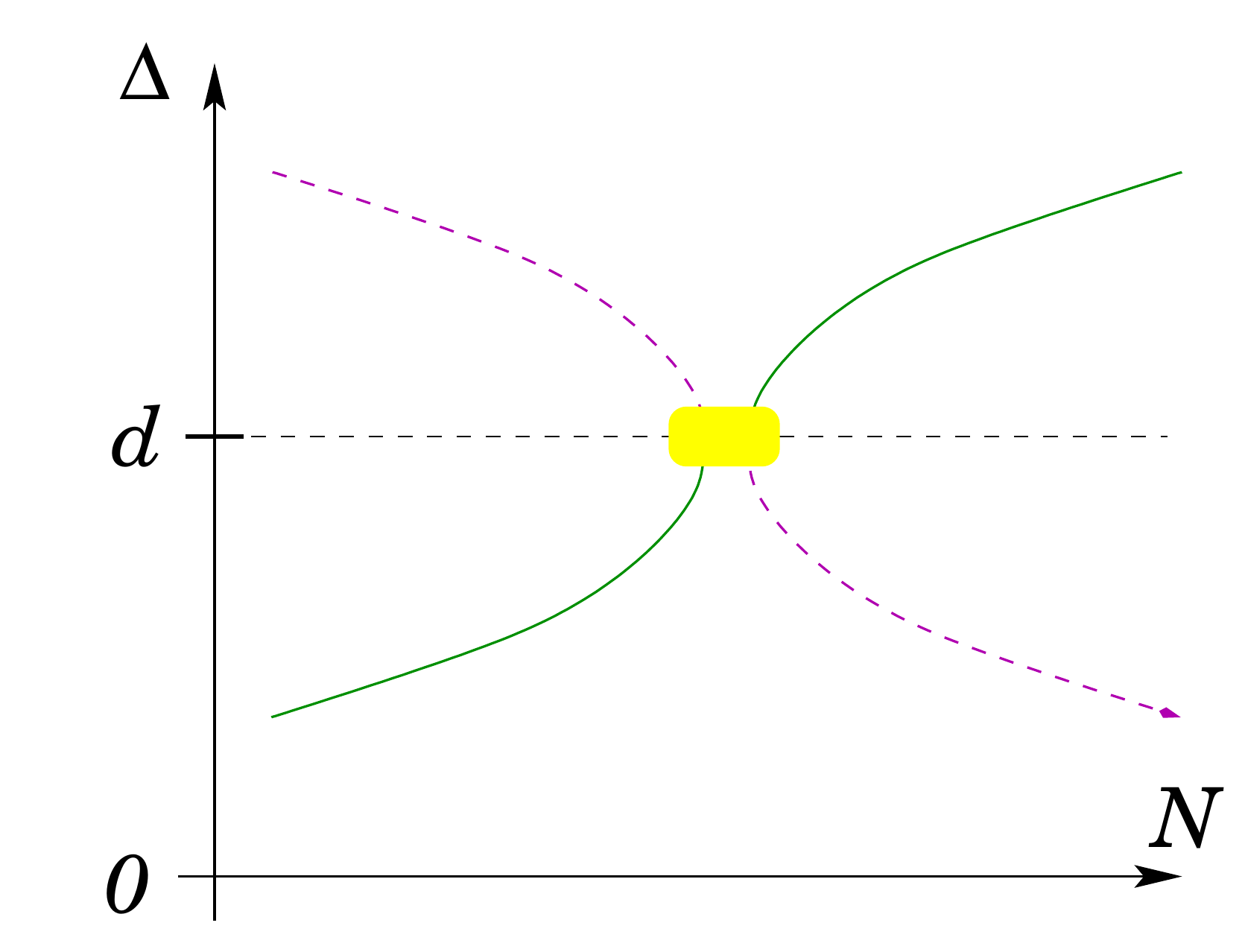}
\end{minipage}
\caption{Different types of phase transitions {\it e.g.} in the $O(N)$ model correspond to very different characteristic behavior
of the scaling dimension of a nearly marginal operator. One common feature, though, is that both the saddle-node (right)
and the transcritical (left) bifurcations require this operator to become marginal at the transition point.}
\label{fig:linvssqrt}
\end{figure}

Thus, in the ordinary 3d $O(N)$ model, the transcritical bifurcation at $N_{\text{crit}}$
implies that the scaling dimension of a slightly irrelevant operator crosses through marginality
in a linear fashion, as illustrated in Figure~\ref{fig:linvssqrt}.
We hope that measuring $\Delta (N) \sim \det (J)$ with sufficient level of precision
can help to reconcile some of the discrepancies in various studies of the behavior
of 3d $O(N)$ model near $N_{\text{crit}}$ and various attempts
to determine this value precisely (which lead to results scattered around $N_{\text{crit}} \approx 3$).

For example, in earlier studies based on $\epsilon$-expansion
it was found that the Wilson-Fisher fixed point is stable at $N=3$, suggesting that $N_{\text{crit}} > 3$.
Then, later studies based on careful resummation of the perturbative series \cite{Shpot:1989sb,Kleinert:1996hy}
computed the eigenvalues of the Jacobian matrix at the Wilson-Fisher fixed point for $N=3$
and concluded that $\det (J) < 0$, {\it i.e.} this fixed point is unstable at $N=3$.
On the other hand, a high precision Monte Carlo simulation \cite{Caselle:1997gf} led to $\det (J) > 0$
for the same problem (Wilson-Fisher fixed point at $N=3$), though the error bars on the smallest
eigenvalue of $J$ were rather high, $d - \Delta = - 0.0007(20)(9)$.
(Here, the first error in parenthesis denotes the statistical uncertainty, while the second error is due to
uncertainty of the critical coupling used in simulations.)
Curiously, the results of Monte Carlo simulation \cite{Caselle:1997gf} show a rather strong asymmetry for
the behavior of $\det (J)$ above and below the critical regime.
See \cite{Pelissetto:2000ek} for further discussion and references therein.\footnote{Note, that our sign conventions
for the eigenvalues of the stability matrix (a.k.a. the Jacobian matrix) follow the standard conventions in dynamical systems.
Some of the physics papers use the opposite sign conventions, motivated by the sign of the beta-function.}

In general --- meaning not only in the $O(N)$ model, but also in other examples --- measuring one of the characteristic
types of behavior \eqref{dtransthree} may require sufficiently high level of precision, especially since control parameters
often take integer values, just like $N$ in the case of the $O(N)$ model.
In some cases, however, recognizing different types of bifurcations may turn out to be extremely easy.
For example, if the same fixed point remains stable throughout the entire neighborhood of $N_{\text{crit}}$,
it is definitely a signature of (\ref{dtransthree}b) illustrated in the lower left panel of Figure~\ref{fig:deformed}.
Or, it may happen that fixed points simply cease to exist for certain (integer) values of $N$;
that would be a smoking gun for the behavior in the lower right of Figure~\ref{fig:deformed}
and scaling dimensions (\ref{dtransthree}a).

This latter possibility is, in fact, realized in a version of the $O(N)$ model analytically continued to $d = 6 - \epsilon$,
which shows a huge ``gap'' between $N_{\text{crit}}^{(\text{C})}$ and $N_{\text{crit}}^{(\text{H})}$ where the higher-dimensional
analogues of the cubic and the Heisenberg fixed points disappear in two independent bifurcations~\cite{Fei:2014yja,Fei:2014xta}:
\begin{eqnarray}
N_{\text{crit}}^{(\text{C})} & = & 1038.266 - 609.840 \epsilon - 364.173 \epsilon^2 + \CO (\epsilon^3) \\
N_{\text{crit}}^{(\text{H})} & = & 1.02145 + 0.03253 \epsilon - 0.00163 \epsilon^2 + \CO (\epsilon^3) \nonumber
\end{eqnarray}
Note, these two curves meet at $\epsilon \approx 1.04664$, which is not unexpected since the two saddle-node bifurcations
at $N_{\text{crit}}^{(\text{C})}$ and $N_{\text{crit}}^{(\text{H})}$ must turn into a transcritical bifurcation in $d=4$,
{\it i.e.} at $\epsilon = 2$, where the ``intermediate phase'' must completely disappear.
However, this also suggests that the higher-loop corrections, which are especially large in the case of $N_{\text{crit}}^{(\text{C})}$,
change the behavior of $N_{\text{crit}}^{(\text{C})} (\epsilon)$ and $N_{\text{crit}}^{(\text{H})} (\epsilon)$
in such a way that the two meet precisely at $\epsilon = 2$ or, equivalently, in $d=4$:
\be
N_{\text{crit}}^{(\text{C})} \vert_{\epsilon = 2} \; = \;
N_{\text{crit}}^{(\text{H})} \vert_{\epsilon = 2} \; = \; 4
\ee
where we also used the fact that $N_{\text{crit}} = 4$ in $d=4$ (or, rather, in $d=4-0$).
It would be interesting to test this prediction numerically or analytically.\footnote{Note, this
is yet another instance of the same principle we encounter over and over again, where alternative approaches provide
information about the RG flow in two different limits or regimes, while methods of dynamical systems ``fill in'' the rest,
at least qualitatively. In the present case, the standard perturbative techniques give us rather detailed information about the flow
in $d=4$ and $d=6$, where the space-time dimension $d$ plays the role of a control parameter, and bifurcation theory
then determines what should happen between these two limiting cases, when $4 < d < 6$.}

It is curious to note that similar intermediate phases appeared in analytical and numerical studies
of gauge theories that will be our next subject,
see {\it e.g.} \cite{Kogut:2002vu} for a lattice study\footnote{One puzzling aspect of the study in \cite{Kogut:2002vu}
is that it finds critical exponents of the chiral phase transition far from mean field theory values.}
of compact QED$_4$ or \cite{Braun:2014wja,Janssen:2016nrm} for functional RG and $\epsilon$-expansion in non-compact QED$_3$.
In particular, the latter proposed that chiral symmetry breaking in non-compact QED$_3$ is separated from
conformal phase transition by a new intermediate phase characterized by a Lorentz-breaking vector
condensate $\langle \bar \Psi \gamma_{\mu} \Psi \rangle \ne 0$ (with $\langle \bar \Psi \Psi \rangle = 0$).
On the other hand, in compact QED$_4$, four-fermion interactions --- which will be the main subject of the following
discussion --- separate the line of second-order chiral symmetry breaking phase transition from
a first-order confinement-deconfinement phase transition controlled by monopole condensation
(where the monopole concentration $\langle M \rangle$ jumps discontinuously).

%%%%%%%%%%%%%%%%%%%%%%%%%%%%%%%%%%%%%%%%%%%%%%%%%%%%%%%%%%%%%%%%%%%%%%%%%%%%%%%%%%%%%%%%%%%%%%%%

\section{Application to QED$_3$}
\label{sec:QED}

During the past 30 years, three-dimensional parity-invariant quantum electrodynamics (QED$_3$) received a lot of attention
for its numerous applications in modern condensed matter physics and its similarities to four-dimensional QCD-like theories.

There are several closely related versions of this theory, and we shall primarily focus on the so-called {\it non-compact} QED$_3$,
which has
no monopoles\footnote{Sometimes this version of QED$_3$ is presented as a theory with non-compact gauge group $G = \R$,
see {\it e.g.} \cite{DiPietro:2015taa}.
The monopole dynamics in the compact and non-compact versions may be different at small values of $N_f$.
Some lattice simulations suggest that, for $N_f > 1$, monopole dynamics does not affect
confining properties of the theory, see {\it e.g.} \cite{Armour:2010fx}.}
and $N_f$ four-component charged massless spinors, $\Psi$ and $\bar \Psi$,
acted upon by the usual $4 \times 4$ Dirac matrices $\gamma_{\mu}$, $\mu=0,1,2$.
Although in three dimensions such spinor representations are reducible, this formulation has some advantages:
it preserves parity and is convenient for extrapolating to four dimensions.
In terms of irreducible complex two-component fermions, the total number of flavors is $2N_f$
and the flavor (``chiral'') symmetry is $SU(2N_f)$; it combines the obvious $SU(N_f)$ symmetry
of 4-component fermions with rotations in the space of irreducible subcomponents of the Dirac spinors.
The latter is generated by ${\bf 1}$, $\gamma_3$, $\gamma_5$, and $i \gamma_3 \gamma_5$,
resulting in a global $U(2)$ symmetry for each four-component spinor.

The Lagrangian of massless QED$_3$ that preserves parity and the $SU(2N_f)$ flavor/chiral symmetry
simply consists of the (gauge covariant) kinetic terms for the gauge field and for the fermions.
Since the gauge coupling $e^2$ has mass dimension one, it sets the scale analogous to $\Lambda_{QCD}$ in four dimensions,
above which QED$_3$ is weakly-interacting. As a result, the theory is asymptotically free for very simple dimensional reasons.
For sufficiently large $N_f$, the screening of fermion fluctuations keeps the coupling small,
so that the theory remains in the disordered massless phase and has a non-trivial IR stable fixed point.
%For a sufficiently large number of fermion flavors the fermion screening of the electric charge overcomes the attractive
%interaction of a fermion and an anti-fermion due to photon exchange.
Starting with the early analysis of Schwinger-Dyson equations \cite{Pisarski:1984dj,Appelquist:1988sr},
the theory is believed to exhibit logarithmic confinement of electric charges and chiral symmetry breaking
when the number of fermion flavors $N_f$ becomes smaller than some critical value $N_f^{\text{crit}}$.
If the fermions acquire dynamical mass the $SU(2N_f)$ symmetry is broken spontaneously to $SU(N_f) \times SU(N_f) \times U(1)$:
\be
SU (2 N_f) \; \to \; SU(N_f) \times SU(N_f) \times U(1)
\label{UUUbreakingQED}
\ee
and $2 N_f^2$ Goldstone bosons appear in the particle spectrum.
In practice, {\it e.g.} in numerical simulations\footnote{On a lattice,
only a subgroup $SU(N_f) \times SU(N_f) \subset SU(2N_f)$ of the chiral/flavor symmetry
is manifest, which is further broken to $SU(N_f)$ either explicitly by $m \ne 0$ or spontaneously by
the chiral chiral condensate $\langle \bar \Psi \Psi \rangle \ne 0$.}, one often studies whether non-zero chiral condensate
is generated by introducing bare fermion mass $m \ne 0$ (which is known to generate $\langle \bar \Psi \Psi \rangle \ne 0$)
and then taking the limit $m \to 0$, see {\it e.g.} \cite{Hands:2002qt}.
Therefore, in our study of massless QED$_3$ it will be convenient to embed it in a larger family of theories,
\be
\CL \; = \; -\frac{1}{4e^2} F_{\mu \nu} F^{\mu \nu}
+ \sum_{a=1}^{N_f} \bar{\Psi}_a (i \gamma^\mu D_\mu - m) \Psi^a + \CL_{\text{4-fermi}}
\label{QED3Lagr}
\ee
which includes a bare mass term and irrelevant four-fermion self-interactions\footnote{Following the standard practice,
we assume the summation over the repeated flavor index $a = 1, \ldots, N_f$.}
\be
\CL_{\text{4-fermi}} \; = \;
\frac{g}{N_f} (\bar \Psi_a \Psi^a)^2 +
\frac{g'}{N_f} (\bar \Psi_a \gamma_{\mu} \gamma_{35} \Psi^a)^2 +
\frac{\lambda}{N_f} (\bar \Psi_a \gamma_{35} \Psi^a)^2 +
\frac{\lambda'}{N_f} (\bar \Psi_a \gamma_{\mu} \Psi^a)^2
\label{QED4fermi}
\ee
As the reader may anticipate from the general theme of this paper, the four-fermion interaction will become
relevant at some point, literally and figuratively.

The extra terms which we added to the massless QED$_3$ Lagrangian can be generated dynamically and, if so,
can break some part of the symmetry. In general, there are two possible mass terms\footnote{Other fermion
bilinears involving $\gamma_3$ and $\gamma_5$ are $SU(2N_f)$ equivalent to these two.}:
$m \bar \Psi \Psi = m (\bar \psi_a \psi^a - \bar \psi_{a+N_f} \psi^{a+N_f})$
and $\tilde m \bar \Psi \gamma_{35} \Psi = \tilde m \bar \psi_i \psi^i$.
In particular, writing these mass terms in terms of complex two-component fermions $\psi_i$, $i = 1, \ldots, 2N_f$,
makes it clear that the first mass term preserves parity and corresponds to a symmetry breaking pattern \eqref{UUUbreakingQED},
whereas the second one preserves $SU(2N_f)$ but breaks parity and time-reversal symmetry.

Likewise, the four-fermion interactions that preserve $SU(2N_f)$ flavor symmetry
and the discrete $C$, $P$, and $T$ symmetries have been completely classified, see {\it e.g.} \cite{Kaveh:2004qa,Braun:2014wja}.
In the flavor-singlet channel, the space of such couplings is two-dimensional,
and the last two terms in \eqref{QED4fermi} could be chosen as its basis
(with all other choices related to it via linear transformations and Fierz identities).
Note, these two 4-fermion operators preserve the same symmetries as $F_{\mu \nu} F^{\mu \nu}$
and, therefore, they all can mix together. As one can anticipate from section \ref{sec:bifurcations},
mixing between these operators will play an important role below in unfolding the bifurcation at $N_f^{\text{crit}}$.
%There are also two independent flavor-non-singlet channels invariant under the full $U(2N_f)$,
%which in the zero-momentum (pointlike) limit become linearly dependent with the flavor-singlet channels.
Relaxing the symmetry to $SU(N_f) \times SU(N_f) \times U(1)$, the space of 4-fermion interactions
becomes four-dimensional, spanned by the linear combinations of the four terms in \eqref{QED4fermi}.

In various special limits the above Lagrangian \eqref{QED3Lagr} reduces to other interesting theories.
For example, the special case of $e=0$ and $g'=\lambda=\lambda'=0$ gives the Gross-Neveu model\footnote{The four-fermion
interaction proportional to $\lambda$ is also similar to the interaction in the Gross-Neveu model;
in fact, it becomes identical to the Gross-Neveu interaction when written in terms of two-component spinors.}
(where a runaway flow for $g$ can be interpreted as dynamical mass generation,
$m \sim \langle \bar \Psi \Psi \rangle$, see {\it e.g.} \cite{Kaveh:2004qa}).
Another special limit, $e=0$ and $g=g'=\lambda=0$, gives the Thirring model.

While in the case of the $O(N)$ model that we discussed in the previous section the debate is whether $N_{\text{crit}} < 3$ or $N_{\text{crit}} > 3$,
in the case of QED$_3$ a lot of attention is centered around $N_f = 2$ and how $N_f^{\text{crit}}$ is positioned relative to it.
The reason, in part, comes from applications to layered condensed matter systems, such as
%graphene \cite{Cortijo:2011aa},
high-$T_c$ cuprate superconductors, surface states of topological insulators \cite{Vafek:2013mpa},
or the unconventional quantum Hall effect in graphene \cite{Gusynin:2005pk}.
%Potential applications to models of high-$T_c$ superconductivity \cite{Dorey:1991kp,Gusynin:2000zb}.
For example, QED$_3$ was proposed to describe the effective theory for the underdoped and non-superconducting phase
of high-$T_c$ superconducting cuprate compounds \cite{Franz:2002qy,Herbut:2002yq}, where the low-energy gapless
quasiparticle excitations at the four nodes on the Fermi surface compose $N_f=2$ four-component Dirac spinors of $QED_3$.
And, if $N_f^{\text{crit}} < 2$ in QED$_3$, then the superconducting phase in these CuO$_2$ layers
is separated from the antiferromagnetic phase by an unconventional non-Fermi-liquid phase (``pseudogap phase'')
whose properties differ from those of the standard Fermi liquid due to non-perturbative anomalous dimensions of the fermions.
On the other hand, if $N_f^{\text{crit}} > 2$, then QED$_3$ predicts a direct phase transition at some non-zero
doping (and $T=0$) from the $d$-wave superconducting phase to the antiferromagnetic phase,
where the chiral condensate of QED$_3$ plays the role of an order parameter for spin density waves.

Therefore, determining the numerical value of $N_f^{\text{crit}}$ is an important problem, for both practical and theoretical reasons.
Even though it has been the subject of active research in the past 30 years, we still do not know what this value is.
Basically, a reader can think of any number between 1 and 10, and there is a good chance this number will appear
as one of the proposed estimates for $N_f^{\text{crit}}$ within $10\%$ accuracy, see Table~\ref{tab:NfQED}.
For all we know, $N_f^{\text{crit}}$ may even be zero, meaning that QED$_3$ flows to a conformal IR fixed point
for all values of $N_f$ and the chiral symmetry breaking \eqref{UUUbreakingQED} does not happen at all~\cite{Karthik:2015sgq}.

\begin{table}
\begin{centering}
\begin{tabular}{|c|c|l|}
\hline
~$\phantom{\int^{\int^\int}} N_f^{\text{crit}} \phantom{\int_{\int}}$~ & ~Method~ & ~Year and Reference \tabularnewline
\hline
\hline
$\phantom{\int^{\int^\int}} \le \frac{3}{2} \phantom{\int_{\int}}$ & thermal free energy & ~~~~~~1999 \cite{Appelquist:1999hr,Appelquist:2004ib} \tabularnewline
\hline
$\phantom{\int^{\int^\int}} 1.5 \phantom{\int_{\int}}$ & lattice simulations & ~~~~~~2008 \cite{Strouthos:2008kc} \tabularnewline
\hline
$\phantom{\int^{\int^\int}} \le 2 \phantom{\int_{\int}}$ & one-loop $\epsilon$-expansion & ~~~~~~2015 \cite{DiPietro:2015taa} \tabularnewline
\hline
$\phantom{\int^{\int^\int}} \le 2 \phantom{\int_{\int}}$ & Hybrid Monte Carlo & ~~~~~~2002 \cite{Hands:2002qt,Hands:2002dv,Hands:2004bh} \tabularnewline
\hline
$\phantom{\int^{\int^\int}} 2.16 \phantom{\int_{\int}}$ & divergence of the chiral susceptibility & ~~~~~~2002 \cite{Franz:2002qc} \tabularnewline
\hline
$\phantom{\int^{\int^\int}} 2.85 \phantom{\int_{\int}}$ & $1/N_f$ expansion & ~~~~~~2016 \cite{Gusynin:2016som} \tabularnewline
\hline
$\phantom{\int^{\int^\int}} 2.89 \phantom{\int_{\int}}$ & $\epsilon$-expansion & ~~~~~~2016 \cite{Herbut:2016ide} \tabularnewline
\hline
$\phantom{\int^{\int^\int}} \frac{32}{\pi^2} \approx 3.24 \phantom{\int_{\int}}$ & Schwinger-Dyson equations & ~~~~~~1984-88 \cite{Pisarski:1984dj,Appelquist:1988sr} \tabularnewline
\hline
$\phantom{\int^{\int^\int}} \le 4 \phantom{\int_{\int}}$ & F-theorem & ~~~~~~2015 \cite{Giombi:2015haa} \tabularnewline
\hline
$\phantom{\int^{\int^\int}} 4 \phantom{\int_{\int}}$ & covariant solutions for propagators & ~~~~~~2004 \cite{Fischer:2004nq} \tabularnewline
\hline
$\phantom{\int^{\int^\int}} 4.3 \phantom{\int_{\int}}$ & Schwinger-Dyson equations & ~~~~~~1996-97 \cite{Maris:1996zg,Aitchison:1997ua} \tabularnewline
\hline
$\phantom{\int^{\int^\int}} 6 \phantom{\int_{\int}}$ & perturbative RG in the large-$N_f$ limit & ~~~~~~2004 \cite{Kaveh:2004qa} \tabularnewline
\hline
$\phantom{\int^{\int^\int}} 5.1 $\ldots$ 6.6 \phantom{\int_{\int}}$ & comparison to the Thirring model & ~~~~~~2007-12 \cite{Christofi:2007ye,Janssen:2012pq} \tabularnewline
\hline
$\phantom{\int^{\int^\int}} 4 \approx N_f^{\chi \text{SB}} \le N_f^{\text{conf}} \le 10 \phantom{\int_{\int}}$ & functional RG & ~~~~~~2014 \cite{Braun:2014wja} \tabularnewline
\hline
\end{tabular}
\par\end{centering}
\caption{\label{tab:NfQED} Search for the critical value of $N_f$ in non-compact QED$_3$.}
\end{table}

Adding to the controversy, there is a wide range of opinions about {\it what} actually happens at $N_f^{\text{crit}}$.
%The analysis of Schwinger-Dyson equations \cite{Pisarski:1984dj,Appelquist:1988sr} suggests
%that, when $N_f < N_f^{\text{crit}}$, the chiral symmetry is broken and the fermion spectrum is gapped.
For example, some lattice simulations \cite{Strouthos:2008kc} suggest a relatively smooth second order phase transition.
Further support for this conclusion comes from the study of 3d Thirring model \cite{Janssen:2012pq},
which has the same global chiral/flavor symmetries and is expected to have $\chi$SB phase transition in the same universality class.
On the other hand, some analytical calculations predict that as $N_f$ approaches $N_f^{\text{crit}}$
the theory undergoes a conformal phase transition \cite{Appelquist:1994ui,Miransky:1996pd,Gusynin:1998kz},
which is a generalization of the infinite order Berezinskii-Kosterlitz-Thouless transition in two dimensions.
Alternatively, it has been suggested that (in the compact version of QED$_3$)
the transition is due to monopole operators reaching the unitarity bound~\cite{Safdi:2012re}.
Yet another proposal \cite{Braun:2014wja} is that a chiral symmetry breaking transition is separate from
the conformal phase transition which, in turn, is due to annihilation of the IR stable fixed point
and another fixed point where the four-fermion interaction of the Thirring model is turned on.

Our goal is to examine the problem from the vantage point of bifurcation theory.
Consider, for example, one-loop RG flow equations from \cite{Kaveh:2004qa}
that describe massless QED$_3$ with large $N_f$ (and $m=0$):
\begin{eqnarray}
\frac{de^2}{d \ln l} & = & e^2 - N_f e^4 + \ldots  \nonumber \\
\frac{d g}{d \ln l} & = & - g - g^2 + 4 e^2 g + 18 e^2 g' + \ldots  \nonumber \\
\frac{d g'}{d \ln l} & = & - g' + {g'}^2 + \frac{2}{3} e^2 g + \ldots \label{KavehRGflow}  \\
\frac{d \lambda}{d \ln l} & = & - \lambda - \lambda^2 + 4 e^2 \lambda + 18 e^2 \lambda' + 9N_f e^4 + \ldots  \nonumber \\
\frac{d \lambda'}{d \ln l} & = & - \lambda' + {\lambda'}^2 + \frac{2}{3} e^2 \lambda + \ldots \nonumber
\end{eqnarray}
where all couplings have been redefined to produce dimensionless quantities,
{\it e.g.} in the case of gauge coupling $\frac{2 e^2}{3 \pi^2 \Lambda} \to e^2$,
similarly $\frac{4g \Lambda}{\pi^2} \to g$, {\it etc}.
Since to the leading order the $\beta$-function for the gauge coupling does not depend on the 4-fermion interactions,
all fixed points have $e^2 = 0$ or $e^2_* = \frac{1}{N_f}$.
The former leads to the Gaussian UV fixed point (all 4-fermion interactions are zero),
to the Gross-Neveu model ($g \ne 0$), to the Thirring model ($\lambda' \ne 0$),
and to their various hybrids and generalizations.
On the other hand, the non-trivial value of the gauge coupling $e^2_* = \frac{1}{N_f}$
leads to interacting CFT in the conformal window $N_f \ge N_f^{\text{crit}}$.

Moreover, as emphasized in \cite{Kaveh:2004qa}, to the one-loop order, the RG flow equations for the 4-fermion
interactions \eqref{KavehRGflow} split into two pairs, in both of which $e^2_* = \frac{1}{N_f}$ can be treated as a parameter.
Using a simple change of variables $e^2_* = \frac{1}{N_f} = \frac{1+x}{6}$,
$g = \frac{3}{2} \lambda_1 - \frac{3}{2} \lambda_2$ and $g' = \frac{1}{6} \lambda_1 + \frac{1}{2} \lambda_2$,
the first pair of the beta-function equations for the symmetry-breaking interactions can be conveniently written as
\begin{eqnarray}
\dot \lambda_1 & = & x \lambda_1 - \frac{13}{12} \lambda_1^2 - \frac{3}{4} \lambda_2^2 + \frac{5}{2} \lambda_1 \lambda_2 + \ldots \\
\dot \lambda_2 & = & - \frac{4+x}{9} \lambda_2  - \frac{2x}{9} \lambda_1
+ \frac{41}{108} \lambda_1^2 + \frac{5}{12} \lambda_2^2 - \frac{13}{18} \lambda_1 \lambda_2 + \ldots \nonumber
\end{eqnarray}
In the range of definition of $x \in (-1 , + \infty)$, the linear stability analysis gives only one critical value, $x_{\text{crit}} = 0$,
near which the anomalous dimension of $\lambda_2$ remains finite, whereas the RG flow equation for $\lambda_1$
exhibits a transcritical bifurcation, {\it cf.} \eqref{transcritical2d}. At this critical value of $x$, the fixed point with
no 4-fermion interactions interchanges its stability properties with the gauged Gross-Neveu fixed point at
$(\lambda_1 , \lambda_2) \simeq \left( \frac{12x}{13} , - \frac{24 x^2}{13 (4+x)} \right)$.
Even though the above leading order RG flow equations indicate otherwise, \cite{Kaveh:2004qa} talks about
annihilation of these two fixed points at $x_{\text{crit}}$, see also \cite{Herbut:2016ide}.

Curiously, while this conlcusion was not fully justified by the approximation used in \cite{Kaveh:2004qa}, it is actually
consistent with our approach based on bifurcation theory. Indeed, as we learned earlier, transcritical bifurcations
never stay in the exact theory with a single parameter (unless there are symmetries protecting them),
and in our present context of QED$_3$ higher-loop corrections and strong coupling effects will ``unfold''
the transcritical bifurcation either as in Figure~\ref{fig:hysttrans} or as in Figure~\ref{fig:deformed}.
Both the higher-order effects (as in Figure~\ref{fig:hysttrans})
and the unfolding with $u<0$ (as in (\ref{dtransthree}a) illustrated in the lower right of Figure~\ref{fig:deformed})
are signalled by a characteristic square-root approach to marginality,
\be
\Delta - d \; \sim \; \sqrt{N_f - N_f^{\text{crit}}}
\label{sqrtMC}
\ee
instead of a linear behavior $\Delta-d \; \sim \; (N_f - N_f^{\text{crit}})$, {\it cf.} Figure~\ref{fig:linvssqrt}.
On the other hand, if the unfolding parameter ends up with a different sign,
namely $u>0$ in the notations of (\ref{dtransthree}b), also illustrated in the lower left of Figure~\ref{fig:deformed},
then the scaling dimensions of nearly marginal operators should exhibit quadratic behavior (with $\Delta_0 > d$):
\be
\Delta - \Delta_0 \; \sim \; \left( N_f - N_f^{\text{crit}} \right)^2
\label{quadMC}
\ee
which is probably less familiar among the three options in (\ref{dtransthree}).
Indeed, such corrections (that lead to unfolding) already show up at the leading order in the second pair of 4-fermion
beta-functions \eqref{KavehRGflow}.
In particular, the term $9N_f e^4$ has the effect of unfolding the transcritical bifurcation in the space of
chiral symmetry preserving couplings $\lambda$ and $\lambda'$,
which otherwise are identical to the RG flow equations for $g$ and $g'$.
%If similar corrections are generated by higher-loop and non-perturbative effects in the first pair of equations for $g$ and $g'$,
%it would lead to unfolding of the transcritical bifurcation at $N_f^{\text{crit}}$, as illustrated in Figure~\ref{fig:deformed},
%and the square-root approach to marginality \eqref{sqrtMC} would be a signature of this behavior.
And, it was stressed already in \cite{Kaveh:2004qa} that, to the next order in the $1 / N_f$ expansion,
the $\beta$-functions for the 4-fermion interactions mix all of the couplings, so the terms leading to unfolding
of the transcritical bifurcation are indeed generated.

It is also instructive to point out that $e^2$, which plays the role of the control parameter in the four last equations of \eqref{KavehRGflow},
affects only lower-order terms. In particular, it affects the structure of the fixed point set in each pair of couplings,
but not the exit set $L \cong I$, which in both cases is determined by the 3-point functions.
We already encountered such two-coupling systems several times in section \ref{sec:Conley}
and from the previous computations summarized in Table~\ref{tab:2dflowsCH} we know that the homological Conley index is trivial,
\be
CH_* (S) \; = \; 0
\ee
The RG flows which realize this must have an even number of fixed points (if the fixed points are isolated)
that can be organized in pairs of fixed points with index \eqref{mudef} equal to $\mu$ and $\mu + 1$,
{\it cf.} examples in Figure~\ref{fig:Type I flow} and in Figure~\ref{fig:Type II flow}.
This agrees with the structure of the phase portraits shown in \cite{Kaveh:2004qa}.

Similarly, we can perform a ``bifurcation diagnostics'' on RG flows obtained by other methods.
For instance, a recent work \cite{DiPietro:2015taa} studied one-loop $\beta$-functions and anomalous dimensions
in QED$_3$ using the $\epsilon$-expansion.
If one is to hope that the quantitative estimate for $N_f^{\text{crit}}$ produced in this analysis is reasonable,
certainly the qualitative features of the analysis must be reliable too.
However, the latter imply that the transition at $N_f^{\text{crit}}$ is a transcritical bifurcation.
Indeed, the RG flow equation for the gauge coupling in \cite{DiPietro:2015taa} is essentially identical to
the first equation in \eqref{KavehRGflow}.
Although the authors of \cite{DiPietro:2015taa} do not write the $\beta$-function for the 4-fermion couplings
\be
\cO_1 = \left(\sum_{a=1}^{N_f} \bar{\Psi}_a \gamma_\mu \Psi^a \right)^2,\;\;\;\; \cO_2 = 6 \left(\sum_{a=1}^{N_f} \bar{\Psi}_a \gamma_{\mu \nu \rho} \Psi^a\right)^2
\ee
%$\left( \bar \Psi_a \gamma_{\mu} \Psi^a \right)^2$ and $\left( \bar \Psi_a \gamma_{[\mu} \gamma_{\nu} \gamma_{\rho]} \Psi^a \right)^2$,
they compute the matrix of anomalous dimensions at the fixed point with $e^2 \ne 0$ and $\lambda_1 = \lambda_2 = 0$.
By adding the classical contributions, one finds the eigenvalues of the stability matrix at the conformal IR fixed point:
\be
d - \Delta_{\text{4-fermi}} \; = \; - \frac{1}{2 N_f} \left( 4N_f +1 \pm 2\sqrt{N_f^2+N_f+25}\right)
\ee
In particular, as noted in \cite{DiPietro:2015taa}, one eigenvalue crosses through marginality at certain value
$N_f^{\text{crit}} = \frac{-1 + \sqrt{298}}{6} \approx 2.71$.
The quantitative estimate for $N_f^{\text{crit}}$ is not as important as the qualitative fact that the crossing is linear in $N_f$,
{\it cf.} Figure~\ref{fig:linvssqrt} (left):
\be
d - \Delta_{\text{4-fermi}} \; \approx \; 0.54  \left( N_f^{\text{crit}} - N_f \right)
\label{linNfQED}
\ee
This implies that the transition is described by a transcritical bifurcation,\footnote{From
Table~\ref{tab:Jacobian} we know that such behavior can be also characteristic of a pitchfork bifurcation.
However, the latter requires cubic $\beta$-functions, whereas the discussion here and in \cite{DiPietro:2015taa}
is only at the level of quadratic terms. So, it can not be a pitchfork bifurcation.}
not the saddle-node bifurcation (a.k.a. fixed point merger / annihilation)!

Again, just like in our earlier discussion, we conclude that this leading order analysis must be {\it qualitatively}
modified at strong coupling, thereby transforming --- or, to use proper terminology, ``unfolding'' --- the transcritical bifurcation
at $N_f^{\text{crit}}$, as illustrated in the lower panels of Figure~\ref{fig:deformed},
and transforming the linear behavior \eqref{linNfQED} into a ``square root law'' \eqref{sqrtMC}
or perhaps even into a more surprising ``quadratic behavior'' \eqref{quadMC}.

The scenario where QED$_3$ fixed point annihilates with another fixed point (QED$_3^*$)
in a merger was advocated in several recent papers, {\it e.g.} in \cite{Giombi:2015haa} using F-theorem
combined with the $\epsilon$-expansion and in \cite{Herbut:2016ide} using another variant of the $\epsilon$-expansion approach.
In fact, one of these studies, namely \cite{Giombi:2015haa}, points out that it can not distinguish between
what we call the saddle-node and the transcritical bifurcation, and poses this as a question.
Here we propose an answer based on bifurcation theory and argue that in QED$_3$
the saddle-node bifurcation always prevails over the transcritical bifurcation.

In order to gain further insight into unfolding of the transcritical bifurcation in QED$_3$,
it is natural to explore a larger class of theories.
Since until so far we restricted our attention to theories with $N_f$ 4-component Dirac fermions,
one such generalization is to allow an arbitrary number, $N_f^{(2)}$, of 2-component Dirac fermions.
Another generalization is to introduce a level-$k$ Chern-Simons term for the gauge field
which, in effect, introduces another control parameter.
These two generalizations are closely related; when $N_f^{(2)}$ is odd,
the ``parity'' anomaly \cite{Redlich:1983dv} requires the Chern-Simons coupling
to be non-zero, which in general has to satisfy
\be
k - \frac{N_f^{(2)}}{2} \; \in \; \Z
\ee
Yet another generalization and another control parameter could be introduced
by considering theories with $U(N_c)$ or $SU(N_c)$ gauge groups, {\it i.e.} variants of QCD$_3$ instead of QED$_3$.
For simplicity, here we discuss only abelian theories.

A generalization of QED$_3$ with a level-$k$ Chern-Simons term and an arbitrary number of 2-component fermions
recently received a lot of attention, in part due to proposed infra-red CFTs
at small values of $N_f^{(2)}$ \cite{Wang:2015qmt,Metlitski:2015eka,Mross:2015idy}
and dualities between them~\cite{Aharony:2015mjs,Karch:2016sxi,Seiberg:2016gmd,Hsin:2016blu}.
Note, the existence of IR fixed points at small number of fermion flavors does not necessarily contradict
a conformal phase transition at higher values of $N_f^{(2)}$; in recent work \cite{Roscher:2016wox}
it was attributed to the fact that theories with small values of $N_f^{(2)}$ have fewer 4-fermion interaction
channels and, therefore, less ``room'' for breaking conformal symmetry in the IR.

{}From the bifurcation theory perspective, the infra-red CFTs at small values of $N_f^{(2)}$ could be
on the same branch of fixed points as the family of weakly coupled CFTs at large $N_f^{(2)}$
(see the lower left panel of Figure~\ref{fig:deformed}),
or they could be on two different branches of fixed points separated by a ``gap''
(as in the lower right panel of Figure~\ref{fig:deformed}).
These two ways of unfolding the transcritical bifurcation correspond to two characteristic types of behavior
of scaling dimensions and, with sufficient level of precision, can be tested either numerically or experimentally.
Namely, the second option can be tested by fitting scaling dimensions to the curve \eqref{sqrtMC};
plus, an integer value of $N_f^{(2)}$ might fall into the ``gap''.
And, in the first option, the scaling dimensions of nearly marginal operators
should exhibit the quadratic behavior \eqref{quadMC} near $N_{f,\text{crit}}^{(2)}$:
\be
|\Delta - d| \; \sim \; \delta + \left( N_f^{(2)} - N_{f,\text{crit}}^{(2)} \right)^2
\ee
with $\delta > 0$.
Indeed, since larger values of $|k|$ tend to increase scaling dimensions (see {\it e.g.} \cite{Klebanov:2011td} for a clear illustration),
it is conceivable that conformality is never lost for $|k|>0$ and any number of fermion flavors
simply because none of the 4-fermi operators crosses through marginality when  $|k|>0$.
In that case, asking for the critical number of flavors is not even the right question and
one should instead focus on the behavior of scaling dimensions.\footnote{In the approach based
on Schwinger-Dyson equations, this may be similar to the scenario in~\cite{Mavromatos:1999jf}.}

One can also use the conjectured dualities to find the scaling dimensions $\Delta$ of the 4-fermion
interactions for various values of $k$ and $N_f^{(2)}$.
For example, a version of QFD$_3$ with the smallest non-trivial $N_f^{(2)}$ --- sometimes called $U(1)_{-1/2}$ theory
or, counting 4-component fermions as in the above discussion, ``QED$_3$ with $N_f = \frac{1}{2}$'' \cite{Roscher:2016wox} ---
has been conjectured to be IR-dual to the critical boson (a.k.a. the $O(2)$ Wilson-Fisher fixed point).
If this duality maps ``$|\phi|^4$'' operator of the bosonic theory to the scalar operator ``$(\bar \psi \psi)^2$'' in the fermionic theory,
then the latter should have the scaling dimension $\Delta \approx 3.8$ when $k = -\frac{1}{2}$ and $N_f^{(2)} = 1$.
Gradually increasing complexity, the next theory with $N_f^{(2)} = 2$ and $k=-1$ is also conformal;
not only it preserves chiral symmetry, it was conjectured that chiral flavor symmetry of this theory
is actually enhanced in the IR to the $SU(2) \times SU(2) = Spin (4)$ symmetry, see {\it e.g.} \cite{Karch:2016sxi,Hsin:2016blu}.
While at present $\Delta ((\bar \psi \psi)^2)$ is not known in this theory, the enhanced quantum symmetry can impose
tight constraints on its value in the conformal bootstrap approach as well as in other methods.
Assuming this self-dual theory is precisely at the critical value $N_{f,\text{crit}}^{(2)} = 2$
where the curve \eqref{dtransunfold} has a turning point,
it is tempting to propose the following bifurcation-inspired fit for the scaling dimensions:
\be
\Delta_{\text{4-fermi}} - d \; \approx \; \frac{1}{N_f^{(2)}} \sqrt{4 u + \left( N_f^{(2)} - N_{f,\text{crit}}^{(2)} \right)^2}
\ee
where $u$ is some constant (presumably, $u \sim |k|^2 + \text{const}$).
This fit would approximate the behavior of scaling dimensions at small $N_f^{(2)}$
and also at large $N_f^{(2)}$, where $\Delta_{\text{4-fermi}} \simeq 4$.
Note, however, that even and odd values of $N_f^{(2)}$ possibly belong to two different families;
in particular, the former makes sense with $k=0$, while the latter requires $k \ne 0$.

As a generalization in a different direction, it would be interesting to apply
the techniques of bifurcation theory and the Conley index theory to close cousins of QED$_3$,
{\it e.g.} to a version with $N_f$ complex scalar fields charged under $U(1)$ gauge group.
This system, also known as the non-compact $\C {\bf P}^{N_f - 1}$ model,
exhibits a marginality crossing that may describe \cite{Nahum:2015jya}
the quantum phase transition between N\'eel antiferromagnet and valence bond solid (VBS).
In the future work we hope to explore phases of this system with the methods of dynamical systems.

%%%%%%%%%%%%%%%%%%%%%%%%%%%%%%%%%%%%%%%%%%%%%%%%%%%%%%%%%%%%%%%%%%%%%%%%%%%%%%%%%%%%%%%%%%%%%%%%

\section{Application to QCD$_4$}
\label{sec:QCD}

\begin{figure}[t!]
\begin{center}
\includegraphics[width=0.8\textwidth]{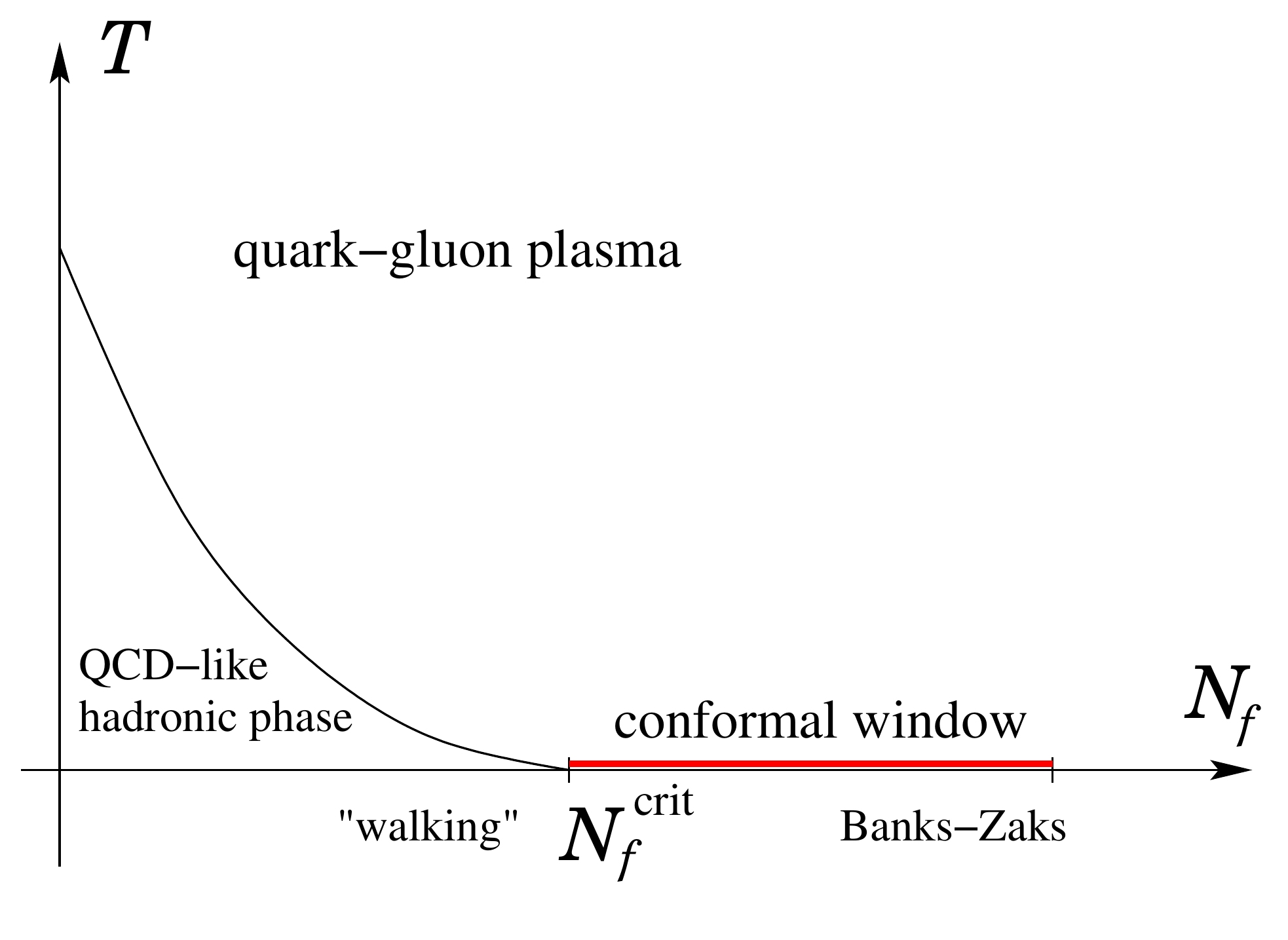}
\end{center}
\caption{\label{fig:QCD} Expected phase diagram of QCD$_4$ as a function of temperature $T$ and the number of flavors $N_f$
(with the number of colors $N_c$ kept fixed).}
\end{figure}

As we come to one of our most interesting examples, the four-dimensional quantum chromodynamics (QCD$_4$),
we encounter a new feature: this theory has {\it two} control parameters, namely the number of colors $N_c$
and the number of flavors $N_f$.

Therefore, codimension-2 bifurcations that in our previous examples could not exist without fine tuning and a symmetry protecting it,
in QCD$_4$ can be generic and require neither fine tuning nor additional symmetries.
Moreover, codimension-1 bifurcations will now appear along lines in the two-dimensional plane of $N_c$ and $N_f$,
where the full structure of RG flows and bifurcations can be much richer than in one-parameter flows.
In particular, mutual intersections of such lines of codimension-1 bifurcations
will result in more interesting codimension-2 bifurcations, {\it etc.}
In order to map out the geography of the $(N_c,N_f)$-plane, it is often convenient consider varying $N_f$
for a fixed number of colors $N_c$. Equivalently, one can vary the ratio
\be
x \; = \; \frac{N_f}{N_c}
\ee
which is particularly useful in the large color and flavor (Veneziano) limit.

Another interesting feature of QCD$_4$ is that the upper end of the conformal window occurs at a finite value of $N_f$.
In particular, in the Veneziano limit, the conformal window looks like
\be
N_f^{\text{crit}} (N_c) \; < \; N_f \; < \; \frac{11}{2} N_c
\ee
or, equivalently, $x_{\text{crit}} < x < \frac{11}{2}$.
This is easy to see already from the perturbative beta-function.
In $SU(N_c)$ gauge theory with $N_f$ quarks in the fundamental representation\footnote{Versions
of the problem with other gauge groups and representations are also interesting,
see {\it e.g.} \cite{DelDebbio:2008zf,DeGrand:2010na,DeGrand:2011qd}
for lattice studies of the theory with quarks in symmetric representations.},
the two-loop beta-function for the gauge coupling $\alpha = \frac{g^2}{(4\pi)^2}$ has the form
\be
\beta_{\alpha} \; = \; \gamma \alpha - b_1 \alpha^2 - b_2 \alpha^3 + \ldots
\ee
where
\begin{eqnarray}
\gamma & = & 0 \nonumber \\
b_1 & = & \frac{2 N_c}{3} (11 - 2 x) \label{bbbQCD4} \\
b_2 & = & \frac{2 N_c^2}{3} \left( 34 - 13 x + \frac{3x}{N_c^2} \right) \nonumber
\end{eqnarray}
expressed in terms of $N_c$ and $x$ (instead of $N_c$ and $N_f$) in order to facilitate applications
to the Veneziano limit as well as to finite values of $N_c$.
Clearly, $\alpha = 0$ is one of the zeros of the beta-function and corresponds to a free UV fixed point when $b_1 > 0$.
In this regime, the theory is asymptotically free and exhibits confining QCD-like behavior when $b_2 > 0$.
On the other hand, for $b_2 < 0$, the RG flow described by this perturbative beta-function
also has an interacting infra-red (Caswell-Banks-Zaks) fixed point \cite{Caswell:1974gg,Banks:1981nn} at
\be
\alpha_* = - \frac{b_1}{b_2} = \frac{1}{N_c} \cdot \frac{11-2x}{ 13x - 34 - \tfrac{3x}{N_c^2} }
\label{BZalpha}
\ee
This fixed point is weakly coupled near the upper edge of the conformal window
and, if \eqref{BZalpha} were also valid at strong coupling, the transition from conformal to confining behavior
would take place where the 2-loop coefficient $b_2 (x)$ changed sign, {\it i.e.} around $x \simeq 2.6$ (for $N_c \ge 10$).
Unfortunately, $\alpha_* (x)$ has a pole there, indicating that the phase transition at the lower edge of the conformal window
is a more interesting strongly-coupled phenomenon.\footnote{Note, a qualitative feature of the perturbative result \eqref{BZalpha}
is that the anomalous dimension of the scalar glueball operator $\text{Tr} (F_{\mu \nu}^2)$,
{\it i.e.} the derivative of the beta-function $\beta_{\alpha}$ at the IR fixed point,
diverges as $N_f$ approaches the critical value from above:
\be
\beta_{\alpha}' \vert_{\alpha = \alpha_*} \; \to \; \infty
\qquad \text{as} \qquad N_f - N_f^{\text{crit}} \to 0^+ \,,
\label{betadiverge}
\ee
a behavior also claimed by a recent lattice study \cite{daSilva:2015vna}.}

The phase transition at the lower edge of the conformal window in QCD$_4$ has been the subject of many analytical and lattice
studies, which lead to a variety of different predictions for the value of $N_f^{\text{crit}}$ and the nature of the phase transition.
For example, the study of Schwinger-Dyson equations
with rainbow (ladder) resummations \cite{Cohen:1988sq,Appelquist:1996dq,Miransky:1996pd,Appelquist:1997fp,Appelquist:1998rb}
suggests that QCD$_4$ is in a hadronic phase (exhibiting confinement and chiral symmetry breaking) below
%chiral symmetry breaking and the dynamical mass generation below $N_f^{\text{crit}}$,
\be
N_f^{\text{crit}} \; = \; N_c \left( \frac{100 N_c^2 - 66}{25 N_c^2 - 15} \right)
\ee
where the order parameter (= the dynamical fermion mass) vanishes continuously as $N_f \to N_f^{\text{crit}}$ from below
and the gauge constant ``walks'' (rather than ``runs'').
The IR spectrum of this hadronic phase is characterized by massless bosonic excitations,
the Nambu-Goldstone bosons associated with the chiral symmetry breaking
\be
SU(N_f)_L \times SU(N_f)_R \; \to \; SU(N_f)_V
\ee
Moreover, both the functional RG approach \cite{Braun:2010qs}
and the holographic models \cite{Alho:2012mh,Jarvinen:2015qaa} for QCD$_4$ in the Veneziano limit (called V-QCD)
indicate that the conformal phase at zero temperature
is continuously connected to the chirally symmetric quark-gluon plasma (QGP) phase at high temperature,
as illustrated in Figure~\ref{fig:QCD}.
While this general picture is in agreement with most lattice simulation (see {\it e.g.} \cite{DelDebbio:2010zz} for a review),
the nature of the phase transition at $N_f^{\text{crit}}$ and the precise value of $N_f^{\text{crit}}$ are much less certain.

For example, Pisarski-Wilczek scenario \cite{Pisarski:1983ms} as well as the jumping scenario \cite{Antipin:2012sm}
might suggest a first order phase transition. Another possibility could be an infinite order BKT-like phase transition
characterized by the exponential Miransky scaling for the dynamical fermion mass ($N_f \le N_f^{\text{crit}}$) \cite{Miransky:1996pd}:
\be
m_{\text{dyn}} \; \sim \; \Lambda e^{- \frac{C}{\sqrt{N_f^{\text{crit}} - N_f}}}
\label{mMiransky}
\ee
It has also been argued \cite{Braun:2010qs,Lombardo:2014mda} that QCD$_4$
is a multi-scale theory when approaching the conformal window from below.
One popular scenario \cite{Gies:2005as,Braun:2006jd} is that the IR stable fixed point in the conformal window of QCD$_4$
annihilates with another fixed point at $N_f^{\text{crit}}$ via what we now can call a saddle-node bifurcation.
We also know from Theorem~\ref{MCtheorem} that this behavior requires an irrelevant operator to cross through
marginality at $N_f^{\text{crit}}$, precisely as anticipated {\it e.g.} in \cite{Appelquist:1997fp,Appelquist:1996dq,Appelquist:1998rb},
so that the instability at the phase transition is triggered by a 4-fermion interaction.
Specifically, for the saddle-node bifurcation, {\it cf.} \eqref{sqrtMC}:
\be
\Delta_{\text{4-fermi}} - d \; \sim \; \sqrt{N_f - N_f^{\text{crit}}}
\ee
and this scenario has also been used in \cite{Kaplan:2009kr} to give further evidence for the Miransky scaling \eqref{mMiransky}.
(Note that, contrary to \eqref{betadiverge}, in this scenario $\beta_{\alpha}'$ remains finite as $N_f$ approaches
the lower end of the conformal window.)

\begin{table}
\begin{centering}
\begin{tabular}{|c|c|l|}
\hline
~$\phantom{\int^{\int^\int}} N_f^{\text{crit}} \phantom{\int_{\int}}$~ & ~Method~ & ~Year and Reference \tabularnewline
\hline
\hline
$\phantom{\int^{\int^\int}} \approx 6 \phantom{\int_{\int}}$ & instanton -- anti-instanton pairs & ~~~~~~1997  \cite{Velkovsky:1997fe} \tabularnewline
\hline
$\phantom{\int^{\int^\int}} \le 7 \phantom{\int_{\int}}$ & lattice simulations & ~~~~~~1991  \cite{Iwasaki:1991mr} \tabularnewline
\hline
$\phantom{\int^{\int^\int}} 6 \le N_f^{\text{crit}} \le 8 \phantom{\int_{\int}}$ & lattice simulations & ~~~~~~2015 \cite{daSilva:2015vna} \tabularnewline
\hline
$\phantom{\int^{\int^\int}} > 8.25 \phantom{\int_{\int}}$ & NSVZ-inspired $\beta$-function & ~~~~~~2007 \cite{Ryttov:2007cx} \tabularnewline
\hline
$\phantom{\int^{\int^\int}} 8 \le N_f^{\text{crit}} \le 12 \phantom{\int_{\int}}$ & lattice simulations & ~~~~~~2007-09 \cite{Appelquist:2007hu,Appelquist:2009ty} \tabularnewline
\hline
$\phantom{\int^{\int^\int}} 8 \le N_f^{\text{crit}} \le 12 \phantom{\int_{\int}}$ & Monte Carlo Renormalization Group (MCRG) & ~~~~~~2009-10 \cite{Hasenfratz:2009ea,Hasenfratz:2010fi} \tabularnewline
\hline
$\phantom{\int^{\int^\int}} 10 \phantom{\int_{\int}}$ & functional RG & ~~~~~~2005 \cite{Gies:2005as} \tabularnewline
\hline
$\phantom{\int^{\int^\int}} 11.58 \phantom{\int_{\int}}$ & 4-loop RG & ~~~~~~2011 \cite{Kusafuka:2011fd} \tabularnewline
\hline
$\phantom{\int^{\int^\int}} < 12 \phantom{\int_{\int}}$ & Highly Improved Staggered Quark (HISQ) action & ~~~~~~2015 \cite{Aoki:2015gea} \tabularnewline
\hline
$\phantom{\int^{\int^\int}} > 12 \phantom{\int_{\int}}$ & staggered lattice fermions & ~~~~~~2011 \cite{Fodor:2011tu} \tabularnewline
\hline
\end{tabular}
\par\end{centering}
\caption{\label{tab:NfQCD} Estimates for $N_f^{\text{crit}}$ in QED$_4$ with $SU(3)$ gauge group ({\it i.e.} $N_c = 3$).}
\end{table}

Anticipating one of the four-fermion operators to cross marginality at $N_f^{\text{crit}}$,
it is natural to consider the Lagrangian for the $SU(N_c)$ gauge theory
with $N_f$ flavors of massless Dirac fermions in the fundamental representation ($j = 1, \ldots, N_f$):
\be
\CL \; = \; - \frac{1}{4 g^2} \text{Tr} F_{\mu \nu} F^{\mu \nu} + i \bar \psi_j D \!\!\!\! \slash \psi^j + \CL_{\text{4-fermi}}
\ee
deformed by the fermion self-interaction terms:
\be
\CL_{\text{4-fermi}} \; = \;
\frac{\lambda_1}{4 \pi^2 \Lambda^2} \CO_1
+ \frac{\lambda_2}{4 \pi^2 \Lambda^2} \CO_2
+ \frac{\lambda_3}{4 \pi^2 \Lambda^2} \CO_3
+ \frac{\lambda_4}{4 \pi^2 \Lambda^2} \CO_4
\label{QCD4fermi}
\ee
where $\Lambda$ is a mass scale introduced to make the couplings $\lambda_i$ dimensionless.
Up to Fierz transformations, there are four independent 4-fermi operators which are invariant under
$SU(N_c)$ gauge symmetry, parity (which acts on fermions as $\psi_L \leftrightarrow \psi_R$),
and $SU(N_f)_L \times SU(N_f)_R$ chiral flavor symmetry \cite{Gies:2003dp,Braun:2005uj,Terao:2007jm}:
\begin{eqnarray} % I doubled all coefficients and V=1, S=2, V1=3, V2=4
\CO_1 & = &
\bar \psi_i \gamma^{\mu} \psi^j \bar \psi_j \gamma_{\mu} \psi^i
+ \bar \psi_i \gamma^{\mu} \gamma_5 \psi^j \bar \psi_j \gamma_{\mu} \gamma_5 \psi^i \nonumber \\
\CO_2 & = &
\bar \psi_i \psi^j \bar \psi_j \psi^i - \bar \psi_i \gamma_5 \psi^j \bar \psi_j \gamma_5 \psi^i \label{QCD4fermop} \\
\CO_3 & = &
(\bar \psi_i \gamma^{\mu} \psi^i)^2 - (\bar \psi_i \gamma^{\mu} \gamma_5 \psi^i)^2 \nonumber \\
\CO_4 & = &
(\bar \psi_i \gamma^{\mu} \psi^i)^2 + (\bar \psi_i \gamma^{\mu} \gamma_5 \psi^i)^2 \nonumber
\end{eqnarray}
Together with the gauge coupling $\alpha$, the RG flow equations in this theory define a dynamical system in the five-dimensional phase space,
whose analysis requires powerful tools such as the Conley index discussed in section~\ref{sec:Conley}.

As a toy model, consider for example the RG flow equations of \cite{Kusafuka:2011fd}:
\begin{eqnarray}
\dot \alpha    & = & -\frac{2}{3}(11 - 2x) \alpha^2 - \frac{2}{3}(34 - 13x) \alpha^3 + 2x \alpha^2 \lambda_1 \nonumber \\
\dot \lambda_1 & = & 2 \lambda_1 + (1 + x) \lambda_1^2 + \frac{x}{4} \lambda_2^2 - \frac{3}{4} \alpha^2 \label{VQCD3flow} \\
\dot \lambda_2 & = & 2 \lambda_2 - 2 \lambda_2^2 + 2x \lambda_1 \lambda_2 - 6 \alpha \lambda_2 - \frac{9}{2} \alpha^2 \nonumber
\end{eqnarray}
which were proposed to describe the Veneziano limit of the five-coupling system that governs the RG flow
in QCD$_4$ with the four-fermi interactions~\eqref{QCD4fermi}.
As usual, in the large $N_c$ limit we rescaled $\alpha \to \frac{1}{N_c} \alpha$
(so that the new coupling $\alpha = \frac{g^2 N_c}{(4\pi)^2}$ is the standard 't Hooft coupling),
$\lambda_{1,2} \to \frac{1}{N_c} \lambda_{1,2}$ in the vector and scalar channels with non-trivial flavor structure,
and $\lambda_{3,4} \to \frac{1}{N_c^2} \lambda_{3,4}$ for the color and flavor singlets.
The RG flow equations for the latter decouple in the Veneziano limit,
thus resulting in a simpler system \eqref{VQCD3flow} that depends on a single control parameter~$x$.

\begin{figure}[h]
\centering
\begin{minipage}{0.45\textwidth}
\centering
\includegraphics[width=0.85\textwidth, height=0.24\textheight]{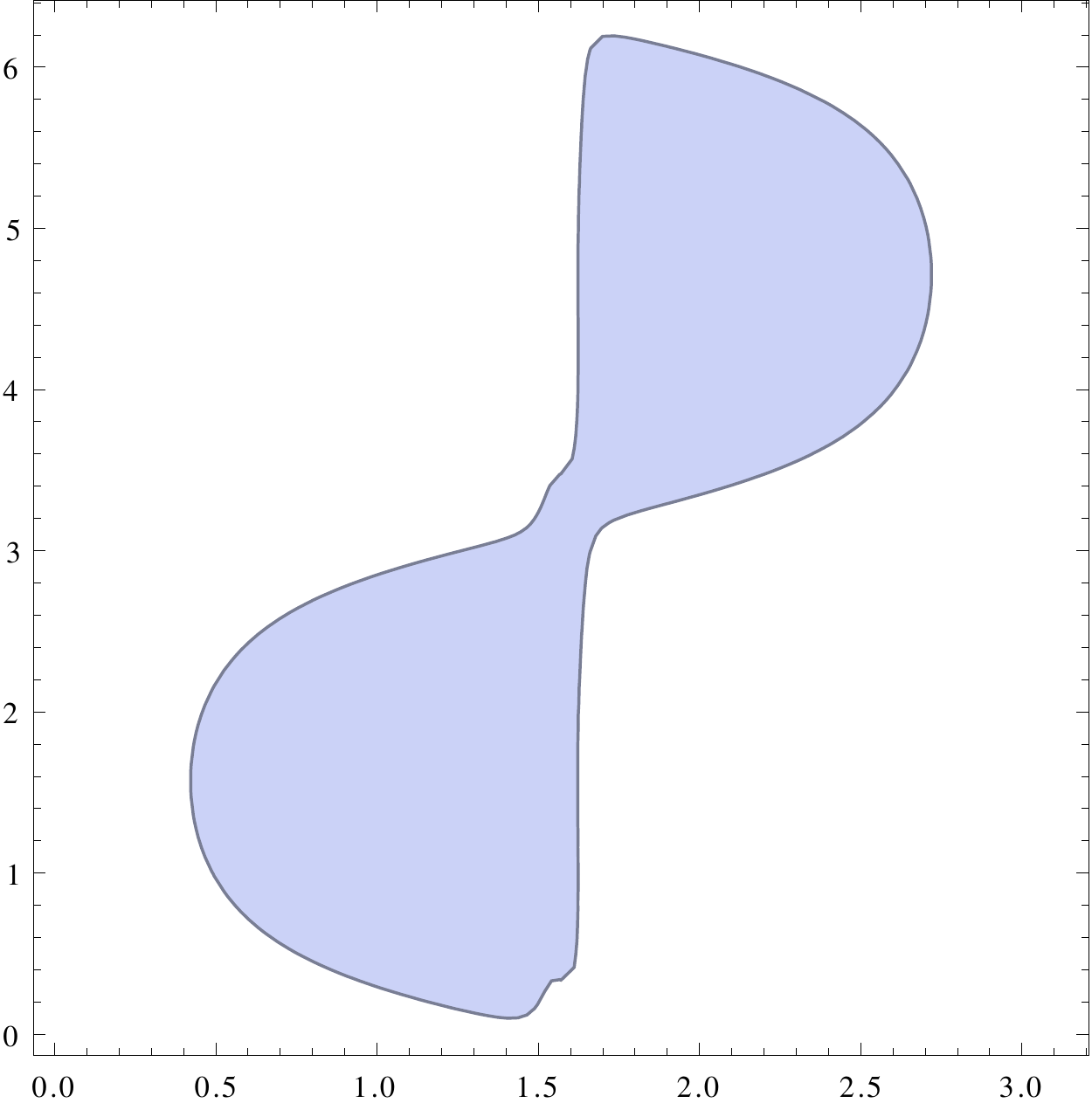}
%\caption{Comment.}
%\label{fig:smallx}
\end{minipage}
\qquad
\begin{minipage}{0.45\textwidth}
\centering
\includegraphics[width=0.9\textwidth, height=0.26\textheight]{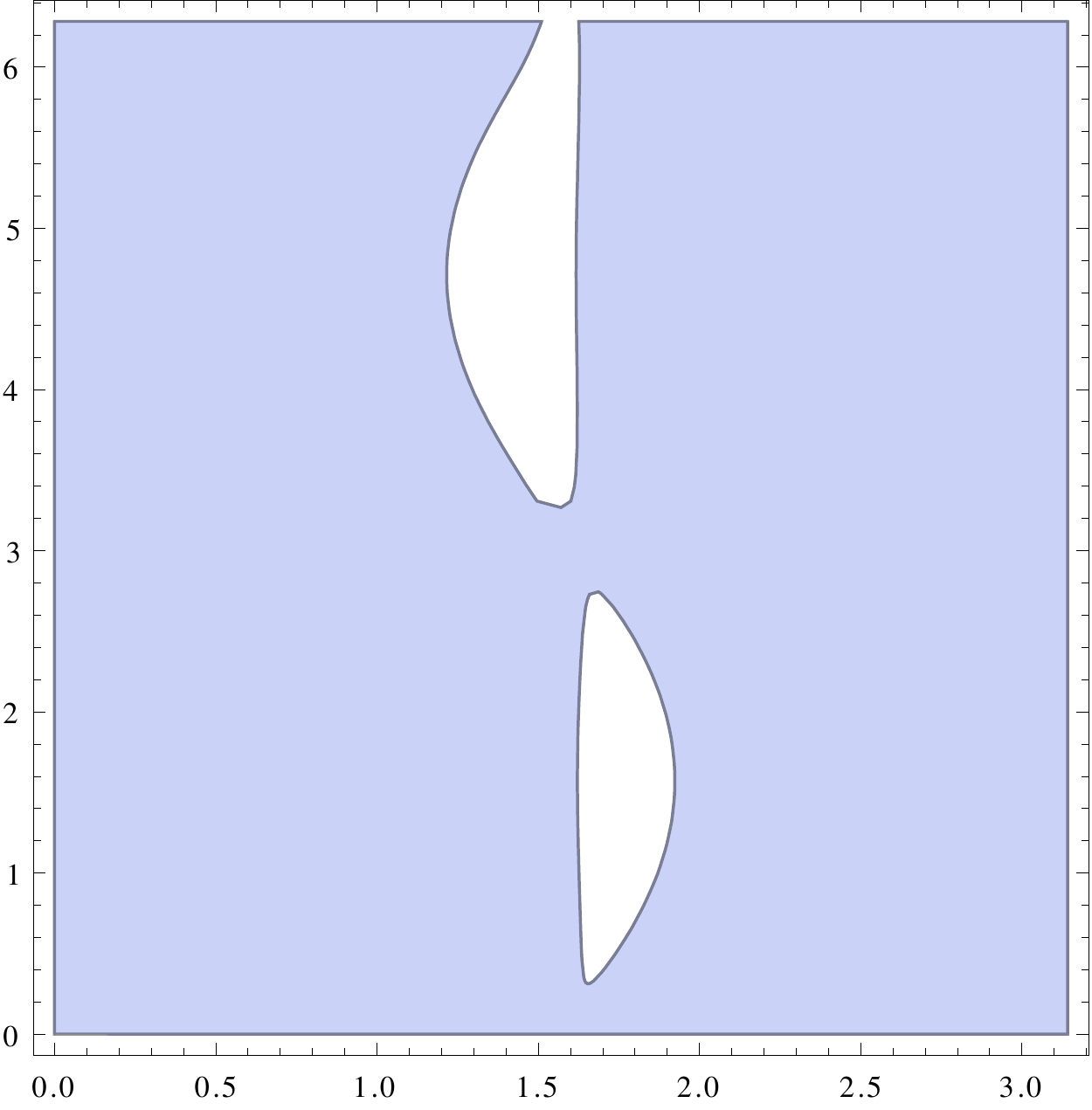}
%\caption{Comment.}
%\label{fig:largex}
\end{minipage}
\caption{A plot of the exist set $L$ in the boundary sphere, $\partial N \cong S^2$,
for the 3-coupling flow \eqref{VQCD3flow} at two different values of the control parameter $x$.
We parametrize the boundary sphere $S^2$ by angles $(\theta, \varphi)$, such that $\theta \in [0, \pi]$ and $\varphi \in [0, 2 \pi)$,
and shown here is the $(\theta, \varphi)$ atlas.
For sufficiently small values of $x$ (smaller than $x \approx 4.7$) the exit set is homeomorphic to
a two-dimensional disk shown on the left panel, whereas for larger values of $x$ (larger than $x \approx 4.7$)
the exit set is homeomorphic to an annulus ($\cong$ sphere with two punctures) shown on the right panel.}
\label{fig:QCDexitset}
\end{figure}

However, even with all of the simplifying assumptions that went into \eqref{VQCD3flow},
it is not easy to find the exact fixed points and all the bifurcations of this system directly.
In Figure~\ref{fig:QCDexitset} we illustrate how the Conley index theory can help with this task.
Specifically, as in \eqref{Lexitvianb}, we construct the exit set $L \subset S^2$ as a set of points
in the boundary of our coupling space, $N \cong D^3$, where $\vec \beta \cdot \vec n$ is positive.
It is curious to note that, in the interesting range of control parameters, $L$ undergoes a topology changing
transition near $x_0 \approx 4.7$:
\be
L \; = \;
\begin{cases}
D^2 \,, & \text{if}~x < x_0 \ldots \\
S^1 \times I \,, & \text{if}~x \ge x_0 \ldots
\end{cases}
\ee
This leads to the change of the homological Conley index $CH_* (S) = H_* (N/L , [L])$:
\be
CH_* (S) \; = \;
\begin{cases}
0 \,, & \text{if}~x < x_0 \ldots \\
\Z [0] \oplus \Z [2]  \,, & \text{if}~x \ge x_0 \ldots
\end{cases}
\ee
indicating that the isolated invariant set $S$ changes, {\it i.e.} some fixed points enter or exit $N$.
In order to get further insight into the structure of $S$, we note that the first equation in \eqref{VQCD3flow},
namely the beta-function for $\alpha$, has a double zero at $\alpha=0$ and a non-trivial solution at
\be
\alpha_* \; = \; \frac{11 - 2 x - 3 x \lambda_1}{13 x - 34}
\ee
which is basically a slight modification of the familiar expression \eqref{BZalpha}.
By looking at the derivative of the beta-function for $\alpha$, it is easy to see that $\alpha_*$ is stable (attractive in the IR)
provided that $\lambda_1$ is sufficiently small at each of the fixed points (as can be verified a posteriori)
and for $x$ in the interesting range, say $3<x<11/2$.
For the purpose of analyzing the invariant set $S$ in the system \eqref{VQCD3flow}, it effectively means that
the problem can be reduced to a two-coupling flow of $(\lambda_1, \lambda_2)$ with $\alpha = \alpha_*$.
This two-coupling system has $N \cong D^2$ and its exit set consists of only one component for all $4<x<11/2$,
indicating that fixed points with $\alpha = \alpha_*$ have $CH_* (S) = 0$.
Furthermore, this ``reduced'' two-coupling flow has a saddle-node bifurcation near $x_{\text{crit}} \simeq 4.05$,
which is basically the result of \cite{Kusafuka:2011fd} re-derived here with the help of the Conley index theory
and the bifurcation theory.

It would be interesting to extend this bifurcation analysis to the entire 5-coupling RG flow of $(\alpha, \lambda_1, \ldots, \lambda_4)$.
Regarded as a family of flows with two control parameters $N_f$ and $N_c$, it is likely to exhibit more interesting types of
bifurcations that we saw in section~\ref{sec:bifurcations}.
We plan to pursue a more detailed study of this interesting possibility in future work.
It would be also interesting to try bifurcation analysis on more general 4d gauge theories
that include scalar fields ({\it e.g.} as in \cite{Rhedin:1998gx}) and matter in other representations of the gauge group.

%%%%%%%%%%%%%%%%%%%%%%%%%%%%%%%%%%%%%%%%%%%%%%%%%%%%%%%%%%%%%%%%%%%%%%%%%%%%%%%

\section{Epilogue: $C$-function and resurgence}
\label{sec:resurgence}

In conclusion we wish to return to some of the questions that motivated our journey.
As we now know, marginality crossing always happens for a reason and usually signals a bifurcation.
For example, in a family of RG flows labeled by $x$,
a marginality crossing in the IR theory $T_{\text{IR}} (x)$ at some value of the parameter $x = x_{\text{crit}}$
often indicates the existence of a nearby fixed point and a local bifurcation listed in Table~\ref{tab:Jacobian}.
A more interesting type of behavior %that motivated our study
occurs when marginality crossing happens
``along the flow'' and does not involve collision of fixed points: in such cases, a violation of \eqref{muuvir}
is a signal of a global bifurcation discussed in section~\ref{sec:MCcrossing} and illustrated in Figure~\ref{fig:MCTTT}.
Both types of behavior can be identified with the help of the Conley index, $\mu$-index \eqref{mudef}, and other quantities.

One of the physically important quantities is the $C$-function, whose behavior along the RG flow was a part of our motivation.
When the strongest form of the $C$-theorem holds, we deal with the steepest descent (gradient) flows
\be
\dot \lambda^i \; = \; - g^{ij} (\lambda) \, \frac{\partial C (\lambda)}{\partial \lambda^j}
\label{gradflow}
\ee
which, in particular, require a positive-definite metric $g_{ij}$ on the space of couplings
and a definition of the function $C(\lambda)$ away from the fixed points.
Under these conditions, a heteroclinic saddle bifurcation illustrated in Figure~\ref{fig:MCTTTpert}
is also known as the Stokes phenomenon \cite{MR1001456}.
It represents a ``phase transition'' under which the RG flow changes topology, while the metric $g_{ij}$
remains positive and non-degenerate throughout the entire transition.

In many of the interesting RG flows, however, we do not have the full access to the function $C(\lambda)$
away from the starting point $T_{\text{UV}}$ and the end-point $T_{\text{IR}}$.
In fact, even the value $C_{\text{IR}}$ at the IR fixed point is often inaccessible,
unless one can use supersymmetry, or expansion in a control parameter $x$, or similar tricks.
Suppose, one of such methods (or combination thereof) provides us with an expression for $C_{\text{IR}}$.
Can this information about the IR value of the $C$-function alone say something about bifurcations?

The answer turns out to be ``yes'', at least when we deal with a family of RG flows
and can say something about $C_{\text{IR}} (x)$ as a function of $x$.
As in \cite{Gukov:2015qea}, the basic idea is that a non-analytic behavior of $C_{\text{IR}} (x)$ is a signal for bifurcations.
For example, if $\beta (\lambda)$ is a gradient flow of the form \eqref{gradflow},
then a heteroclinic saddle bifurcation illustrated in Figure~\ref{fig:MCTTTpert}
will cause $C_{\text{IR}} (x)$ to ``jump'' as the steepest descent trajectory hits another critical point of $C (\lambda)$.
After all, the IR end-point of the flow trajectory starting
at a given UV theory (denoted by $T_1$ in Figure~\ref{fig:MCTTTpert})
is very different before and after the bifurcation.
Similarly, other types of bifurcations lead to different types of non-analytic behavior in $C_{\text{IR}} (x)$.
For example, at a transcritical bifurcation $C_{\text{IR}} (x)$ itself is continuous, but its derivative in
general is not, because $T_{\text{IR}} (x)$ goes to a different branch at $x = x_{\text{crit}}$.

In order to explain this in more detail consider, for example, three-dimensional theories with $\CN=2$ supersymmetry,
where the value of the $C$-function (usually called $F$) at the IR fixed point (but not throughout the flow)
can be determined \cite{Jafferis:2010un} by locally maximizing the free energy $F = - \log |Z|$,
\be
\partial_{\Delta} \log |Z| \; = \; 0 \,,
\label{Zmin}
\ee
where $Z$ is the 3-sphere partition function.
The latter, in turn, can be reduced to a matrix model by means of supersymmetric localization.
For example, in the $\CN=2$ SQCD with gauge group $SU(2)$ and $N_f$ fundamental flavors,
it is given by a single, one-dimensional integral
\be
Z \; = \; \int_{- \infty}^{+ \infty} dz \; \sinh^2 (2\pi z) \;
e^{ N_f [\ell (1 - \Delta + iz) + \ell (1 - \Delta - iz)] }
\label{3dN2SQCD}
\ee
where
\begin{eqnarray}
\ell (z) & = & - z \log \left( 1 - e^{2\pi i z} \right)
+ \frac{i}{2} \left( \pi z^2 + \frac{1}{\pi} \text{Li}_2 (e^{2\pi i z}) \right)
- \frac{i\pi}{12} \label{lfunction} \\
\ell' (z) & = & - \pi z \cot (\pi z) \nonumber
\end{eqnarray}
Evaluating the integral \eqref{3dN2SQCD} in the large $N_f$ limit,
the extremization \eqref{Zmin} leads to the $\frac{1}{N_f}$-expansion of the partition function,
\be
Z \; \simeq \; N_f^{-3/2} \; e^{- N_f \log 2} \; \sum_{n=0}^{\infty} a_n N_f^{-n}
\qquad \text{as} \quad N_f \to \infty
\label{3DN2pert}
\ee
where the ``perturbative'' coefficients $a_n$ can be computed numerically to the desired accuracy,
see {\it e.g.} \cite{Klebanov:2011td}.

Similarly, in the case of $\CN=4$ SQED with $N_f$ charged ``flavors'' the partition function
is also given by a single integral, which can be evaluated explicitly and does not require
the extremization \eqref{Zmin} (since $\Delta$ is fixed by the $\CN=4$ supersymmetry):
\begin{eqnarray}
Z & = & \frac{1}{2^{N_f}} \int_{- \infty}^{+ \infty} \frac{dz}{\cosh^{N_f} (\pi z)}
\; = \; \frac{2^{-N_f} \Gamma (\tfrac{N_f}{2})}{\sqrt{\pi} \Gamma (\tfrac{N_f + 1}{2})} \label{3dN4SQED} \\
& \simeq & \frac{1}{\sqrt{N_f}} \; e^{- N_f \log 2} \; \sum_{n=0}^{\infty} a_n N_f^{-n}
\qquad \text{as} \quad N_f \to \infty
\nonumber
\end{eqnarray}
Note, both \eqref{3DN2pert} and \eqref{3dN4SQED} have
the form of the perturbative expansion of the partition function in complex Chern-Simons TQFT,
where $N_f$ plays the role of the Chern-Simons level $k$
or, equivalently, $\hbar = \frac{2\pi i}{N_f}$ is the usual perturbative expansion parameter
(see {\it e.g.} \cite{Gukov:2016njj} for recent work and references therein).
This connection is actually not too surprising in view of the 3d-3d correspondence;
in fact, both of our examples are particular limits of the so-called ``Lens space theory''
related to the equivariant Verlinde formula~\cite{Gukov:2015sna}.

Not only in these examples (which we use for concreteness), but also more generally,
there are many parallels between $\hbar$-expansion in complex Chern-Simons theory and
$\frac{1}{N_f}$-expansion in 3d $\CN=2$ gauge theories with many flavors.
In particular, as we illustrate next, many salient features of the resurgent analysis in complex
Chern-Simons theory \cite{Gukov:2016njj} carry over directly to the $\frac{1}{N_f}$-expansion
of the partition function and $F_{\text{IR}}$ in 3d $\CN=2$ theories.
Specifically, starting with the asymptotic series like \eqref{3DN2pert} or \eqref{3dN4SQED},
\be
Z_0 \; = \; e^{- N_f S_0} \sum_{n=0}^{\infty} a_n \, N_f^{-n-\delta}
\qquad \text{as} \quad N_f \to \infty
\label{Zpert}
\ee
we expect it to be completed by resurgence (Borel resummation)
to the exact partition function\footnote{In fact, this is a slightly simplified form of
the more general expression given in eq. (2.24) of \cite{Gukov:2016njj}, which will not be needed here.}
(that makes sense even for complex values of $N_f$):
\be
Z \; = \; e^{- F_{\text{IR}}}
\; = \; \sum_{\alpha} n_{\alpha} e^{- N_f S_{\alpha}} \sum_{n=0}^{\infty} a_n^{\alpha} \, N_f^{-n-d_{\alpha}}
\label{Ztransseries}
\ee
where $\alpha=0$ labels the original contribution \eqref{Zpert} (with $d_0 \equiv \delta$, {\it etc.}).
The {\it transseries parameters} $n_{\alpha}$ are piecewise constant functions of $\theta = \text{arg} (N_f)$
and experience jumps along the Stokes lines, see {\it e.g.} \cite{mm-lectures,Dorigoni:2014hea} for a nice introduction.
In addition, different branches of solutions to \eqref{Zmin} may cross.
In general, this crossing happens when there are two (or more) solutions to \eqref{Zmin}, $\Delta_1 (x)$ and $\Delta_2 (x)$,
such that at $x = x_{\text{crit}}$:
\be
\text{Re} \; F(\Delta_1) \; = \; \text{Re} \; F(\Delta_2)
\qquad \text{at} \quad x = x_{\text{crit}}
\ee
If $\Delta_1 (x_{\text{crit}}) = \Delta_2 (x_{\text{crit}})$, then we are dealing with a local bifurcation,
otherwise it is sign of a global bifurcation.

Comparing \eqref{Zpert} to \eqref{3DN2pert} and \eqref{3dN4SQED}, it is clear that $S_0 = \log 2$ in both of these examples.
Then, the ``action'' $S_{\alpha}$ of the next transseries in \eqref{Ztransseries},
{\it i.e.} the one with the smallest absolute value of $A = S_{\alpha} - S_0$,
can be determined by the growth rate of the ``perturbative'' coefficients $a_n$,
which for a Gevrey order-1 asymptotic series like \eqref{3DN2pert} or \eqref{3dN4SQED} is expected to be
\be
|a_n| \; \sim \; \frac{\Gamma (n+\delta)}{|A|^n}
\qquad \text{as} \quad n \to \infty
\label{angrowth}
\ee
For example, it is easy to verify numerically that the asymptotic series in \eqref{3dN4SQED}, where $\delta = \frac{1}{2}$,
indeed has this expected behavior with $\log \frac{1}{|A|} \approx - 1.14$.
This matches perfectly the exact value $A = i \pi$.
Indeed, both integrals \eqref{3dN2SQCD} and \eqref{3dN4SQED} have the form
\be
Z \; = \; \int_{- \infty}^{+ \infty} dz \; e^{ - N_f V(z,N_f)}
\label{ZviaV}
\ee
where $V(z,N_f) = \log 2 + \log \cosh (\pi z)$ in the case of 3d $\CN=4$ SQED.
Critical points of this ``potential'', {\it i.e.} the saddle points of the integral \eqref{ZviaV},
are located at $iz_{\alpha} \in \Z$ and yield
\be
S_{\alpha} - S_0 \; = \;
\begin{cases}
i \pi \,, & \text{if}~iz_{\alpha} \in 2\Z + 1 ~~\text{(odd)} \\
0 \,, & \text{if}~iz_{\alpha} \in 2\Z ~~\text{(even)}
\end{cases}
\label{SSN4}
\ee
Even though the potential is slightly more complicated for $\CN=2$ SQCD \eqref{3dN2SQCD},
\be
V \; = \; - \ell (1 - \Delta + iz) - \ell (1 - \Delta - iz) + \CO (\tfrac{1}{N_f})
\ee
in the large $N_f$ limit it also has critical points at $z_{\alpha} = i \alpha$, $\alpha \in \Z$,
and the extremization \eqref{Zmin} gives $\Delta = \frac{1}{2}$ in this limit.
The saddle point at $z=0$ is what gives rise to the ``perturbative'' $\frac{1}{N_f}$-expansion \eqref{3DN2pert},
whereas the other saddle points of the integral \eqref{ZviaV} have the ``instanton action'':
\be
S_{\alpha} - S_0 \; = \; - i \pi \alpha^2
\qquad\qquad (z_{\alpha} = i \alpha,~\alpha \in \Z)
\label{SSN2}
\ee
Since the instanton with the smallest absolute value of $A = S_{\alpha} - S_0$ has $|A| = \pi$, from \eqref{angrowth}
we predict the asymptotic behavior of the coefficients $a_n$ in \eqref{3DN2pert}:
\be
|a_n| \; \sim \; \frac{\Gamma (n+\tfrac{3}{2})}{\pi^n}
\qquad \text{as} \quad n \to \infty
\ee
where we also used the fact that \eqref{3DN2pert} has $\delta = \frac{3}{2}$ in the conventions of \eqref{Zpert}.
It would be interesting to test this prediction either numerically,
by computing the coefficients $a_n$ order by order in \eqref{3dN2SQCD}, or analytically.

Notice many close parallels with the volume conjecture and resurgence in complex Chern-Simons theory \cite{Gukov:2016njj}.
Namely, just like in Chern-Simons theory\footnote{The instanton action $A$ corresponds to $- 2\pi i \ell_*$ in \cite{Gukov:2016njj}.
In particular, $\ell_*$ is real-valued for $SU(2)$ or $SL(2,\R)$ flat connections on $M_3$.
When $\ell_*$ has non-zero imaginary part, {\it i.e.} when $S_{\alpha} - S_0$ has non-trivial real part,
one needs to replace a crude estimate \eqref{angrowth} with a more refined analysis, as in section 5.3 of \cite{Gukov:2016njj}.}
on a Seifert manifold $M_3$, the instanton factor $e^{- N_f (S_{\alpha} - S_0)}$
is a ``pure phase'' in our examples \eqref{SSN4} and \eqref{SSN2},
both of which are ``non-chiral'' (in the terminology of \cite{Klebanov:2011td}).
Moreover, the tower of critical points $z_{\alpha}$ on the imaginary axis is analogous to lifting the flat connections
on $M_3$ to the universal cover, so that for integer values of $N_f$ (that plays the role of ``level'' in Chern-Simons TQFT)
these lifts give the same result.
In particular, summing over these contributions often requires a regularization that, when done carefully,
leads to a more refined version of \eqref{angrowth} which no longer requires the absolute value on $a_n$
and can even give the subleading asymptotics.
But when $N_f$ is analytically continued to non-integer (complex) values,
these lifts give distinct contributions to \eqref{Ztransseries} and can be easily ``seen''.

What we described so far is only a small part of the powerful arsenal of the resurgent analysis,
namely the part which has to do with the first transseries or the first singularity near the origin of the Borel plane.
In order to get a full picture about the analytic structure of the $C$-function $F_{\text{IR}}$
as a function of $N_f$, one needs to produce the full portrait of singularities in the Borel plane.
In practice, this means constructing (the analytic continuation of) the Borel transform from
the ``perturbative'' coefficients $a_n$ of the power series \eqref{Zpert}:
\be
\tilde{BZ}_0 (\xi) \; = \; \sum_{n=0}^{\infty} \frac{a_n}{\Gamma (n+\delta)} \xi^{n+\delta-1}
\ee
In our examples, this function is expected to have poles along the imaginary axis in the Borel $\xi$-plane,
whose locations $\xi_{\alpha} = S_{\alpha} - S_0$ are given in \eqref{SSN4} and \eqref{SSN2}, respectively
(and whose residues we did not compute).
Then, the exact partition function \eqref{Ztransseries} can be recovered from the directional Borel resummation,
\be
\CS_{\theta} Z_0 \; = \; \int_{e^{-i \theta} \R_+} d \xi \, \tilde{BZ}_0 (\xi) \, e^{- \xi N_f}
\ee
What is the structure of singularities in the Borel plane for 3d $\CN=2$ SQCD?
That could be a good subject for another paper.

%%%%%%%%%%%%%%%%%%%%%%%%%%%%%%%%%%%%%%%%%%%%%%%%%%%%%%%%%%%%%%%%%%%%%%%%%%%%%%%%%%%%%%%%%%%%%%%%%%%%%%%%%%%%%%%%%%%%%

\section*{Acknowledgements}

It is a pleasure to thank O.~Aharony, D.~Gross, E.~Kiritsis, I.~Klebanov, N.~Nekrasov, V.~Rychkov, N.~Seiberg,
R.~Shrock, D.~Sullivan, R.~Sundrum, G.~Torroba, N.~Warner, R.~Wijewardhana and E.~Witten for useful discussions and comments,
and V.~Lysov for collaboration during the early stages of this project
%for his comments on the manuscript
and for his help with Figures~\ref{fig:Type I flow}-\ref{fig:Type IV flow}.
We also thank the anonymous referee for many insightful comments
and gratefully acknowledge the warm hospitality of SCGP during the 2016 summer workshop,
as well as participants and organizers of the GGI workshop (May 23 - July 8) and ``Strings 2016'' conference (August 1-5)
where preliminary results of this work were presented.
This material is based upon work supported by the U.S. Department of Energy, Office of Science,
Office of High Energy Physics, under Award Number DE-SC0011632.
This work is also supported in part by the ERC Starting Grant no. 335739 ``Quantum fields and knot homologies''
funded by the European Research Council under the European Union Seventh Framework Programme.

%%%%%%%%%%%%%%%%%%%%%%%%%%%%%%%%%%%%%%%%%%%%%%%%%%%%%%%%%%%%%%%%%%%%%%%%%%%%%%%%%%%%%%%%%%%%%%%%%%%%%%%%%%%%%%%%%%%%%

%\clearpage
\bibliography{RGflows}{}
\bibliographystyle{utphys}

\end{document}